\newcommand\surveyArea{13,211}
\newcommand\compLimitFull{3.8}
\newcommand\DESOverlapArea{4566}

\newcommand\HSCOverlapArea{469}

\newcommand\KiDSOverlapArea{825}

\newcommand\DECaLSOverlapArea{10,822} 

\newcommand\totalCandidates{8878}
\newcommand\totalFixedSNRCandidates{7407}
\newcommand\totalConfirmed{4195}

\newcommand\totalNew{868}
\newcommand\totalHighZ{222}

\newcommand\percentageSpecRedshifts{39.3}
\newcommand\minRedshift{0.04}
\newcommand\medianRedshift{0.52}
\newcommand\medianRedshiftNew{0.75}
\newcommand\maxRedshift{1.91}

\newcommand\numKnownLenses{210}
\newcommand\numPossNewLenses{67}

\documentclass[twocolumn]{aastex63}

\reportnum{DES-2020-0547}
\reportnum{FERMILAB-PUB-20-458-AE}

\usepackage{color}

\mathchardef\mhyphen="2D

\shorttitle{ACT: A Catalog of $>4000$ SZ-Selected Galaxy Clusters}
\shortauthors{ACT Collaboration}

\begin{document}

\title{The Atacama Cosmology Telescope: A Catalog of $> 4000$ Sunyaev-Zel'dovich Galaxy Clusters}

\correspondingauthor{Matt Hilton}
\email{hiltonm@ukzn.ac.za}


\author[0000-0002-8490-8117]{M. Hilton}
\affiliation{Astrophysics Research Centre, University of KwaZulu-Natal, Westville Campus, Durban 4041, South Africa}
\affiliation{School of Mathematics, Statistics \& Computer Science, University of KwaZulu-Natal, Westville Campus, Durban 4041, South Africa}
\author[0000-0002-8149-1352]{C. Sif\'{o}n}
\affiliation{Instituto de F\'{i}sica, Pontificia Universidad Cat\'{o}lica de Valpara\'{i}so, Casilla 4059, Valpara\'{i}so, Chile}
\author{S. Naess}
\affiliation{Center for Computational Astrophysics, Flatiron Institute, New York, NY 10010, USA}
\author{M. Madhavacheril}
\affiliation{Centre for the Universe, Perimeter Institute for Theoretical Physics, Waterloo, ON, Canada N2L 2Y5}
\author{M. Oguri}
\affiliation{Research Center for the Early Universe, University of Tokyo, Tokyo 113-0033, Japan}
\affiliation{Department of Physics, University of Tokyo, Tokyo 113-0033, Japan}
\affiliation{Kavli Institute for the Physics and Mathematics of the Universe (Kavli IPMU, WPI), University of Tokyo, Chiba 277-8582, Japan}
\author{E. Rozo}
\affiliation{Department of Physics, University of Arizona, Tucson, AZ 85721, USA}
\author{E. Rykoff}
\affiliation{Kavli Institute for Particle Astrophysics \& Cosmology, P. O. Box 2450, Stanford University, Stanford, CA 94305, USA}
\affiliation{SLAC National Accelerator Laboratory, Menlo Park, CA 94025, USA}

\author{T. M. C. Abbott}
\affiliation{Cerro Tololo Inter-American Observatory, NSF's National Optical-Infrared Astronomy Research Laboratory, Casilla 603, La Serena, Chile}
\author[0000-0002-0298-4432]{S. Adhikari}
\affiliation{Kavli Institute for Particle Astrophysics \& Cosmology, P. O. Box 2450, Stanford University, Stanford, CA 94305, USA}
\affiliation{Department of Physics, Stanford University, 382 Via Pueblo Mall, Stanford, CA 94305, USA}
\affiliation{SLAC National Accelerator Laboratory, Menlo Park, CA 94025, USA}
\author{M. Aguena}
\affiliation{Departamento de F\'isica Matem\'atica, Instituto de F\'isica, Universidade de S\~ao Paulo, CP 66318, S\~ao Paulo, SP, 05314-970, Brazil}
\affiliation{Laborat\'orio Interinstitucional de e-Astronomia - LIneA, Rua Gal. Jos\'e Cristino 77, Rio de Janeiro, RJ - 20921-400, Brazil}
% ...




\begin{abstract}
We present a catalog of \totalConfirmed{} optically confirmed Sunyaev-Zel'dovich (SZ) selected galaxy clusters
detected with signal-to-noise~$>4$ in \surveyArea{} deg$^2$ of sky surveyed by the Atacama Cosmology Telescope (ACT). 
Cluster candidates were selected by applying a multi-frequency matched filter to 98 and 150\,GHz maps
constructed from ACT observations obtained from 2008--2018, and confirmed 
using deep, wide-area optical surveys. The clusters span the redshift range 
$\minRedshift{} < z < \maxRedshift{}$ (median $z = \medianRedshift{}$). 
The catalog contains \totalHighZ{} $z > 1$ clusters, and a total 
of \totalNew{} systems are new discoveries. 
Assuming an SZ-signal vs. mass scaling relation calibrated
from X-ray observations, the sample has a 90\% completeness mass limit of $M_{\rm 500c} > \compLimitFull{} \times 10^{14}$\,$M_{\sun}$, 
evaluated at $z = 0.5$, for clusters detected at 
signal-to-noise ratio $>5$ in maps filtered at an angular scale of 2.4\arcmin{}. 
The survey has a large overlap with deep optical weak-lensing surveys that are being used to calibrate
the SZ-signal mass-scaling relation, such as the 
Dark Energy Survey (\DESOverlapArea{} deg$^2$), the Hyper Suprime-Cam Subaru Strategic Program (\HSCOverlapArea{} deg$^2$),
and the Kilo Degree Survey (\KiDSOverlapArea{} deg$^2$).
We highlight some noteworthy objects in the sample, including potentially projected systems; 
clusters with strong lensing features; clusters with active central galaxies or star formation;
and systems of multiple clusters that may be physically associated. The cluster catalog will be 
a useful resource for future cosmological analyses, and studying the evolution of the intracluster medium
and galaxies in massive clusters over the past 10\,Gyr.
\end{abstract}





\defcitealias{Hasselfield_2013}{H13}
\defcitealias{Hilton_2018}{H18}
\defcitealias{Arnaud_2010}{A10}
\defcitealias{Naess_2020}{N20}


\section{Introduction} 
\label{sec:intro}

%

The thermal Sunyaev-Zel'dovich effect \citep[SZ; e.g., ][]{SZ_1970, SZ_1972} is well established as a method 
for constructing approximately mass-limited samples of galaxy clusters, independently of redshift. The
SZ effect arises through the inverse Compton scattering of cosmic microwave background (CMB) photons by
electrons within the hot gas atmospheres of galaxy clusters 
\citep[see reviews by][]{Birkinshaw_1999, Carlstrom_2002, Mroczkowski_2019}. This leads to a spectral distortion in sight
lines towards clusters, such that at frequencies below 220\,GHz, clusters appear as ``cold spots'' in the
mm-wave sky, while at frequencies above 220\,GHz, they appear as ``hot spots.'' The amplitude of the SZ signal
scales with the mass of the cluster.

The unique power of SZ-selected cluster surveys to detect all of the massive structures in the Universe
regardless of their distance from the observer has driven the development of ``blind'' SZ surveys that
constrain cosmological parameters through measuring the evolution of the cluster mass function
\citep[e.g.,][]{Vanderlinde_2010, Sehgal_2011, Hasselfield_2013, Reichardt_2013, Planck_XX_2013, PlanckPSZ2Cosmo_2016, Bocquet_2019}.
SZ cluster surveys over large areas of sky have been conducted by the 
South Pole Telescope \citep[SPT; e.g.,][]{Williamson_2011, Bleem_2015, Bleem_2020, Huang_2020},
the \textit{Planck} satellite mission \citep{Planck_XXIX_2013, PlanckPSZ2_2016},
and the Atacama Cosmology Telescope \citep[ACT;][]{Marriage_2011, Hasselfield_2013, Hilton_2018}. 
Collectively, since the first blind SZ detections by SPT \citep{Staniszewski_2009}, 
these surveys have detected approximately 2300 clusters with redshift measurements to date.

In this paper we present the first cluster catalog derived from observations using the Advanced ACTPol receiver
\citep{Henderson_2016, Ho_2017, Choi_2018}, combining this with all observations by ACT from 2008--2018
(\citealt{Naess_2020}, N20 hereafter; details of previous generations of ACT instrumentation
can be found in \citealt{Fowler_2007}, \citealt{Swetz_2011}, and \citealt{Thornton_2016}).
This is the first ACT cluster catalog to use multi-frequency data (98 and 150\,GHz) in its construction. The SZ cluster 
search area covers \surveyArea{} deg$^2$, and we have optically confirmed and measured redshifts for
\totalConfirmed{} clusters out of \totalCandidates{} candidates detected with signal-to-noise ratio (SNR) $>4$. 
The cluster catalog is publicly available in \texttt{FITS} Table format at the NASA 
Legacy Archive for Microwave Background Data (LAMBDA) as part of the fifth ACT data release
(ACT DR5\footnote{\url{https://lambda.gsfc.nasa.gov/product/act/actpol_prod_table.cfm}}). 
Table~\ref{tab:FITSTableColumns} describes the contents of the catalog. 

The structure of this paper is as follows. In Section~\ref{sec:ACTSZ}, we describe the ACT maps used in this
work, the SZ cluster detection algorithm, and our process for estimating cluster masses from the SZ signal.
In Section~\ref{sec:Redshifts}, we explain how we optically confirmed SZ detections as galaxy clusters and
assigned their redshifts, making use of deep wide-area optical/IR surveys in conjunction with our own 
follow-up observations. In Section~\ref{sec:Catalog}, we present the statistical properties of the cluster 
catalog, and compare it with previous work by the ACT collaboration. We discuss our catalog in comparison with 
other cluster samples in Section~\ref{sec:Discussion}. We present a summary in Section~\ref{sec:Summary}.

We assume a flat cosmology with $\Omega_{\rm m}=0.3$, $\Omega_\Lambda=0.7$, and $H_0=70$~km~s$^{-1}$~Mpc$^{-1}$ 
throughout. We quote cluster mass estimates ($M_{\rm 500c}$) within a spherical radius that encloses an average 
density equal to 500 times the critical density at the cluster redshift ($R_{\rm 500c}$). 
All magnitudes are on the AB system \citep{Oke_1974}, unless stated otherwise.

\section{ACT Observations and SZ Cluster Candidate Selection}
\label{sec:ACTSZ}

\subsection{98 and 150 GHz Observations and Maps}
\label{sec:ACTObs}

\startlongtable
\begin{deluxetable*}{llp{120mm}}
\small
\tablecaption{Description of the columns in the \texttt{FITS} Table format cluster catalog, 
available from LAMBDA (\url{https://lambda.gsfc.nasa.gov/product/act/actpol_prod_table.cfm}).
The Symbol column provides a mapping between column names and symbols used in the text and 
figures of this article.}
\label{tab:FITSTableColumns}
\tablewidth{0pt}
\decimals
\tablehead{
\colhead{Column}       &
\colhead{Symbol} &
\colhead{Description} \\
}
\startdata
\texttt{name} & \nodata & Cluster name in the format ACT-CL JHHMM.m$\pm$DDMM.\\
\texttt{RADeg} & \nodata & Right Ascension in decimal degrees (J2000) of the SZ detection by ACT.\\
\texttt{decDeg} & \nodata & Declination in decimal degrees (J2000) of the SZ detection by ACT.\\
\texttt{SNR} & SNR & Signal-to-noise ratio, optimized over all filter scales.\\
\texttt{y\_c} & $y_{0}$ & Central Comptonization parameter ($10^{-4}$) measured using the optimal matched filter template (i.e., the one that maximizes SNR). Uncertainty column(s): \texttt{err\_y\_c}.\\
\texttt{fixed\_SNR} & SNR$_{2.4}$ & Signal-to-noise ratio at the reference 2.4\arcmin{} filter scale.\\
\texttt{fixed\_y\_c} & $\tilde{y}_0$ & Central Comptonization parameter ($10^{-4}$) measured at the reference filter scale (2.4\arcmin{}). Uncertainty column(s): \texttt{fixed\_err\_y\_c}.\\
\texttt{template} & \nodata & Name of the matched filter template resulting in the highest SNR detection of this cluster.\\
\texttt{tileName} & \nodata & Name of the ACT map tile (typically with dimensions 10\,deg\,$\times 5$\,deg) in which the cluster was found.\\
\texttt{redshift} & $z$ & Adopted redshift for the cluster. The uncertainty is only given for photometric redshifts. Uncertainty column(s): \texttt{redshiftErr}.\\
\texttt{redshiftType} & \nodata & Redshift type (\texttt{spec} = spectroscopic, \texttt{phot} = photometric).\\
\texttt{redshiftSource} & \nodata & Source of the adopted redshift (see Table~\ref{tab:RedshiftSources}).\\
\texttt{M500c} & $M_{\rm 500c}^{\rm UPP}$ & $M_{\rm 500c}$ in units of $10^{14}\,M_{\sun}$, assuming the UPP and \citet{Arnaud_2010} scaling relation to convert SZ signal to mass. Uncertainty column(s): \texttt{M500c\_errPlus, M500c\_errMinus}.\\
\texttt{M500cCal} & $M_{\rm 500c}^{\rm Cal}$ & $M_{\rm 500c}$ in units of $10^{14}\,M_{\sun}$, rescaled using the richness-based weak-lensing mass calibration factor of $0.71 \pm 0.07$ (see Section~\ref{sec:catalogProperties}). Uncertainty column(s): \texttt{M500cCal\_errPlus, M500cCal\_errMinus}.\\
\texttt{M200m} & $M_{\rm 200m}^{\rm UPP}$ & $M_{\rm 200}$ with respect to the mean density, in units of $10^{14}\,M_{\sun}$, converted from $M_{\rm 500c}$ using the \citet{Bhattacharya_2013} c-M relation. Uncertainty column(s): \texttt{M200m\_errPlus, M200m\_errMinus}.\\
\texttt{M500cUncorr} & $M_{\rm 500c}^{\rm Unc}$ & $M_{\rm 500c}$ in units of $10^{14}\,M_{\sun}$, assuming the UPP and \citet{Arnaud_2010} scaling relation to convert SZ signal to mass, uncorrected for bias due to the steepness of the cluster mass function and intrinsic scatter. Uncertainty column(s): \texttt{M500cUncorr\_errPlus, M500cUncorr\_errMinus}.\\
\texttt{M200mUncorr} & $M_{\rm 200m}^{\rm Unc}$ & $M_{\rm 200}$ with respect to the mean density, in units of $10^{14}\,M_{\sun}$, converted from $M_{\rm 500c}$ using the \citet{Bhattacharya_2013} c-M relation, uncorrected for bias due to the steepness of the cluster mass function and intrinsic scatter. Uncertainty column(s): \texttt{M200mUncorr\_errPlus, M200mUncorr\_errMinus}.\\
\texttt{footprint\_DESY3} & \nodata & Flag indicating if the cluster falls within the DES Y3 footprint.\\
\texttt{footprint\_HSCs19a} & \nodata & Flag indicating if the cluster falls within the HSC-SSP S19A footprint (assuming the full-depth full-color HSC-SSP mask).\\
\texttt{footprint\_KiDSDR4} & \nodata & Flag indicating if the cluster falls within the KiDS DR4 footprint.\\
\texttt{zCluster\_delta} & $\delta$ & Density contrast statistic measured at the \texttt{zCluster} photometric redshift. Uncertainty column(s): \texttt{zCluster\_errDelta}.\\
\texttt{zCluster\_source} & \nodata & Photometry used for \texttt{zCluster} measurements (see Section~\ref{sec:zCluster}).
One of: DECaLS (DR8), KiDS (DR4), SDSS (DR12).\\
\texttt{RM} & \nodata & Flag indicating cross-match with a redMaPPer-detected cluster in the SDSS footprint \citep{Rykoff_2014}.\\
\texttt{RM\_LAMBDA} & $\lambda$ & Optical richness measurement for the redMaPPer algorithm in the SDSS footprint. Uncertainty column(s): \texttt{RM\_LAMBDA\_ERR}.\\
\texttt{RMDESY3} & \nodata & Flag indicating cross-match with a redMaPPer-detected cluster in the DES Y3 footprint \citep[for details of the redMaPPer algorithm applied to DES data, see][]{Rykoff_2016}.\\
\texttt{RMDESY3\_LAMBDA\_CHISQ} & $\lambda$ & Optical richness measurement for the redMaPPer algorithm in the DES Y3 footprint. Uncertainty column(s): \texttt{RMDESY3\_LAMBDA\_CHISQ\_E}.\\
\texttt{CAMIRA} & \nodata & Flag indicating cross-match with a CAMIRA-detected cluster in the HSCSSP S19A footprint \citep[for details of the CAMIRA algorithm, see][]{Oguri_2014, Oguri_2018}.\\
\texttt{CAMIRA\_N\_mem} & \nodata & Optical richness measurement for the CAMIRA algorithm in the HSCSSP S19A footprint.\\
\texttt{opt\_RADeg} & \nodata & Alternative optically-determined Right Ascension in decimal degrees (J2000), from a heterogeneous collection of measurements (see \texttt{opt\_positionSource}).\\
\texttt{opt\_decDeg} & \nodata & Alternative optically-determined Declination in decimal degrees (J2000), from a heterogeneous collection of measurements (see \texttt{opt\_positionSource}).\\
\texttt{opt\_positionSource} & \nodata & Indicates the source of the alternative optically-determined cluster position. One of: \texttt{AMICO} (position from the AMICO cluster finder; \citealt{2019MNRAS.485..498M}), \texttt{CAMIRA} (position from the CAMIRA cluster finder; \citealt{Oguri_2018}), \texttt{RM, RMDESY3, RMDESY3ACT} (position from the redMaPPer cluster finder, in SDSS, DES Y3, or DES Y3 using the ACT position as a prior; \citealt{Rykoff_2014, Rykoff_2016}), \texttt{Vis-BCG} (brightest central galaxy (BCG) position from visual inspection of available optical/IR imaging; this work), \texttt{WHL2015} (position from \citealt{WH_2015}).\\
\texttt{notes} & \nodata & If present, at least one of: \texttt{AGN?} (central galaxy may have color or spectrum indicating it may host an AGN); \texttt{Lensing?} (cluster may show strong gravitational lensing features); \texttt{Merger?} (cluster may be a merger); \texttt{Star formation?} (a galaxy near the center may have blue colors which might indicate star formation if it is not a line-of-sight projection). These notes are not comprehensive and merely indicate some systems that were identified as potentially interesting during visual inspection of the available optical/IR imaging.\\
\texttt{knownLens} & \nodata & Names of known strong gravitational lenses within 2 Mpc projected distance of this cluster (comma delimited when there are multiple matches).\\
\texttt{knownLensRefCode} & \nodata & Reference codes (comma delimited when there are multiple matches) corresponding to the entries in the \texttt{knownLens} field. See Table~\ref{tab:lensCodes} to map between the codes used in this field and references to the corresponding lens catalog papers.\\
\texttt{warnings} & \nodata & If present, a warning message related to the redshift measurement for this cluster (e.g., possible projected system).\\
\enddata
\end{deluxetable*}

\begin{figure*}
\includegraphics[width=\textwidth]{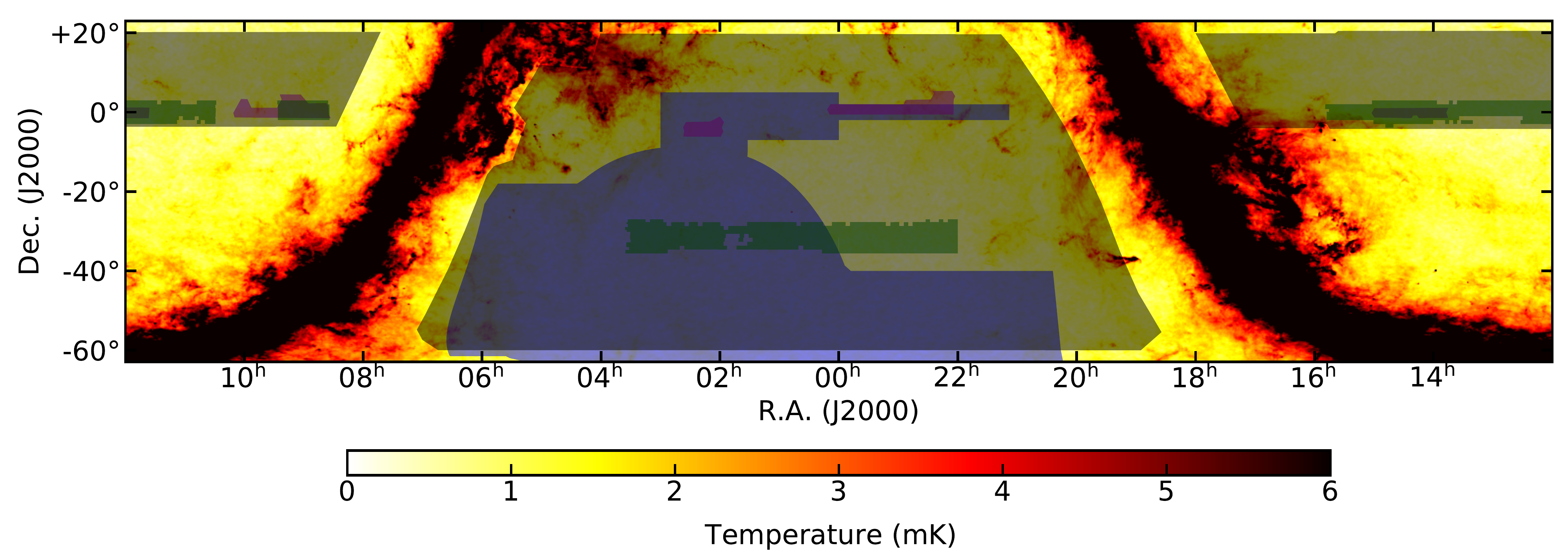}
\caption{The ACT DR5 cluster search area (shaded in gray; covering \surveyArea\,deg$^{2}$ after masking), 
overlaid on the \textit{Planck} 353\,GHz map, which is sensitive to thermal emission by dust. 
The footprints of deep and wide optical surveys that will provide weak-lensing mass calibration of the cluster sample are 
highlighted: DES (blue); HSC (magenta); and KiDS (green).}
\label{fig:surveyArea}
\end{figure*}

\begin{figure*}
\includegraphics[width=\textwidth]{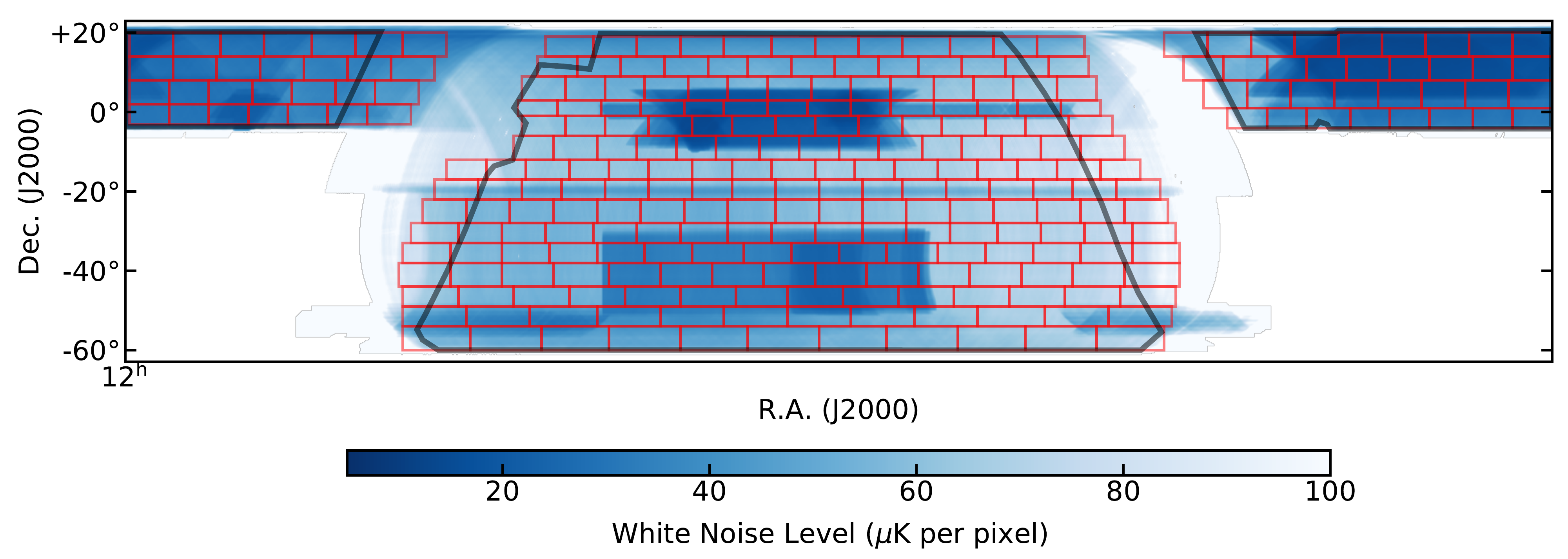}
\caption{A map of the white noise level in the 150\,GHz map, as produced by the map maker. The 98 and 150\,GHz
maps, which cover 18,000\,deg$^2$ in total, are broken into tiles (marked in red) before filtering, 
for the reasons outlined in Section~\ref{sec:SZDetection}. The black outline marks the cluster search region, 
before the dust and point source masks are applied.}
\label{fig:tiles}
\end{figure*}

The ACT experiment saw first light in 2007, and since 2016 has been observing with its third generation receiver, 
Advanced ACTPol \citep[AdvACT;][]{Henderson_2016}. AdvACT consists of three detector arrays containing dichroic,
dual polarization horn-coupled Transition Edge Sensor (TES) bolometers, observing at 98, 150, and 220\,GHz,
with 27 and 39\,GHz channels added in the 2020 season.
For this work, we use data from only the 98 and 150\,GHz channels, which have approximate beam FWHM 2.2\arcmin{}
and 1.4\arcmin{} respectively. 

The cluster search was performed on co-added maps containing ACT data obtained between 2008--2018 (made
available to the community as ACT DR5). The ACT maps 
for the 2008--2016 observing seasons are publicly available on the LAMBDA website, with seasons 2013--2016
being processed for ACT DR4 \citep[][]{Aiola_2020, Choi_2020}. 
ACT DR5 contains co-added maps that incorporate 
early versions of the 2017--2018 data (\citetalias{Naess_2020}), and unlike previous ACT data releases, 
includes observations taken during daylight hours. 
These maps have not been subjected
to the full battery of tests needed for precision measurements of the CMB power spectrum, and may contain
gain errors at the level of a few per cent. They are, however, much deeper over a much
wider area than the maps used in the ACT DR4 analysis. More than 12,000 deg$^2$ (91\% of the
\surveyArea{} deg$^2$ cluster search area) has noise
level $< 30$\,$\mu$K-arcmin at 150\,GHz (\citetalias{Naess_2020}).

The co-added maps used in this work were produced in a two-step procedure 
(described in detail in \citetalias{Naess_2020}). Individual
maximum likelihood maps were first made for each observing season, frequency, and detector array, following
the procedures described in \citet{Dunner_2013} and \citet{Aiola_2020}. These maps were 
then combined into a single map per frequency, convolved to a common beam, by breaking each map into a series
of tiles and weighting by a noise model constructed from the hitcount-modulated 2d noise 
power spectrum for each tile.

The co-added maps cover a sky area of approximately 18,000\,deg$^2$. However, several thousand square 
degrees correspond to low Galactic latitudes ($|\,b\,| < 20$\,deg), where either the level of dust emission is
high (making cluster detection in our mm-wave maps problematic), or the stellar density is high (making 
optical confirmation and redshift measurements difficult or impossible). Therefore, we defined the cluster
search area (plotted over the \textit{Planck} 353\,GHz map in Fig.~\ref{fig:surveyArea}) to exclude 
such regions. We also mask dusty regions within the cluster search region, defined as pixels
with temperature $>0.004$\,K (in CMB temperature units) in the \textit{Planck} 353\,GHz map.
We initially masked the locations of point sources detected in the ACT 150\, GHz map using circular
regions with radii in the range 3--12\arcmin{}, depending on the amplitude of the 
source at 150\,GHz. After visual inspection of the filtered maps (see Section~\ref{sec:visualInspect}),
we found it necessary to mask some regions that were not captured by the above procedures. Typically these
were cases where our automated procedure to define source masking had not selected a large enough masking 
radius. We subsequently masked the locations of all sources with 150\,GHz flux density $> 10$\,mJy
(approximately 11,000 objects) using circles with radius 320\arcsec{}, except for those sources located
within 9\arcmin{} of bright clusters (with $\tilde{y}_0 > 1 \times 10^{-4}$; $\tilde{y}_0$ is our 
chosen SZ observable, defined in Section~\ref{sec:SZMasses} below), which are not masked
(because ``ringing'' around bright clusters can be detected as spurious sources).
After masking, the cluster search area is \surveyArea{}\,deg$^2$.

\subsection{Cluster Detection}
\label{sec:SZDetection}

We search for clusters using a multi-frequency matched filter \citep[e.g.,][]{Melin_2006, Williamson_2011}, 
applied to the 98 and 150\,GHz maps,
\begin{equation}
\label{eq:MFMF}
\psi(k_x, k_y, \nu_i) = A \sum_j \textbf{N}^{-1}_{ij}(k_x, k_y) f_{\rm SZ}(\nu_j) S (k_x, k_y, \nu_j) \, .
\end{equation}
where $\psi$ is the filter, ($k_x$, $k_y$) denote the spatial frequencies in the horizontal and vertical directions
in the maps, $\textbf{N}$ is the noise covariance between the maps at different frequencies $\nu$, $S$ is a beam-convolved signal template, and $A$ is a normalization factor chosen such that, when applied to a set of maps containing a beam-convolved cluster signal (in temperature units), the matched filter returns the central Comptonization parameter (see Section~\ref{sec:SZMasses} below). We use the non-relativistic form for the spectral dependence of the SZ
effect given by
\begin{equation}
\label{eq:fSZ}
f_{\rm SZ} = x \frac{e^x+1}{e^x-1} - 4 \, ,
\end{equation}
where $x = h \nu / k_B T_{\rm CMB}$. We adopt 97.8\,GHz and 149.6\,GHz as the thermal SZ-weighted band centers for the 98 and 150\, GHz maps analyzed here. These are the median values of the SZ-weighted band centers of the individual detector arrays; in practice the effective band centers vary slightly by position on the sky - see the Appendix of \citetalias{Naess_2020} - with uncertainty $\approx 1$\,GHz on arcmin scales.

We use the map itself to form the noise covariance $\textbf{N}$, as the maps are dominated by the CMB on large scales, and white noise on small scales, rather than by the thermal SZ signal. Note that the filter is 2d in Fourier space, in order to account for the anisotropic noise that arises due to the scan pattern of ACT \citep[e.g.,][]{Marriage_2011}, which varies according to position on the sky. However, the signal template $S$ is axisymmetric. We fill holes in the map created by point source masking (see Section~\ref{sec:ACTObs}) with a heavily smoothed version of the map itself prior to Fourier transforming.

As in previous ACT cluster searches \citep{Hasselfield_2013, Hilton_2018}, throughout this work we model the cluster signal using the Universal Pressure Profile \citep[UPP;][\citetalias{Arnaud_2010} hereafter]{Arnaud_2010}, which is convolved with the appropriate ACT beam for each frequency to form the signal template $S$. To improve the detection efficiency for clusters with different angular sizes, we create a set of 16 matched filters, corresponding to $M_{\rm 500c} \in \{ (1, 2, 4, 8) \times 10^{14}$\,$M_\sun \}$ and $z \in \{0.2, 0.4, 0.8, 1.2\}$.

The ACT maps cover approximately 18,000\,deg$^2$ and the noise level in the maps varies considerably as
a function of position on the sky (see Fig.~\ref{fig:tiles}). In addition, the maps are produced in plate carr\'{e}e 
projection (\texttt{CAR} in the terminology of FITS world coordinate systems; \citealt{Calabretta_2002}), which 
leads to distortion away from the celestial equator as the solid angle covered by a pixel changes with declination. 
Therefore, we break the maps into a set of 280 tiles, each with approximate dimensions 
10\,deg\,$\times 5$\,deg (right ascension $\times$ declination), and construct a different set of matched filters 
for each tile. 
Fig.~\ref{fig:tiles} shows the layout of the tiles on the 150\,GHz white noise level map (as produced by the map maker).
Since we apodize each tile before Fourier transforming when constructing $\textbf{N}$, each tile is extended with a one 
degree wide border that overlaps with its neighbours.

\begin{figure*}
\centering
\includegraphics[width=174mm]{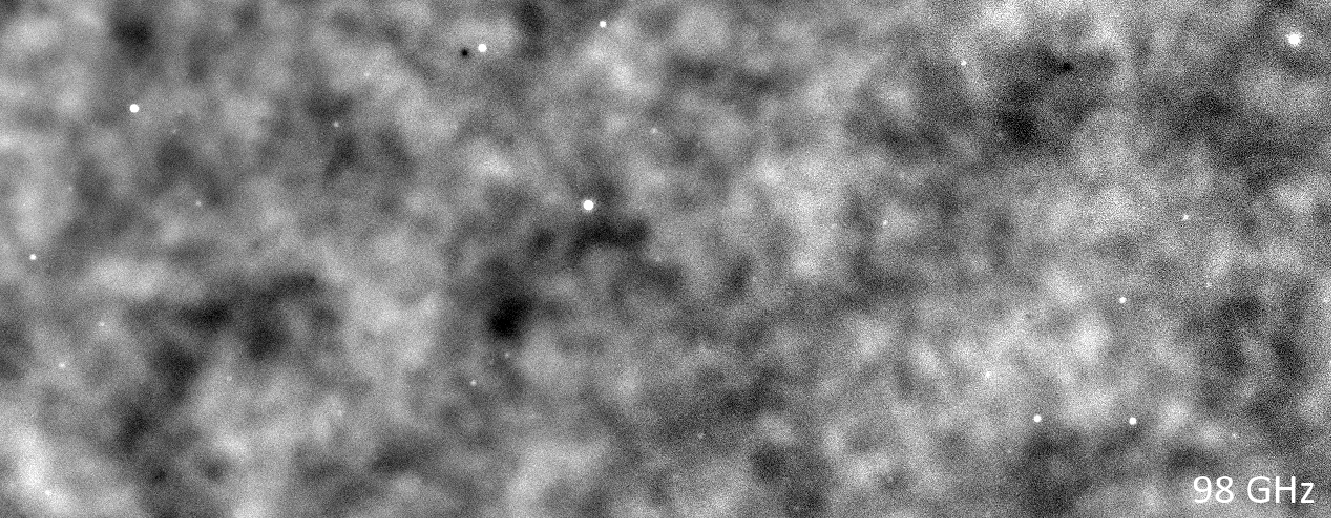}
\includegraphics[width=174mm]{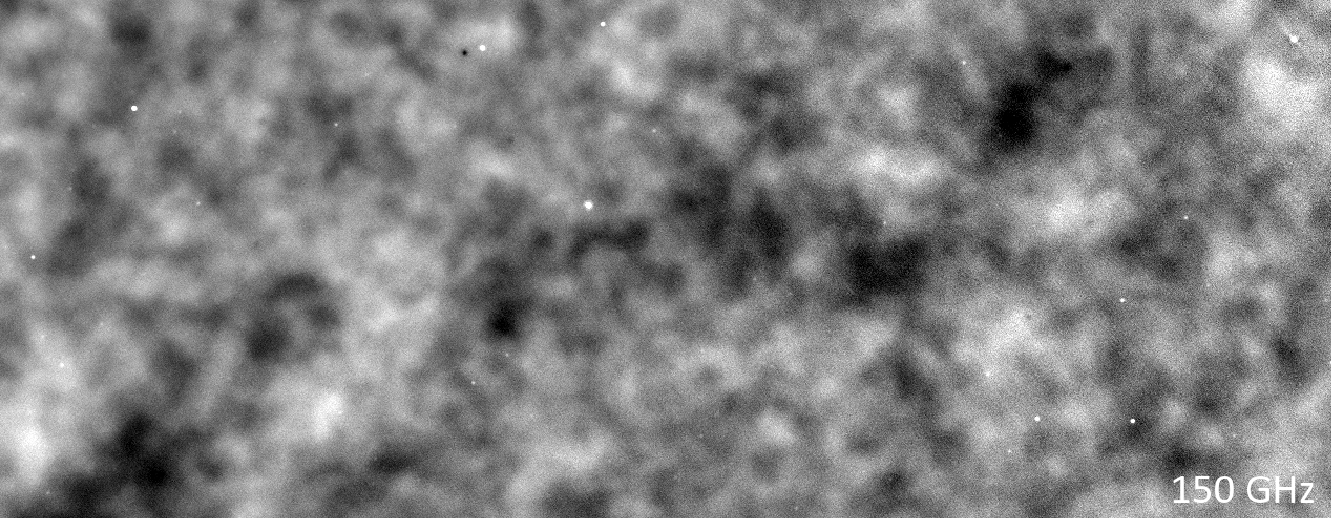}
\includegraphics[width=174mm]{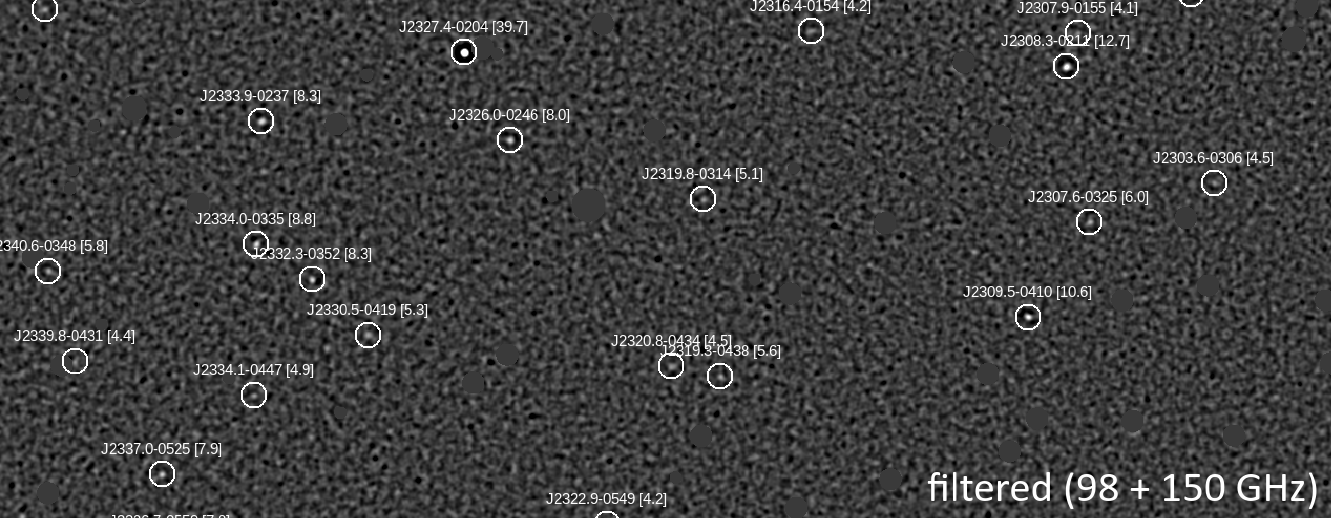}
\caption{A comparison between the unfiltered 98 and 150\,GHz maps, and the filtered signal-to-noise map, for an
approximately $10$\,deg$ \times 4$\,deg patch of sky. In the unfiltered maps, clusters appear as decrements
(dark spots) in the map. Point sources appear as white spots, and CMB fluctuations dominate at large
angular scales. In the filtered signal-to-noise map, clusters appear as white spots (marked with white
circles to guide the eye; the number given in brackets is SNR$_{2.4}$), and point sources
have been masked. The brightest object visible is the $z = 0.70$ cluster ACT-CL J2327.4-0204 
(center left, near the top left of the image), which is a SNR$_{2.4} = 39.7$ detection.}
\label{fig:filteredMap}
\end{figure*}

To select cluster candidates, we construct a signal-to-noise ratio (SNR) map for each filtered tile, and in turn, make
a segmentation map that identifies peaks with SNR~$ > 4$. We estimate the noise map in a similar way
to that used in \citet{Hilton_2018}, by dividing each tile into square 
40$\arcmin$ cells and measuring the 3$\sigma$-clipped standard deviation in each cell, taking into account
masked regions\footnote{This method can underestimate the noise level within a 40$\arcmin$ cell if it straddles
an abrupt, large change in the map depth, resulting in spurious candidates along such features. This can be corrected by
binning the filtered maps according to the weight maps produced by the map maker, and will be implemented for the next
version of the catalog.}. This accounts for variations in depth within each tile.
Finally, we apply the cluster search area mask shown in Fig.~\ref{fig:surveyArea}, and apply the dust and point source 
masks (see Section~\ref{sec:ACTObs}). Fig.~\ref{fig:filteredMap} shows a comparison between the unfiltered
98 and 150\,GHz maps, and a filtered map, after the application of all the above procedures.

\begin{figure}
\includegraphics[width=\columnwidth]{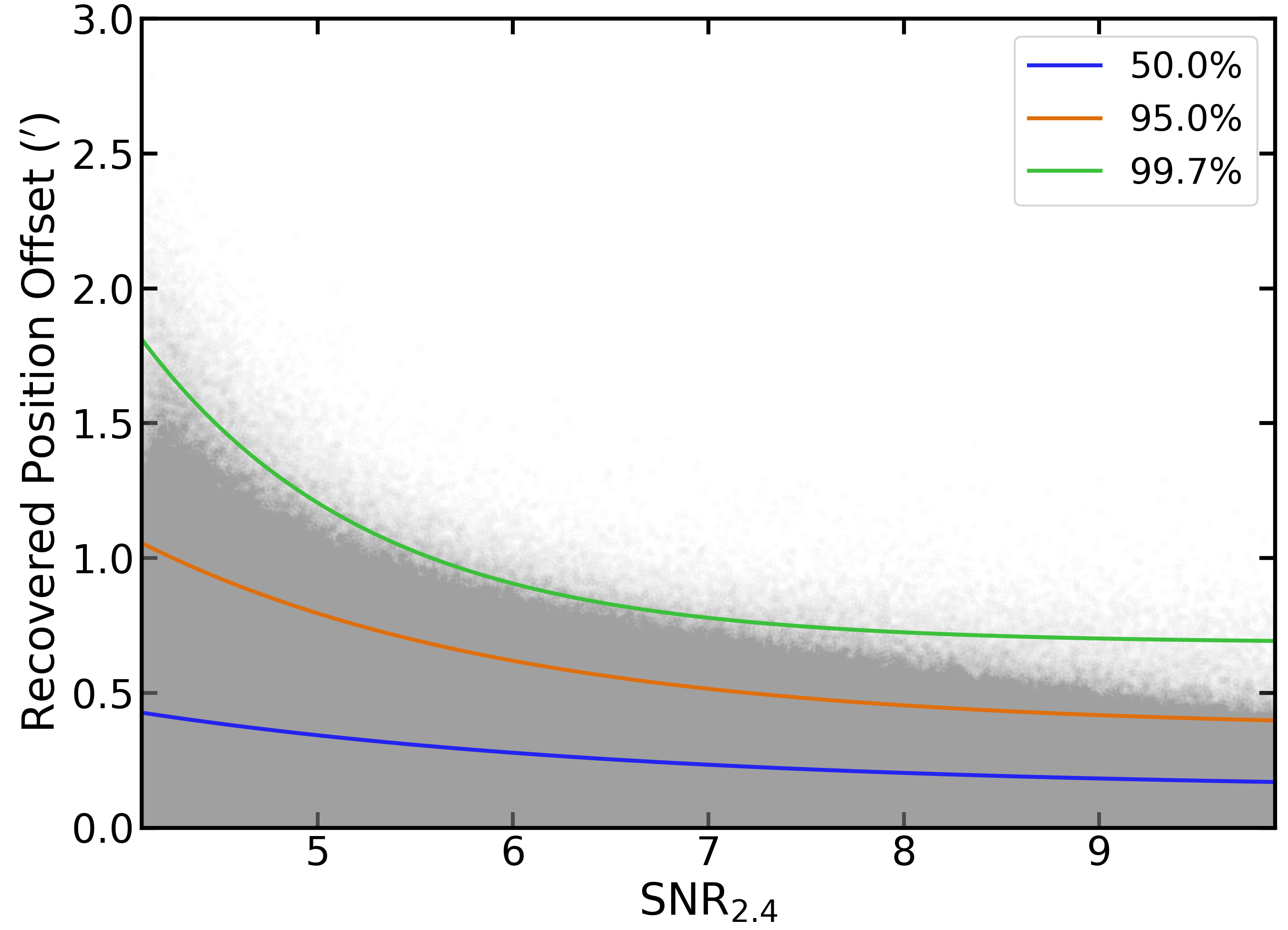}
\caption{Accuracy of position recovery for injected UPP-model clusters, as a function of SNR$_{2.4}$.
The offset with respect to the original input cluster position is plotted on the vertical axis. The 
gray points show the offsets recovered for individual model clusters. The solid lines show model fits 
of the form given in equation (\ref{eq:posRecModel}) that enclose the 50, 95, and 99.7 percentiles.}
\label{fig:posRecTest}
\end{figure}

\begin{figure}
\includegraphics[width=\columnwidth]{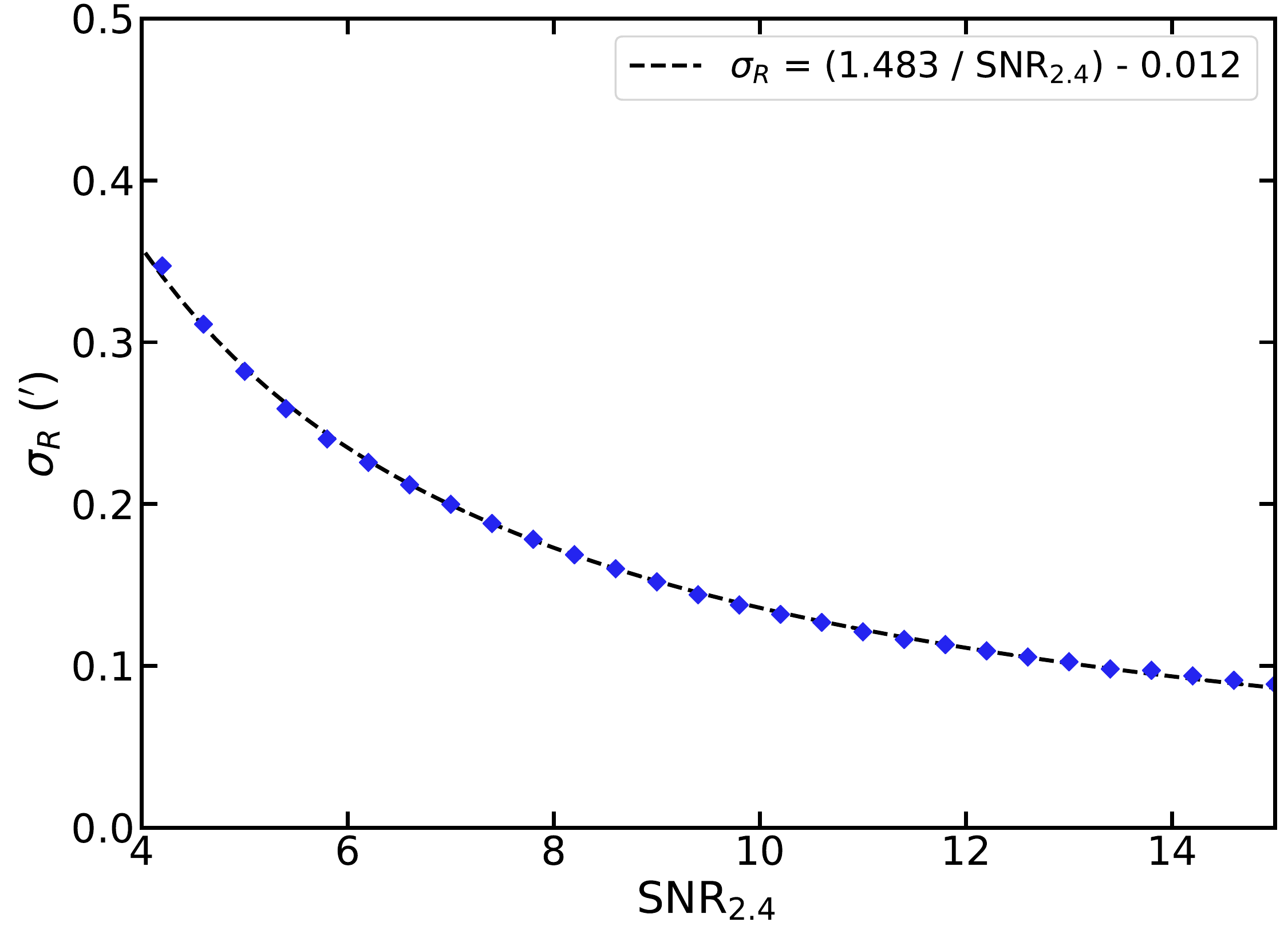}
\caption{Results of fitting the distribution of recovered position offsets obtained from source
insertion simulations using the Rayleigh distribution (equation~\ref{eq:rayleigh}). The simple
model shown is a good description of how the scale parameter $\sigma_R$ changes as a function
of SNR$_{2.4}$.}
\label{fig:rayleigh}
\end{figure}

We assemble the final catalog of cluster candidates from a set of catalogs extracted from each SNR map
for each filter scale in each tile, using a similar procedure to \citet{Hilton_2018}. We use a 
minimum detection threshold of a single pixel with SNR~$ > 4$ in any filtered map, and adopt the
location of the center-of-mass of the SNR~$ > 4$ pixels in each detected object in the filtered map as the 
coordinates of each cluster candidate. We then create a final master candidate list by cross-matching the catalogs 
assembled at each cluster scale using a 1.4$\arcmin$ matching radius.
Objects in the regions that overlap between tiles are removed by applying a mask; the tiles are defined
such that each non-overlapping pixel in a tile maps to a unique pixel in the pixelization of the
original monolithic map.
We adopt the maximum SNR across all filter scales for each candidate as the `optimal' SNR 
detection. However, as in \citet{Hasselfield_2013} and \citet{Hilton_2018}, we also use a single reference 
filter scale (chosen to be $\theta_{\rm 500c} = 2.4\arcmin$; see Section~\ref{sec:SZMasses} below) at which
we measure the cluster SZ signal and signal-to-noise ratio. Throughout this work we use SNR to refer to 
the `optimal' signal-to-noise ratio (maximized over all filter scales), and SNR$_{2.4}$ for the 
signal-to-noise ratio measured at the fixed 2.4$\arcmin$ filter scale. 
The final catalog contains \totalCandidates{} SNR~$>4$ candidates selected from a survey area of
\surveyArea{}\,deg$^2$.

We checked the accuracy of recovered cluster positions by injecting simulated clusters into the 
maps and re-running the filtering and cluster detection procedures, taking care
to remove objects corresponding to real cluster candidates from the resulting catalogs.
The injected clusters are UPP models with uniformly distributed amplitudes and sizes selected from 
$\theta_{500c} (\arcmin{}) \in \{7.8, 4.2, 2.4, 1.5\}$. More than 5.7 million model clusters 
with $4 < $~SNR$_{2.4} < 20$ are recovered from these simulations. We fit a model of the form
\begin{equation}
\label{eq:posRecModel}
r = A e^{-{\rm{SNR}_{2.4}}/B} + C\, , 
\end{equation}
where $r$ specifies the distance (in arcmin) between input and recovered model cluster positions within
which some percentile of the objects are found, and $A$, $B$, and $C$ are fit parameters. 
Fig.~\ref{fig:posRecTest} shows this model plotted over the position recovery data for the 50, 95, and 99.7
percentiles. The radial distance within which 99.7\% of the model clusters are recovered is specified by 
a model with $A = 38.1$, $B = 1.16$, and $C = 0.69$. We use this model for cross matching cluster
candidates against external catalogs (see Section~\ref{sec:visualInspect} below).
Note that the accuracy of position recovery depends on cluster scale, with larger scale
clusters having less accurately recovered positions, but for our purposes an average over
several scales is sufficient.

For some applications (e.g., stacking on cluster positions), it is useful to model the positional
uncertainty using the Rayleigh distribution, i.e.,
\begin{equation}
\label{eq:rayleigh}
P(r, \sigma_R) = \frac{r}{\sigma_R} \exp{\left(-r^2 / 2 \sigma_R^2 \right)}\,,
\end{equation}
where $r$ is the distance between the true and recovered cluster position, and $\sigma_R$ is the
scale parameter for the distribution. We fitted models of the form given in equation~(\ref{eq:rayleigh})
to the distribution of recovered position offsets obtained from the source insertion simulations
described above, binned by SNR$_{2.4}$. Fig.~\ref{fig:rayleigh} shows the resulting measurements of
$\sigma_R$ as a function of SNR$_{2.4}$, together with a simple model that captures how $\sigma_R$
changes with SNR$_{2.4}$.

We assess the number of false positive detections in the candidate
list as a function of SNR$_{2.4}$ by running the cluster detection algorithm over sky simulations
that are free of cluster signal. We use the map based
simulations\footnote{\url{https://github.com/simonsobs/map_based_simulations/}} 
developed for Simons Observatory \citep{SimonsObs_2019} for this purpose (Zonca et al., in prep.). 
We create maps at 93 and 145\,GHz (the available bandpasses in the simulations are slightly different to ACT)
on the \citetalias{Naess_2020} pixelization, containing a realization of the CMB plus the 
Cosmic Infrared Background (CIB) as implemented in
WebSky\footnote{\url{https://mocks.cita.utoronto.ca/index.php/WebSky_Extragalactic_CMB_Mocks}}
\citep{Stein_2020}. 
Since a complete model suitable for generating random realizations of the noise in
the \citetalias{Naess_2020} maps is not currently available 
(and there are no splits of the \citetalias{Naess_2020} maps), we add white noise to the simulated maps 
following the levels in the \citetalias{Naess_2020} inverse variance maps. This means that the false
positive rate inferred from these maps will be slightly optimistic, but as shown in 
Section~\ref{sec:purity}, it is a reasonable match to the purity of the real cluster sample as assessed
from regions with deep optical data. We apply all the same masks to the signal-free simulated maps
as are used on the real maps, so the resulting catalog is drawn from a simulated survey 
with exactly the same area as the real ACT DR5.

\begin{figure}
\includegraphics[width=\columnwidth]{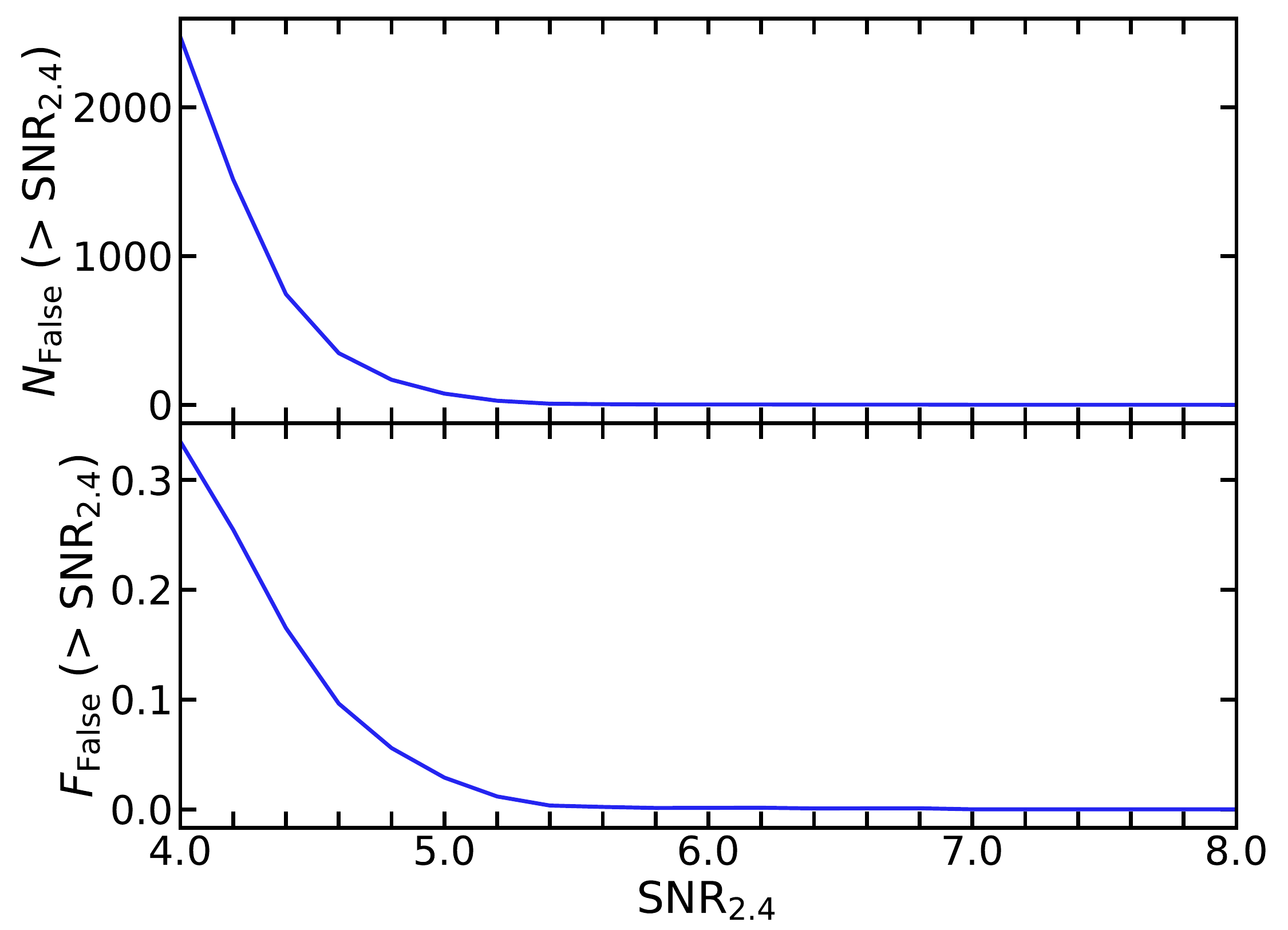}
\caption{The number of false positive detections ($N_{\rm False}$; upper panel) above
a given SNR$_{2.4}$ cut resulting from running the cluster finder on signal-free simulated
maps with the same survey area, masks, and pixelization as the real \citetalias{Naess_2020} 
maps. The lower panel shows the fraction of false positives ($F_{\rm False}$) expected in the 
real ACT DR5 candidate list above a given SNR$_{2.4}$ cut. As shown in Section~\ref{sec:purity},
the simple simulations used here are a reasonable match to the fraction of clusters recovered 
in regions where deep optical observations are used for confirmation.}
\label{fig:falsePositives}
\end{figure}

The upper panel of Fig.~\ref{fig:falsePositives} shows the number of detections in the 
signal-free simulation ($N_{\rm False}$) as a function of SNR$_{2.4}$ cut. For SNR$_{2.4} > 4$,
we find there are 2471 false detections, falling to 75 for SNR$_{2.4} > 5$, and 2 for SNR$_{2.4} > 6$. 
For comparison, there are \totalFixedSNRCandidates{}
SNR$_{2.4} > 4$ candidates in the real candidate list (note that the full candidate list is not
provided with this paper; we release only the catalog of optically confirmed clusters).
Assuming that $N_{\rm False}$ is a reasonable 
estimate of the false positive rate in the real cluster candidate list, we can estimate
the fraction of false positives as $F_{\rm False} = N_{\rm False} / N_{\rm Total}$, where
$N_{\rm Total}$ is the number of objects in the real ACT DR5 candidate list. This is shown in the
lower panel of Fig.~\ref{fig:falsePositives}. We find that $F_{\rm False} = 0.34$ for
SNR$_{2.4} > 4$, 0.03 for SNR$_{2.4} > 5$, and 0.001 for SNR$_{2.4} > 6$. 
Note that while these figures are a survey-wide average, we find little difference if we
repeat this exercise considering only deeper parts of the map (e.g., $F_{\rm False}$ differs by
$<2$\% if we compare the footprint that overlaps with HSCSSP, where the ACT observations are deepest,
with the whole survey footprint).
We caution that these
figures represent lower limits to the contamination rate in the candidate list, as 
the simulations used here do not capture all of the possible noise sources in the real maps.
We compare $1-F_{\rm False}$ to the fraction of optically confirmed clusters in Section~\ref{sec:purity}.

\subsection{Cluster Characterization}
\label{sec:SZMasses}

In this work we continue to use the same approach to characterizing the SZ signal and its relation to mass
as introduced in \citet{Hasselfield_2013} and used in the ACTPol cluster search \citep{Hilton_2018}. Briefly,
we choose to characterize the SZ signal and survey completeness by selecting a single reference filter 
scale of angular size $\theta_{\rm 500c} = 2.4$\arcmin{}, which corresponds to a UPP-model cluster with mass 
$M_{\rm 500c} = 2 \times 10^{14}$\,$M_{\sun}$ at $z = 0.4$ (close to the median redshift of the sample) 
for our fiducial cosmology. This avoids inter-filter noise bias, where local noise variations
(e.g., the presence of CMB cold spots near candidates) can affect estimates of the cluster signal (and size)
based on the maximal signal-to-noise filter scale \citep[see the discussion in][]{Hasselfield_2013}.
However, we note that since the cluster finder still maximizes SNR over location on the sky,
there is still a small positive bias in the recovered SNR values 
($\approx 7$\% at SNR$_{2.4} = 4.0$; see, e.g., \citealt{Vanderlinde_2010}). 

For a map filtered at the fixed 2.4\arcmin{} reference scale, we assume that the cluster central Compton parameter 
$\tilde{y}_{0}$ is related to mass through
\begin{equation}
\tilde{y}_{0} = 10^{A_0} E(z)^2 \left( \frac{M_{\rm 500c}}{M_{\rm pivot}} \right)^{1+B_0} Q(\theta_{\rm 500c}) f_{\rm rel} (M_{\rm 500c}, z) \, ,
\label{eq:y0}
\end{equation}
where $10^{A_0} = 4.95 \times 10^{-5}$ is the normalization, $B_0 = 0.08$,  $M_{\rm pivot} = 3 \times 10^{14}$\,$M_{\sun}$, $Q(\theta_{\rm 500c})$ is the filter mismatch function 
($\theta_{\rm 500c} = R_{\rm 500c}/D_{\rm A}$, where $D_{\rm A}$
is the angular diameter distance), and $f_{\rm rel}$ is a relativistic correction.
$E(z)$ describes the evolution of the Hubble parameter with redshift, i.e., 
$E(z) = \sqrt{\Omega_{\rm m}(1+z)^3 + \Omega_{\Lambda}}$.
The parameter values for $10^{A_0}$, $B_0$ and $M_{\rm pivot}$ are equivalent to 
the \citetalias{Arnaud_2010} scaling relation,
which was calibrated using X-ray observations. While this will typically result in masses that are 
lower than those calibrated against weak-lensing measurements 
\citep[e.g.,][in the case of ACTPol]{Miyatake_2019}, we choose to use
the \citetalias{Arnaud_2010} relation here to ease comparison with our previous work.
We also provide an alternative set of masses, rescaled via a richness-based
weak-lensing calibration procedure, as described in Section~\ref{sec:catalogProperties}.

The function $Q(\theta_{\rm 500c})$ in equation~(\ref{eq:y0}) accounts for the mismatch between the 
size of a cluster with a different mass and redshift to the reference model used to define the matched 
filter (including the effect of the beam) and in turn $\tilde{y}_{0}$ (see Section~3.1 of 
\citealt{Hasselfield_2013}; Section~2.3 of \citealt{Hilton_2018}). Since
we break the map into tiles and construct a filter for each tile (Section~\ref{sec:SZDetection}),
each tile has its own $Q(\theta_{\rm 500c})$ function.

We implement the relativistic correction $f_{\rm rel}$ applied in equation~(\ref{eq:y0}) differently in 
this work compared to previous ACT cluster surveys, which were based solely on 150\,GHz data
rather than the 98 and 150\,GHz maps analyzed here. The size of $f_{\rm rel}$ depends on frequency, and
is up to 1\% larger at 150\,GHz than 98\,GHz for very massive 
clusters ($M_{\rm 500c} \approx 8 \times 10^{14}$\,$M_{\sun}$). We use the \citet{Arnaud_2005} 
mass--temperature relation to convert $M_{\rm 500c}$ to temperature at a given cluster redshift, and 
then apply the formulae of \citet{Itoh_1998} to calculate $f_{\rm rel}$ at each frequency. The 
filter $\psi$ defined in equation~(\ref{eq:MFMF}) returns the value $\tilde{y}_{0}$ when 
applied to a set of multi-frequency maps, weighting the contribution of each map to the 
returned SZ signal according to both the spectral dependence of the SZ signal 
(equation~\ref{eq:fSZ}) and the noise in the map. We use these weights, which differ from 
tile to tile, to estimate an average $f_{\rm rel}$ for each cluster. The overall impact of 
the relativistic correction is small (approximately 3\% for the median mass of the ACT DR5 cluster sample).

Equation~(\ref{eq:y0}) cannot be inverted to obtain the mass $M_{\rm 500c}$, due to the steepness of the cluster mass
function and the presence of intrinsic log normal scatter $\sigma_{\rm int}$ in $\tilde{y}_{0}$ about the mean 
relation defined by equation~(\ref{eq:y0}). We adopt $\sigma_{\rm int} = 0.2$ throughout this work,
based on the results of numerical simulations \citep[see][]{Hasselfield_2013}. Given a cluster redshift 
measurement, mass estimates are extracted by computing the posterior probability
\begin{equation}
P( M_{\rm 500c} | \tilde{y}_{0}, z) \propto P ( \tilde{y}_{0} | M_{\rm 500c}, z) P(M_{\rm 500c} | z) \, ,
\label{eq:PM500}
\end{equation}
where $P(M_{\rm 500c} | z)$ is the halo mass function at redshift $z$, for which we use the fitting formulae of \citet{Tinker_2008}, as implemented in the Core Cosmology Library v2.1 \citep[\texttt{CCL}\footnote{\url{https://github.com/LSSTDESC/CCL}};][]{Chisari_2019}. We assume $\sigma_8 = 0.80$ for such calculations throughout this work.
We account for the uncertainties on both $z$ and $\tilde{y}_{0}$ in calculating $P( M_{\rm 500c} | \tilde{y}_{0}, z)$,
and adopt the maximum of the $P( M_{\rm 500c} | \tilde{y}_{0}, z)$ distribution as the cluster 
$M_{\rm 500c}$ estimate. The uncertainties quoted on these masses are $1\sigma$ error bars that do not take into account any uncertainty on the scaling relation parameters.

The mass estimates obtained through equations~(\ref{eq:y0}) and (\ref{eq:PM500}) are referred to as $M^{\rm UPP}_{\rm 500c}$ throughout this work. For comparison with some 
other works (e.g., the \textit{Planck} PSZ2 catalog; \citealt{PlanckPSZ2_2016}), it is sometimes necessary 
to neglect the Eddington bias correction (done by equation~\ref{eq:PM500}) that accounts for the steepness of the cluster mass function and intrinsic scatter (see the discussion in \citealt{Battaglia_2016} and \citealt{Hilton_2018}). We label these `uncorrected' masses as $M^{\rm Unc}_{\rm 500c}$.

\begin{figure}
\includegraphics[width=\columnwidth]{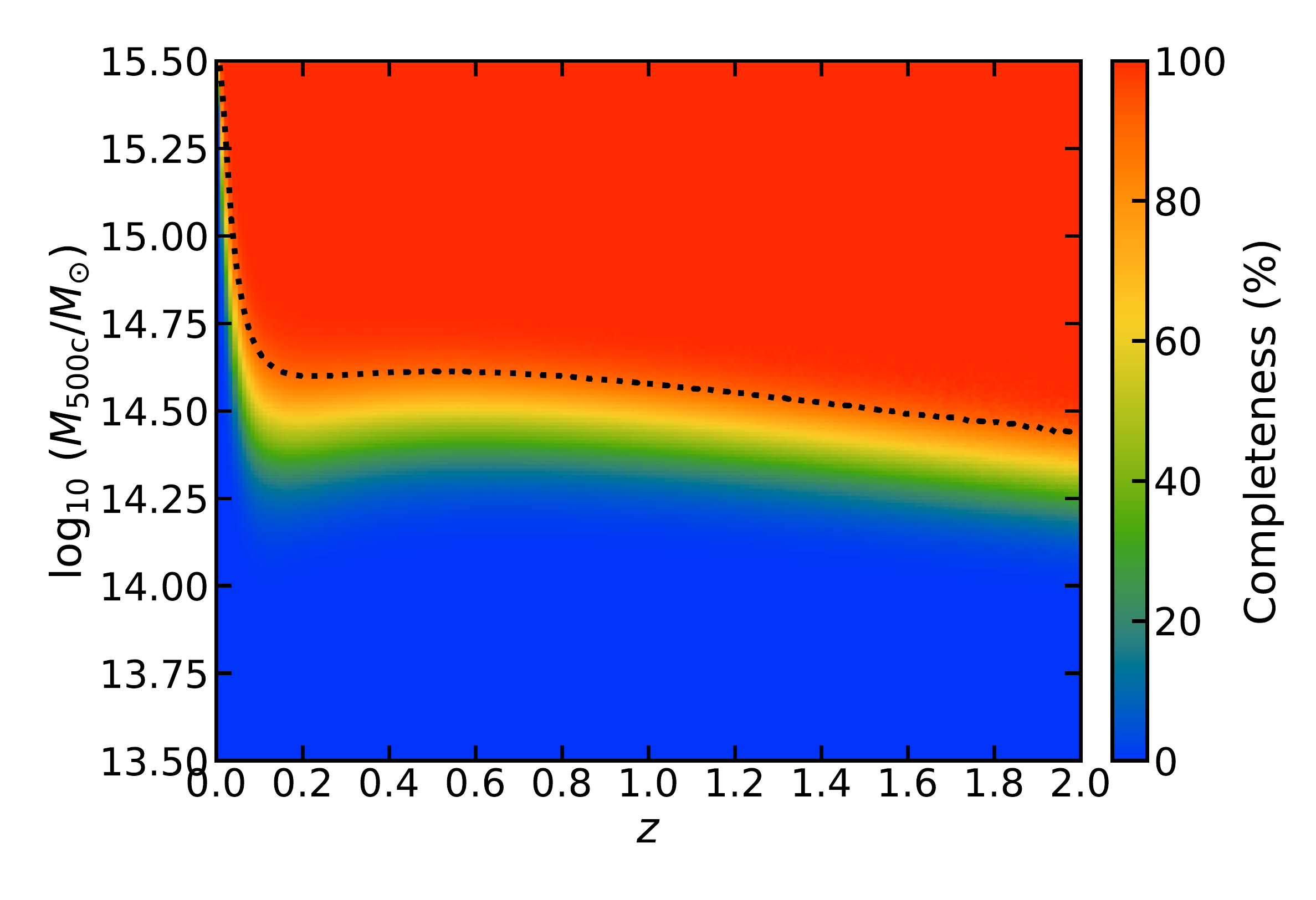}
\caption{The completeness for SNR$_{2.4} >5$ as a function of redshift, 
in terms of $M^{\rm UPP}_{\rm 500c}$, over the full \surveyArea{}\,deg$^2$ survey footprint.
The \citet{Tinker_2008} halo mass function and \citet{Arnaud_2010} scaling relation
are assumed (see Section~\ref{sec:SZCompleteness}). The dashed black contour marks the 
90\% completeness limit.}
\label{fig:completeness}
\end{figure}

\begin{figure*}
\includegraphics[width=\textwidth]{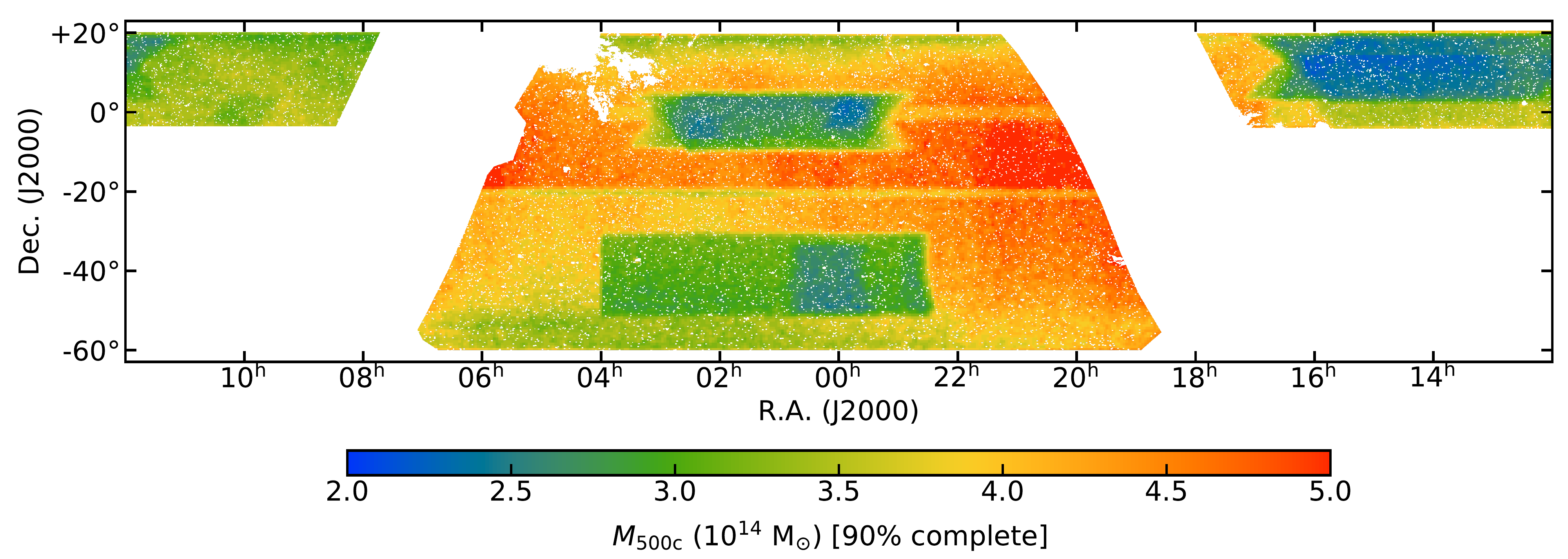}
\caption{Map of the 90\% completeness mass limit for SNR$_{2.4} >5$ as a function of redshift, 
in terms of $M^{\rm UPP}_{\rm 500c}$, evaluated at $z = 0.5$ (the median redshift of the
detected clusters). The variation is driven by the ACT observing strategy.}
\label{fig:completenessMap}
\end{figure*}

\begin{figure*}
\centering
\includegraphics[width=\columnwidth]{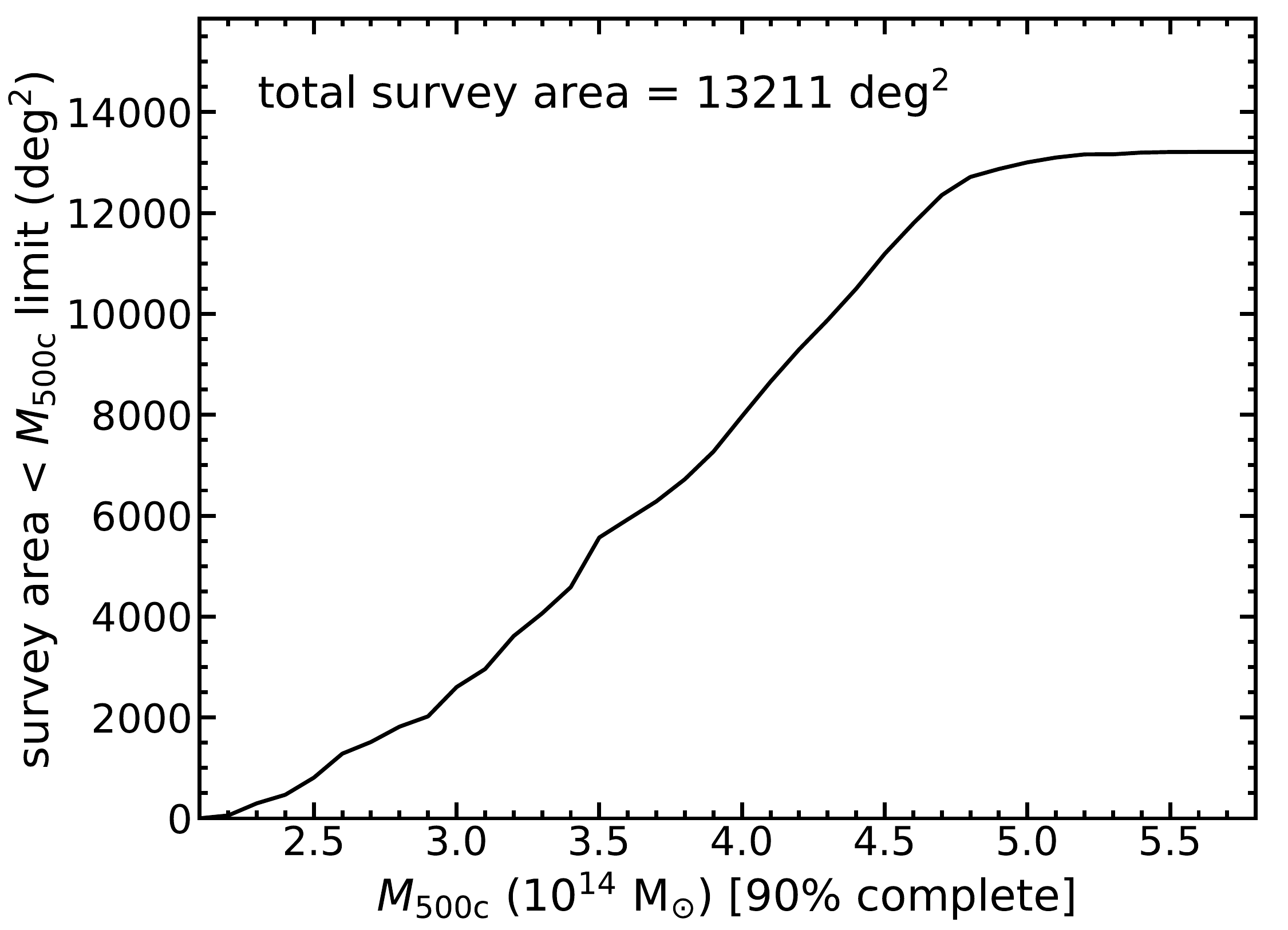}
\includegraphics[width=\columnwidth]{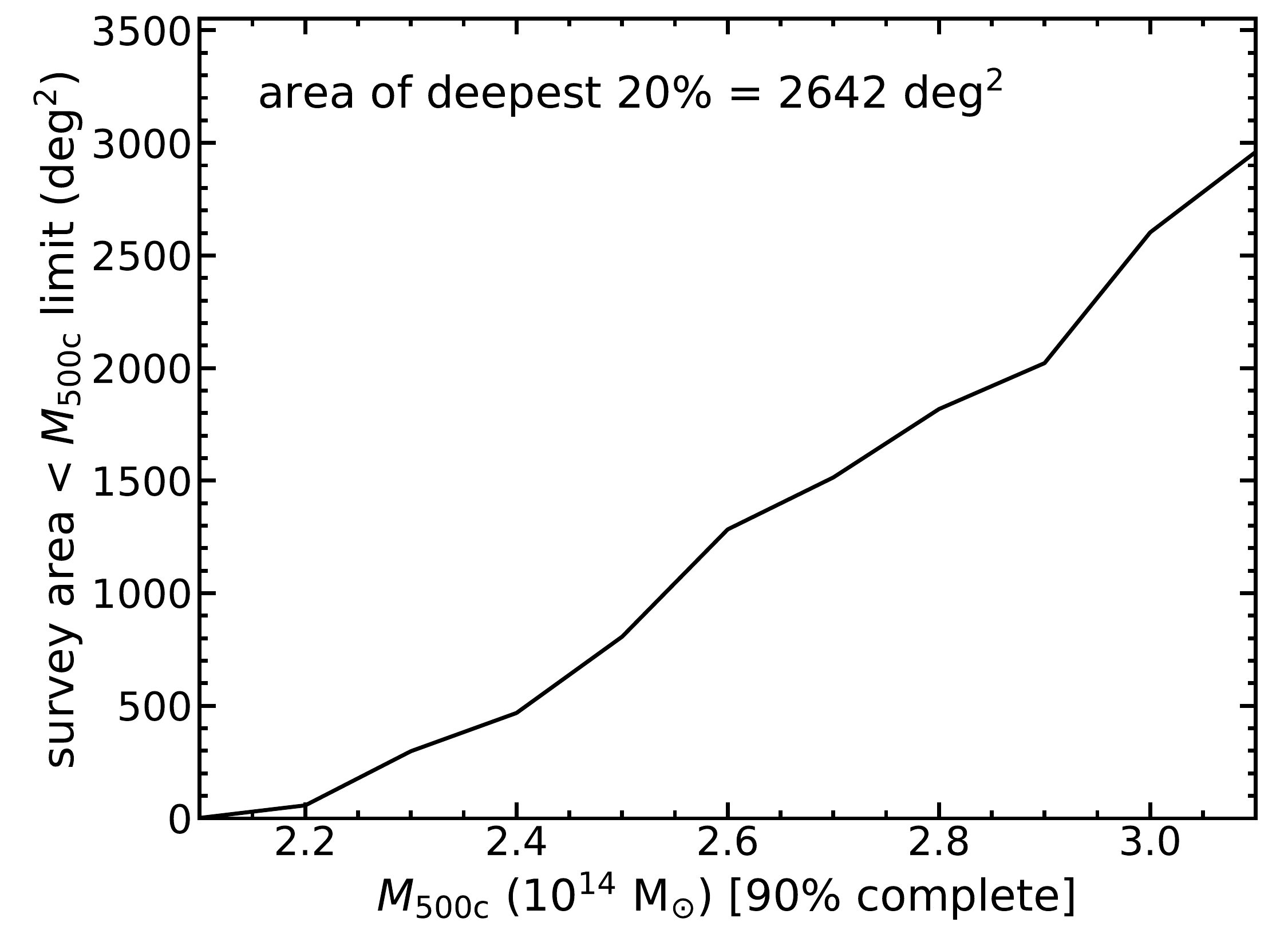}
\caption{Mass sensitivity in terms of $M^{\rm UPP}_{\rm 500c}$, evaluated at $z = 0.5$, as a 
cumulative function of area, for the whole survey (left), and for the deepest 20\% (right).
}
\label{fig:cumulativeArea}
\end{figure*}

\subsection{Survey Completeness}
\label{sec:SZCompleteness}

We estimate the completeness of the survey in terms of mass using mock catalogs generated through Monte Carlo
simulations. For speed, the calculations are performed on a redshift grid, covering the range $0 < z < 2$
in steps of size $\Delta z = 0.01$. At each redshift step, we make 2 million draws from 
the \citet{Tinker_2008} halo mass function, above a minimum halo mass of 
$M_{\rm500c} > 8 \times 10^{13}$\,$M_{\sun}$ (i.e., well below the expected mass limit).
We then calculate the true value of $\tilde{y}_{0}$ for each of the randomly drawn halo masses using
equation~(\ref{eq:y0}), assuming the scaling relation parameters derived from \citetalias{Arnaud_2010}.
Here we apply the appropriate filter mismatch function ($Q$) for the tile each
mock cluster is located in, and apply the relativistic correction as described in Section~\ref{sec:SZMasses}. 
We then add Gaussian-distributed random noise to $\tilde{y}_{0}$, according to the level estimated in the
$\tilde{y}_{0}$ noise map, and finally we add log-normal scatter 
to $\tilde{y}_{0}$ with size $\sigma_{\rm int} = 0.2$ (see Section~\ref{sec:SZMasses}). After repeating
this for each redshift step and each map tile (see Section~\ref{sec:SZDetection}), 
we have assembled an oversampled mock catalog containing true masses, redshifts, 
and mock $\tilde{y}_{0}$ values (and their uncertainties) over the full ACT DR5 cluster search area that 
extends well below the mass selection limit. We then
project this catalog onto a ($\log_{10} M_{\rm 500c}, z$) grid, and estimate the completeness as the
fraction of the mock clusters in each ($\log_{10} M_{\rm 500c}, z$) bin that are above a chosen 
SNR$_{2.4}$ detection threshold. We repeat this process 1000 times, taking the average as the estimate
of the overall survey completeness.

Fig.~\ref{fig:completeness} shows the 90\% completeness limit as a function of redshift in 
terms of $M^{\rm UPP}_{\rm 500c}$ for SNR$_{2.4} > 5$ over the full \surveyArea{}\,deg$^2$
survey area. Evaluated at $z = 0.5$ (approximately the median redshift of the cluster sample;
see Section~\ref{sec:Redshifts}), we estimate that the cluster catalog is 90\% complete
for $M^{\rm UPP}_{\rm 500c} > \compLimitFull{} \times 10^{14}\,M_{\sun}$. The survey is
slightly more sensitive to lower mass clusters than this in areas that overlap with
the DES, HSC, and KiDS optical surveys ($M^{\rm UPP}_{\rm 500c} > 3.6 \times 10^{14}\,M_{\sun}$).
This statement relates only to the noise levels in the ACT maps in the regions of overlap,
i.e., no optical information is used in deriving estimates of the survey mass limit.

There is a fairly large spatial variation in the mass completeness limit across the map, 
as shown in Fig.~\ref{fig:completenessMap}, which is driven by the ACT observing strategy.
Fig.~\ref{fig:cumulativeArea} shows the cumulative survey area as a function of the estimated
mass completeness limit. Almost all of the vastly-increased survey area reaches a lower mass
limit than previous ACT cluster surveys \citep{Marriage_2011, Hasselfield_2013, Hilton_2018}. 
Clusters with masses in the range $2.1 < M^{\rm UPP}_{\rm 500c} / 10^{14}\,M_{\sun} < 3.1$ can be detected
in the deepest 20\% of the survey, corresponding to an area of 2634\,deg$^{2}$ -- more than 
double the area searched in \citet{Hilton_2018}, and larger than the area searched in the
SPT-SZ survey \citep{Bleem_2015, Bocquet_2019}.

\section{Optical/IR Follow-up and Redshifts}
\label{sec:Redshifts}

In this Section we describe the process of optical/IR confirmation of SZ-detected candidates as 
clusters of galaxies. The redshifts assigned to objects in the cluster catalog come from a variety of sources,
because the ACT DR5 cluster search area is not covered by a single, deep optical/IR survey.
We have attempted to obtain as many reliable redshift estimates as possible, given the data available.
We provide details on each of the redshift sources in Section~\ref{sec:redshiftSources} below.
Section~\ref{sec:visualInspect} summarizes the process of cross matching the cluster candidate list 
against external catalogs, visual inspection of the available optical/IR data, and the process by which
we adopted a single redshift measurement for each cluster. We comment on redshift follow-up completeness
and the purity of the cluster candidate list in Section~\ref{sec:purity}.

\begin{figure*}
\centering
\includegraphics[width=58mm]{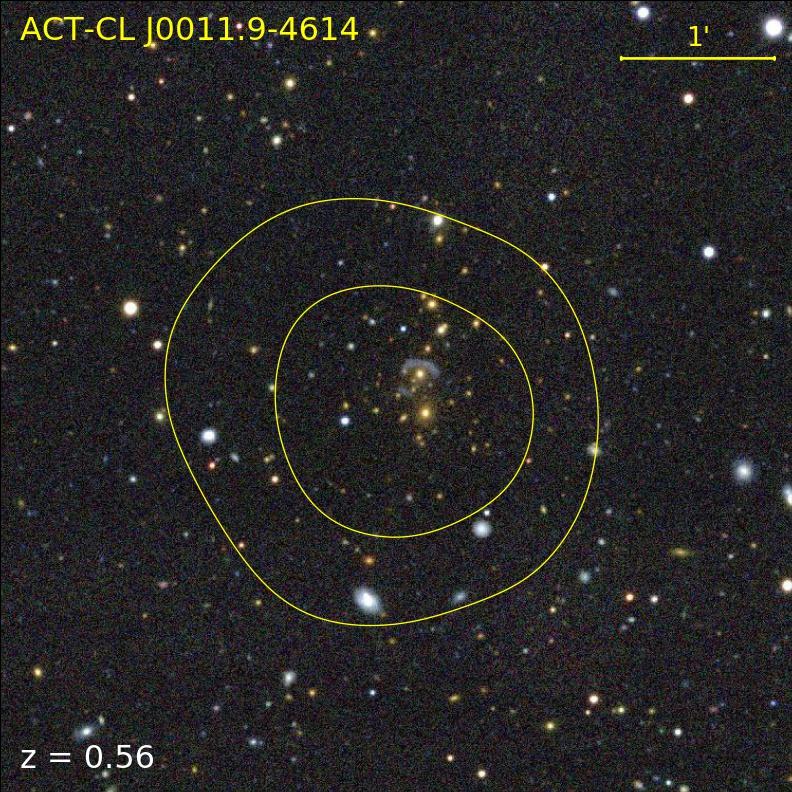}
\includegraphics[width=58mm]{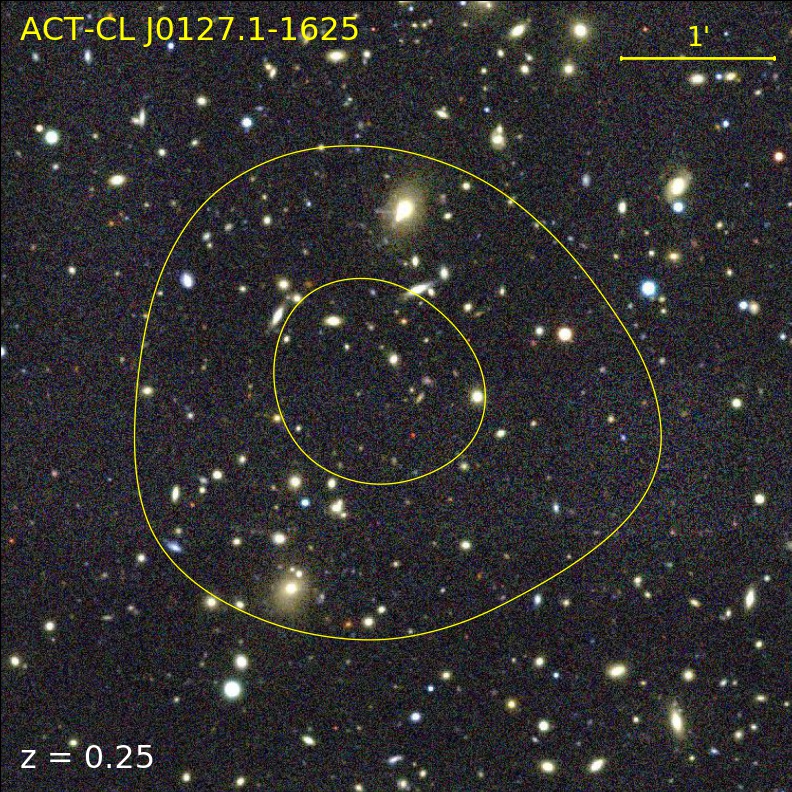}
\includegraphics[width=58mm]{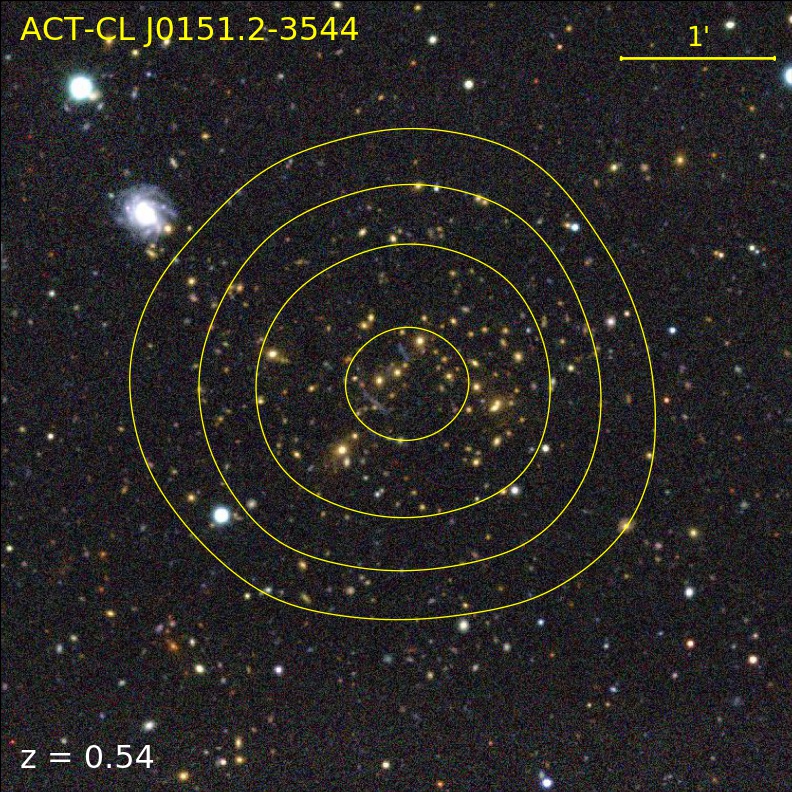}
\includegraphics[width=58mm]{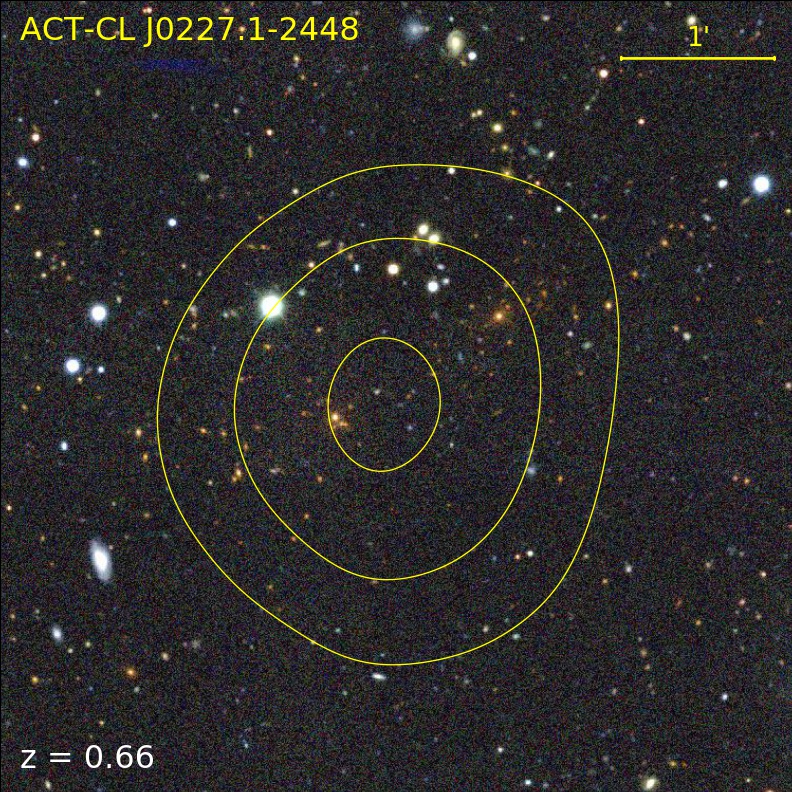}
\includegraphics[width=58mm]{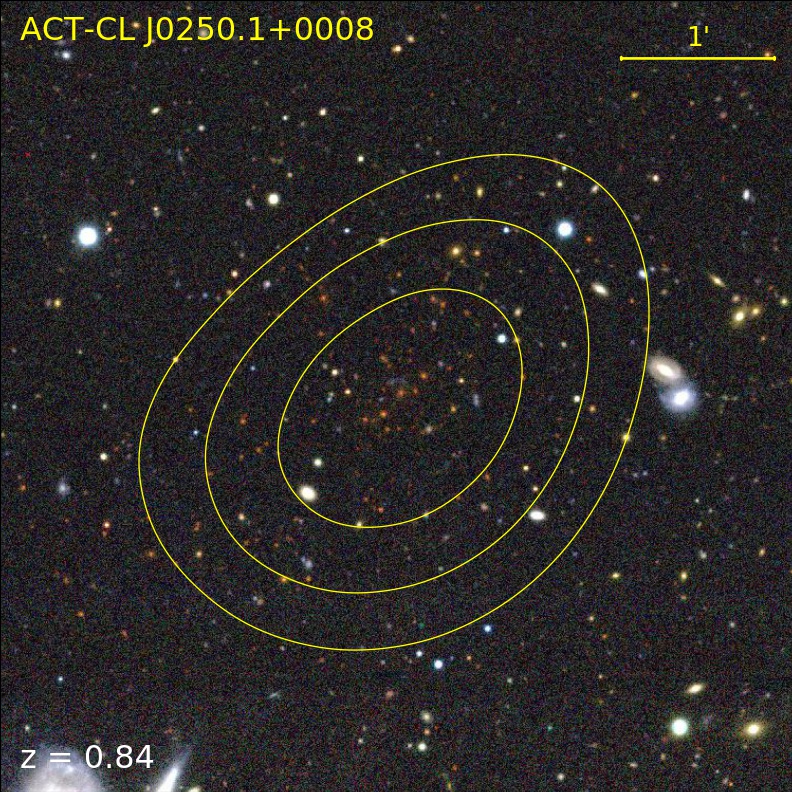}
\includegraphics[width=58mm]{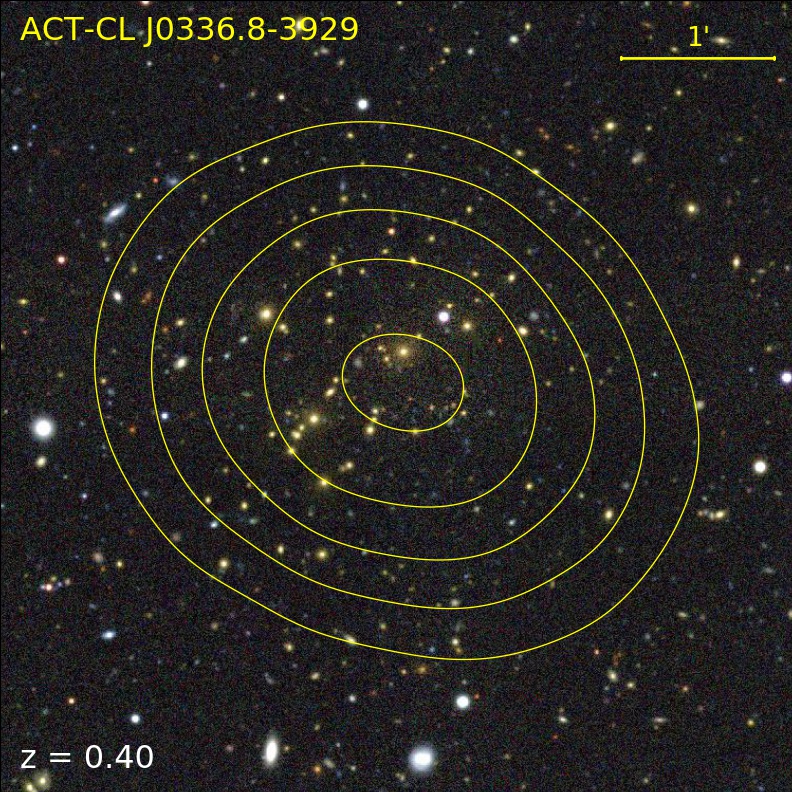}
\caption{Example DES $gri$ images of ACT DR5 clusters at various redshifts confirmed using redMaPPer. Each
image is 5\arcmin{} on a side, with North at the top, East at the left. The contours mark signal-to-noise
ratio in the ACT map filtered at the reference 2.4\arcmin{} scale. The lowest level shown corresponds to 
3$\sigma$ significance, and each subsequent level is 2$\sigma$ higher.}
\label{fig:RMDESY3}
\end{figure*}

\subsection{Redshift Sources}
\label{sec:redshiftSources}


\subsubsection{Large Public Spectroscopic Surveys}
\label{sec:publicSpec}
The large ACT DR5 survey area overlaps with several large public spectroscopic surveys. In this work we made 
use of 2dFLenS \citep{2016MNRAS.462.4240B}, OzDES \citep{2017MNRAS.472..273C}, SDSS DR16 \citep{2020ApJS..249....3A},
and VIPERS \citep{2018AandA...609A..84S}. We cross matched the cluster candidate list against each of these
surveys in turn, and estimated cluster redshifts using an iterative procedure similar to that used in 
\citet{Hilton_2018}. For each cluster in the list, we first select only galaxies with secure spectroscopic
redshifts located within a projected distance of 1\,Mpc from the cluster SZ position. 
We then iteratively
estimate the cluster redshift using the biweight location estimator \citep[e.g.,][]{Beers_1990},
keeping only galaxies with peculiar velocities within 3000\,km\,s$^{-1}$ 
of the cluster redshift estimated at each iteration.
In some iterations, there may be no galaxies found within these peculiar velocity limits 
(e.g., on rare occassions where the redshift distribution is bimodal). In these cases, we disregard the
peculiar velocity cut, and take the median of all the galaxy redshifts as the cluster redshift estimate, 
before beginning the next iteration.
This procedure typically converges within a couple of iterations.

SDSS DR16 provides the vast majority of spectroscopic redshifts assigned to clusters in the final catalog (1123),
followed by 2dFLens (56), OzDES (3), and VIPERS (2). Note that following visual inspection of optical imaging
(Section~\ref{sec:visualInspect}), we rejected 56 cases of erroneous redshift estimates produced by the 
above automated procedure in favour of a ``manually assigned'' spectroscopic redshift (e.g., based on an obvious brightest
central galaxy). These objects are flagged in the \texttt{warnings} field of the cluster catalog (see Table~\ref{tab:FITSTableColumns}).

\begin{figure*}
\centering
\includegraphics[width=58mm]{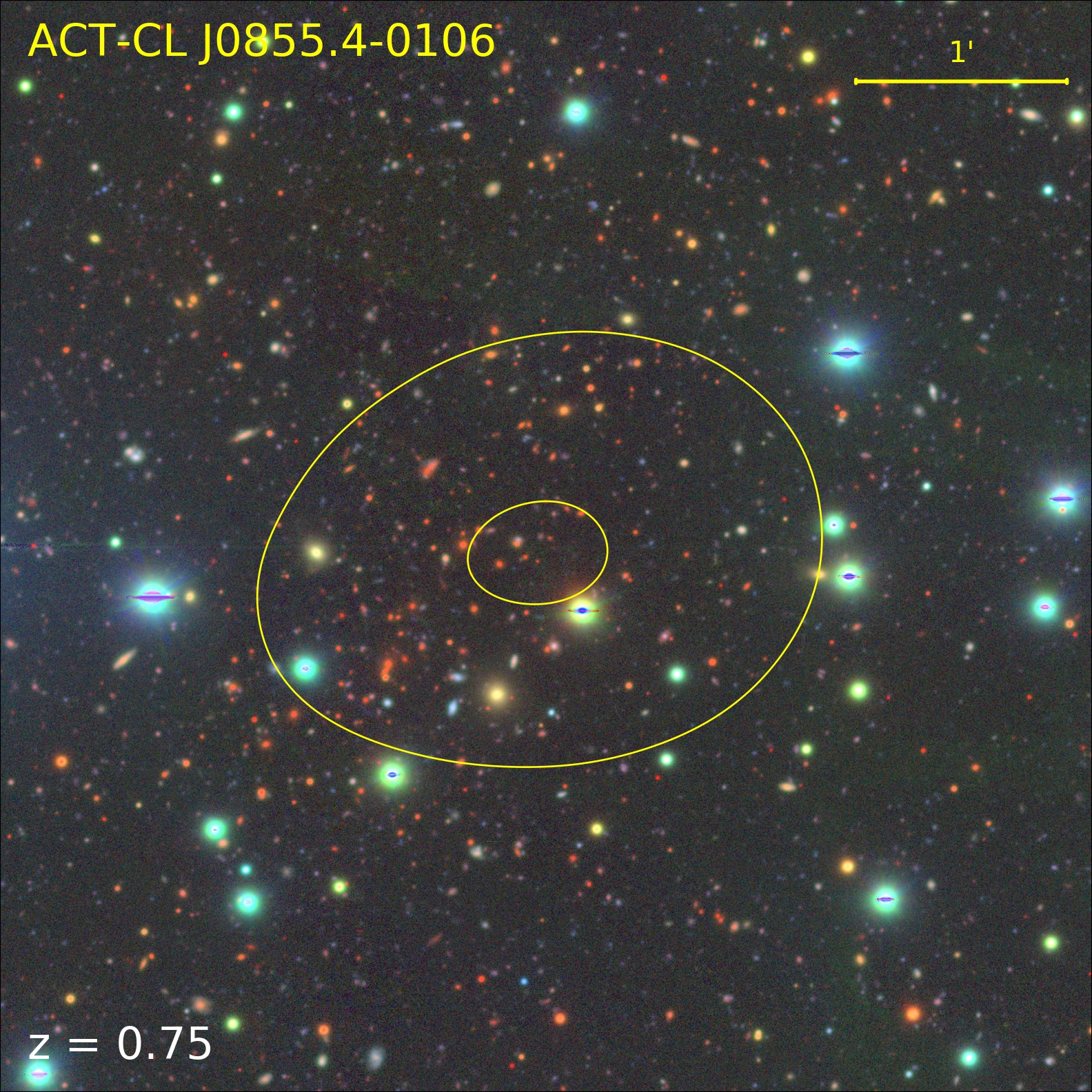}
\includegraphics[width=58mm]{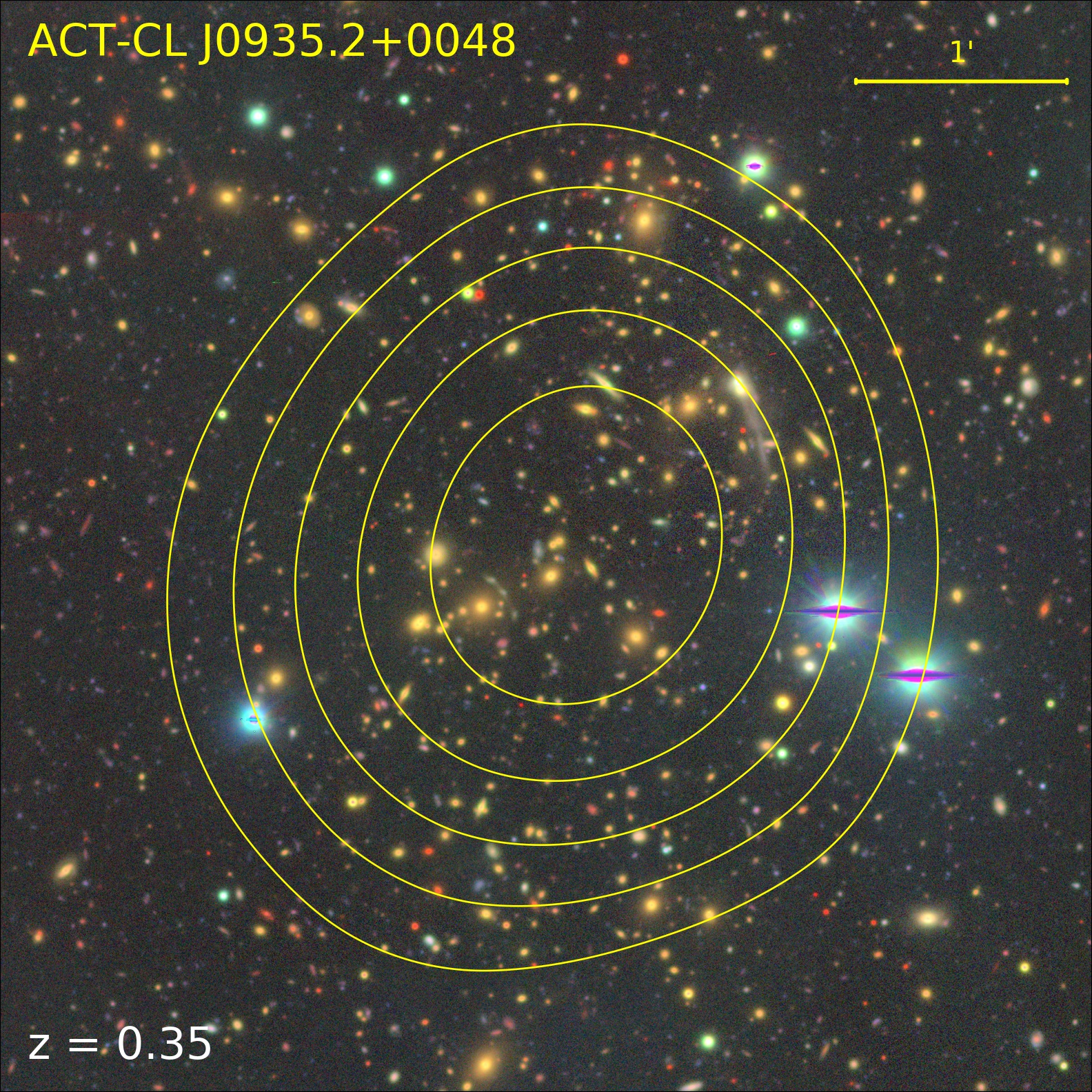}
\includegraphics[width=58mm]{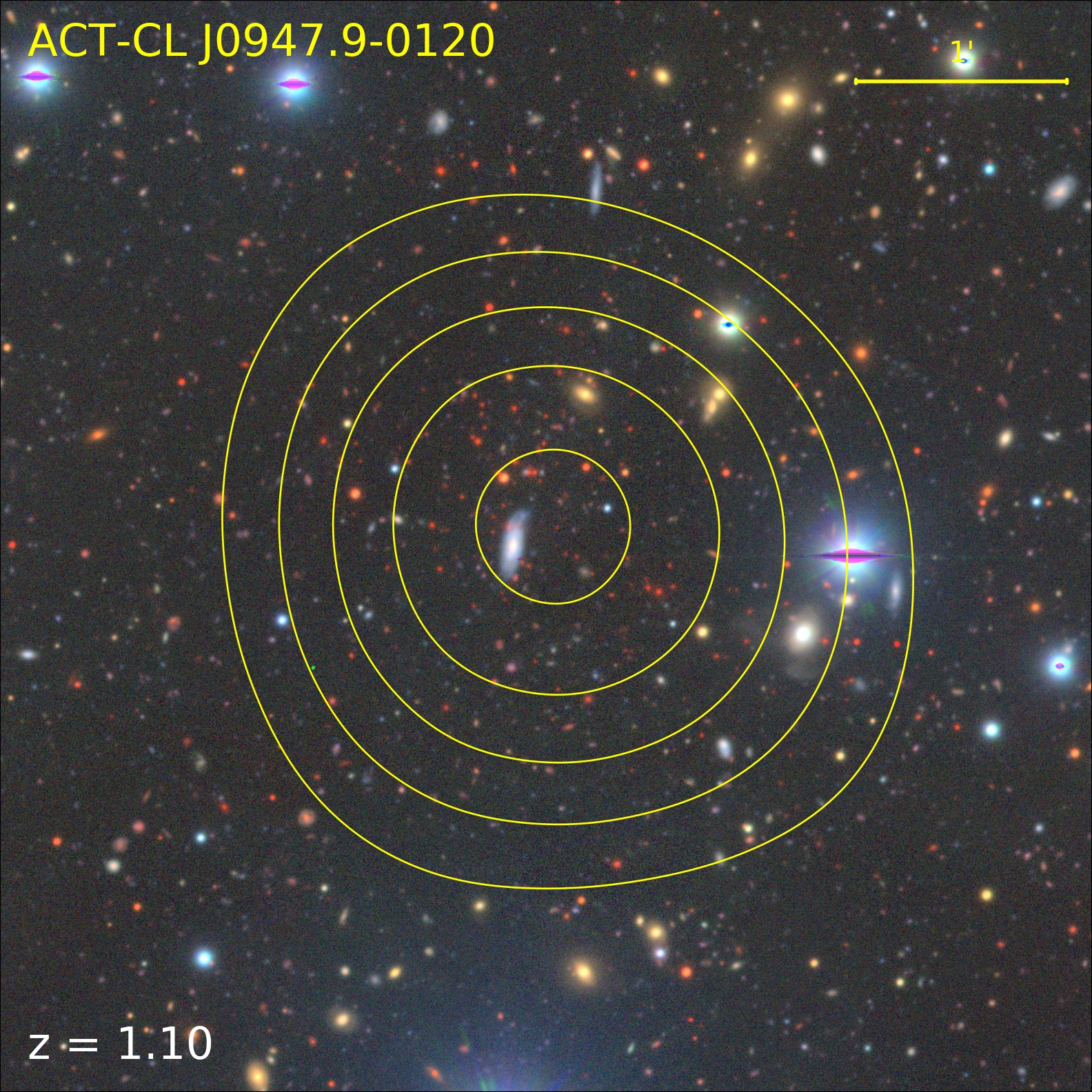}
\includegraphics[width=58mm]{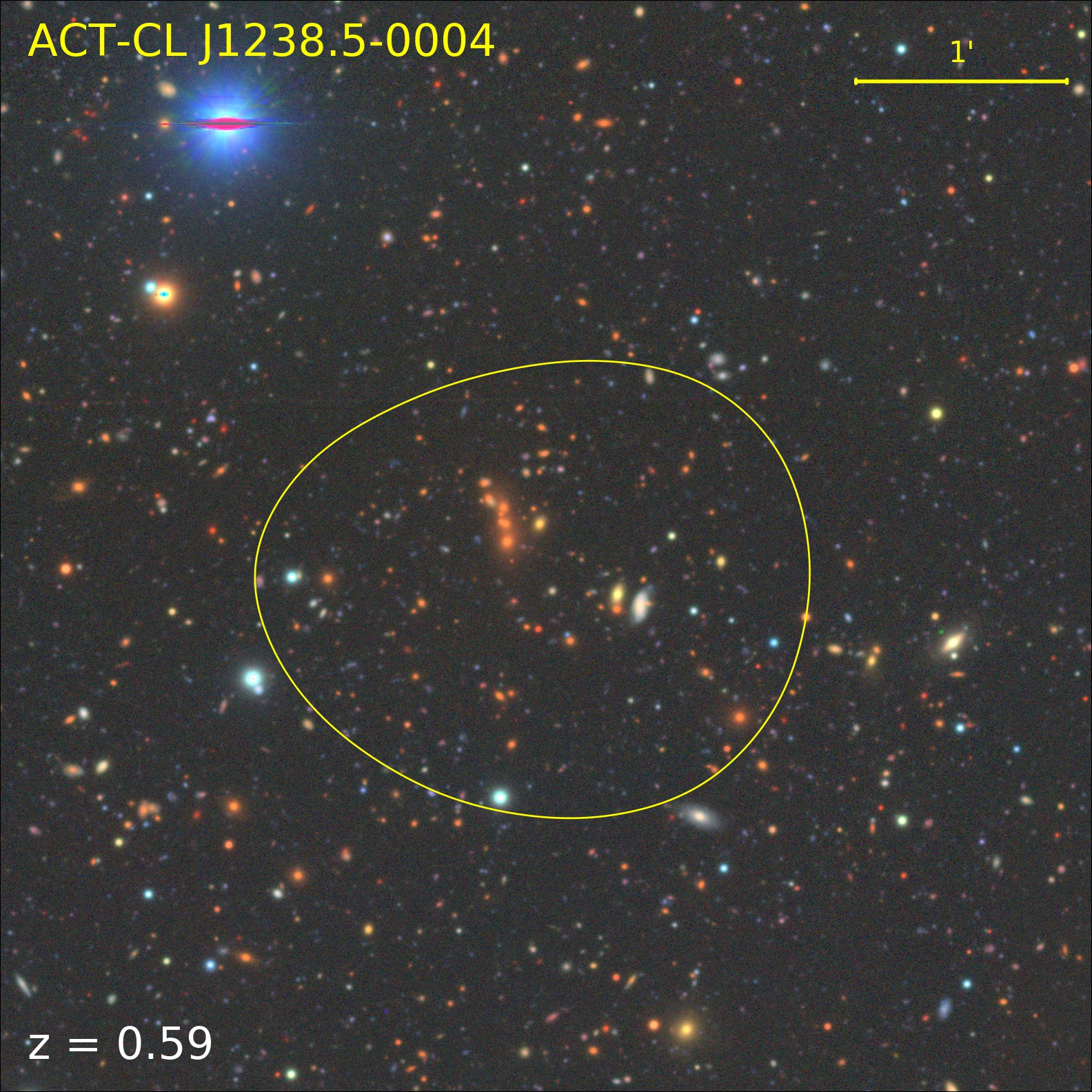}
\includegraphics[width=58mm]{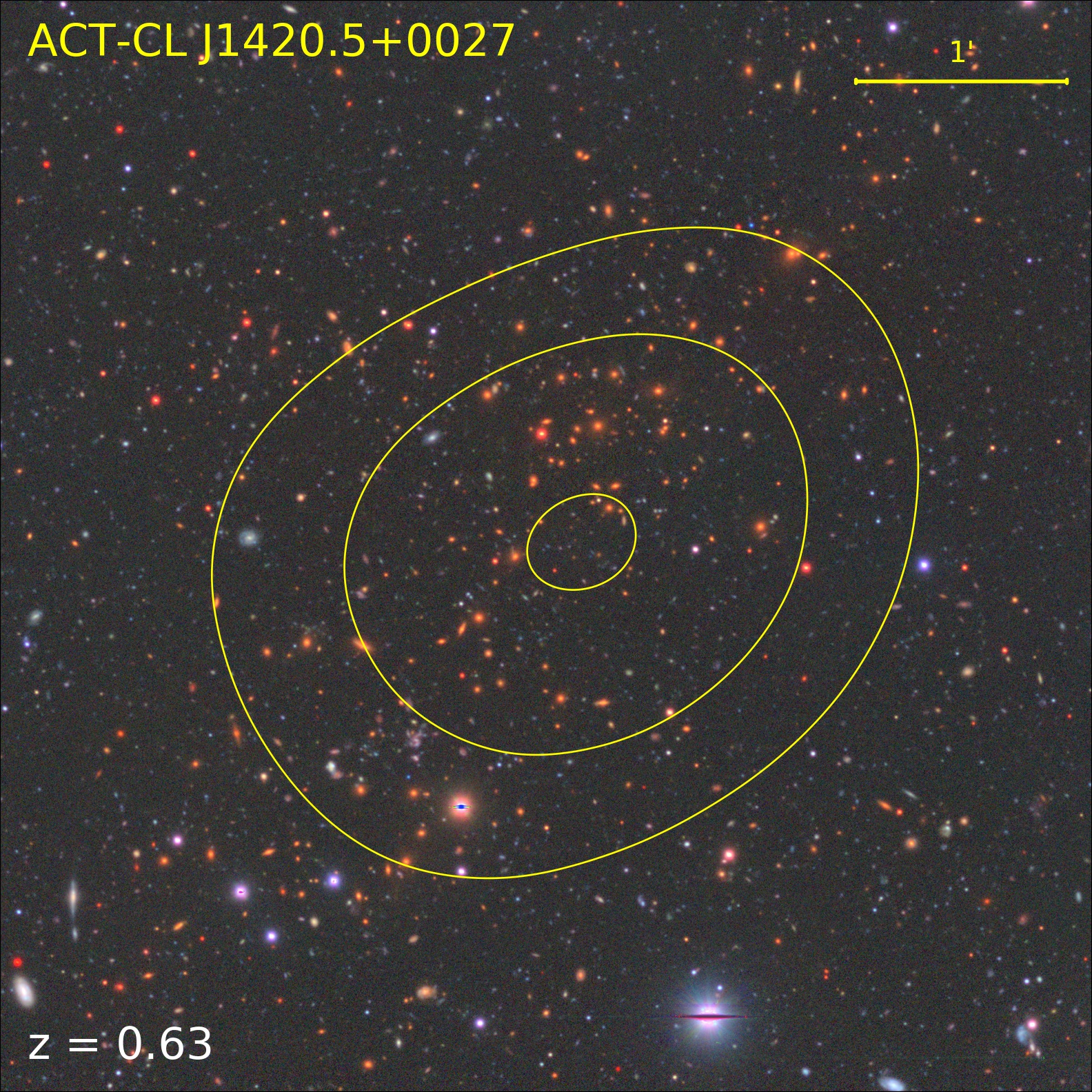}
\includegraphics[width=58mm]{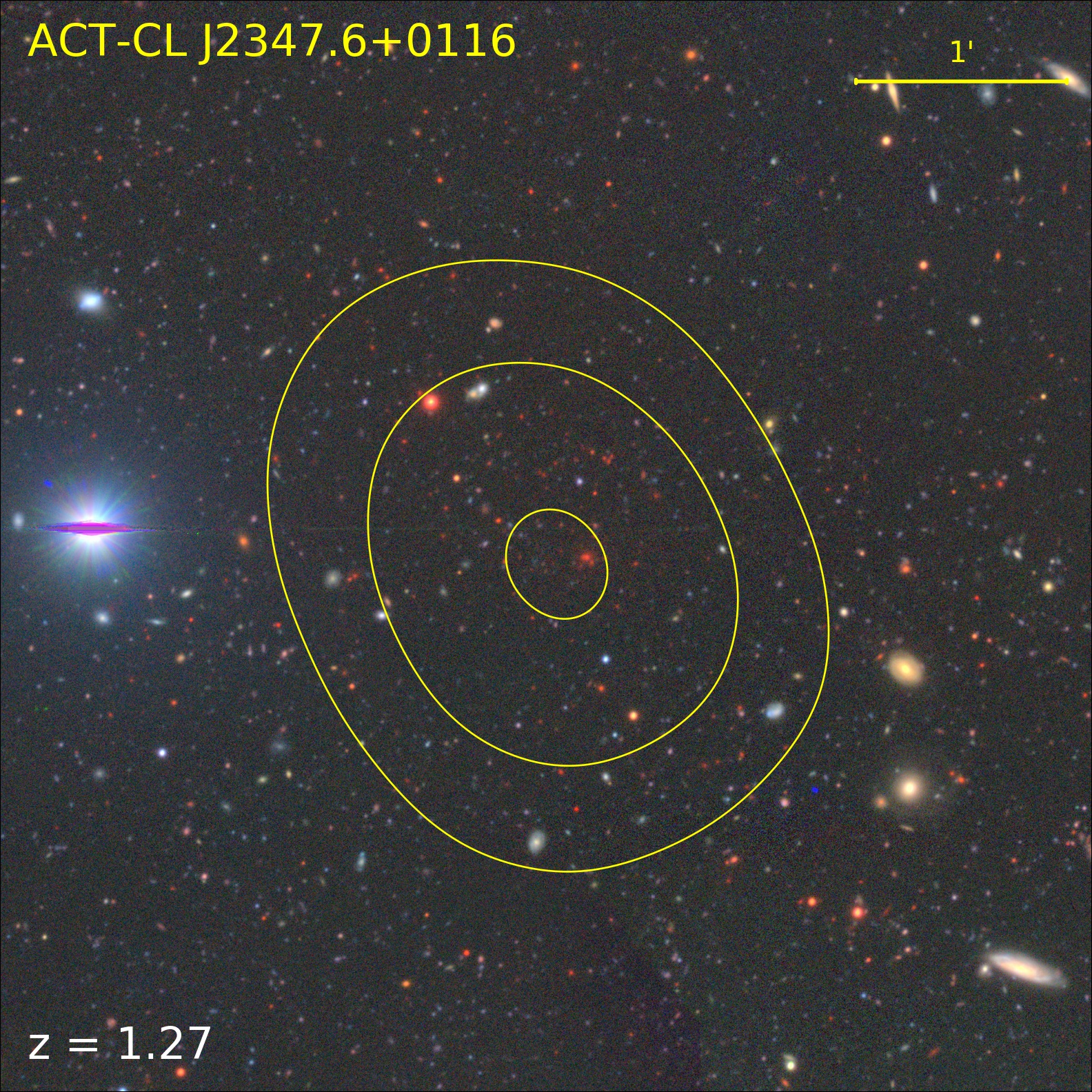}
\caption{Example $gri$ images of ACT DR5 clusters at various redshifts, confirmed using HSC imaging
and the CAMIRA optical cluster finder. Each image is 5\arcmin{} on a side, with North at the top, 
East at the left. The ACT signal-to-noise contours have the same scaling as in Fig.~\ref{fig:RMDESY3}.}
\label{fig:CAMIRA}
\end{figure*}

\subsubsection{Photometric Redshifts From RedMaPPer}
\label{sec:RM}

The cluster search area has a large overlap with SDSS (in equatorial regions) and \DESOverlapArea{}\,deg$^2$
in common with the deep $griz$ imaging provided by DES - i.e., almost all of the DES footprint 
(see Fig.\ref{fig:surveyArea}). The DES data used in this
work come from the first three years of observations (referred to throughout this paper as ``DES Y3''),
for which the imaging and photometric catalogs are publicly available as DES DR1\footnote{\url{https://des.ncsa.illinois.edu/releases/dr1}} \citep{Abbott_2018}. 

RedMaPPer is an optical
red-sequence based cluster finding algorithm that was applied to $ugriz$ SDSS data \citep{Rykoff_2014}, and
has subsequently been developed to run on DES photometry \citep{Rykoff_2016}. In SDSS, redMaPPer is able to
find clusters out to $z \approx 0.5$, while the increased depth of DES allows it to find clusters out to 
$z \approx 0.9$. One of the key features of redMaPPer is its optical richness measurement ($\lambda$), which
has been shown to scale with cluster mass \citep[e.g.,][]{Simet_2017, McClintock_2019}.
The photometric redshift estimates provided by redMaPPer are very accurate, with $\sigma_z/(1+z) < 0.02$
over the full redshift range probed in each survey.

In this work we use the public SDSS redMaPPer catalog \citep[v6.3;][]{Rykoff_2014} and a new redMaPPer
catalog based on the DES Y3 photometry (v6.4.22), containing 33,654 clusters. Both catalogs contain 
only $\lambda > 20$ systems; at this richness, only $5-7$\% of the clusters are expected to be projections 
along the line of sight \citep{Rykoff_2014, Rykoff_2016}. Since the SZ-selected ACT DR5 cluster catalog may
contain clusters at high redshift ($z > 0.8$) that may not be found by redMaPPer alone, we also ran 
redMaPPer in `scanning mode', using the prior information of the ACT cluster candidate positions. 
We found that there is 5\% chance association probability of detecting a $\lambda > 20$ system by 
using redMaPPer in this mode, from a test based on a mock ACT DR5 catalog containing $>93,000$ random 
positions within the DES Y3 footprint, generated from the $\tilde{y}_0$ noise map. 
Note that this represents
the average chance association probability; it is possible that this quantity varies with redshift
\citep[see the treatment in][]{Klein_2019}.
\citet{Bleem_2020} applied the redMaPPer scanning mode to the SPT Extended Cluster Survey (SPT-ECS),
and report a similar chance association probability to that which we find between ACT DR5 and redMaPPer.

We adopted redMaPPer redshifts for 1433 clusters in the ACT DR5 catalog (256 from SDSS, 1023 from DES Y3, and
a further 154 from the `scanning mode' run in DES Y3). This is the most from any of the redshift sources
used in this work. Fig.~\ref{fig:RMDESY3} shows some example images of clusters confirmed using redMaPPer in DES Y3.

\subsubsection{Photometric Redshifts from CAMIRA}
\label{sec:CAMIRA}
The Hyper Suprime-Cam Subaru Strategic Program (HSC-SSP) is a deep optical $grizy$ survey 
reaching to depths fainter than 26th magnitude in the $r$-band \citep{Miyazaki_2018, Aihara_2018}. The 
HSC-SSP full-depth full-color (FDFC) footprint corresponding to observations up to the 2019A semester has \HSCOverlapArea{}\,deg$^2$ of overlap with the ACT DR5 cluster search area, as shown in Fig.~\ref{fig:surveyArea}.

\begin{figure*}
\includegraphics[width=\textwidth]{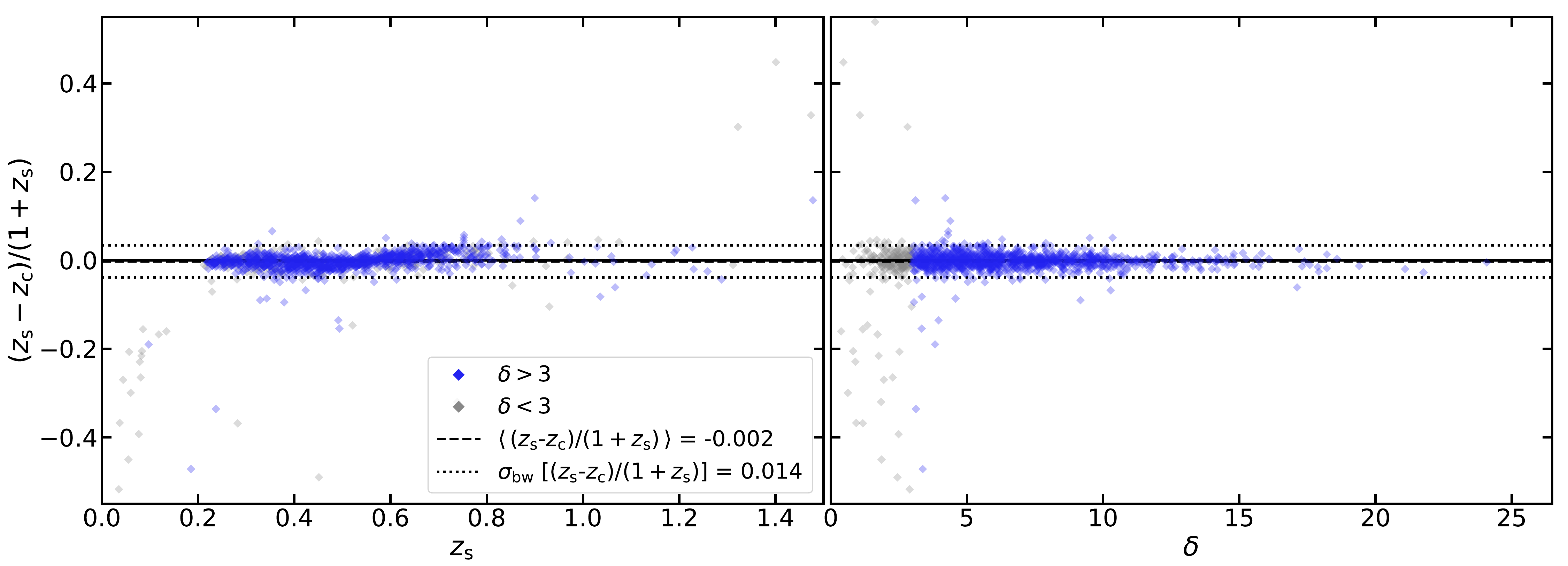}
\caption{The left panel shows a comparison of spectroscopic redshifts ($z_s$) with zCluster photometric redshifts ($z_c$)
based on DECaLS DR8 photometry. More than 98\% of the objects are recovered within $\Delta z/(1+z_{\rm s}) < 0.05$ of the 
spectroscopic redshift. Note that a bias correction of the form $z = z_{\rm c}+0.02(1+z_{\rm c})$ has been
applied to the photometric redshifts. The right panel illustrates how the scatter in the photometric
redshifts varies with the $\delta$ statistic. Objects with $\delta > 3$ are highlighted; many objects with
$\delta$ below this threshold have accurate redshift estimates, but the scatter is much larger.}
\label{fig:zClusterDECaLSComparison}
\end{figure*}

An optical cluster finding algorithm named CAMIRA 
\citep[Cluster finding Algorithm based on Multi-band Identification of Red-sequence gAlaxies;][]{Oguri_2014}, 
which is similar to redMaPPer but was developed independently, has been run on the HSC data.
Here we use the CAMIRA cluster catalog based on HSC-SSP S19A photometry; note that the CAMIRA
cluster search uses a less conservative mask and covers slightly more area than the FDFC mask.
The photometric redshift estimates 
provided by CAMIRA have low scatter ($\sigma_z/(1+z) = 0.008$ at $z < 1.1$; $\sigma_z/(1+z) \approx 0.02$ for $z > 1.1$), 
and reach to $z \approx 1.4$ 
\citep[higher than the $z < 1.1$ limit in the S16A catalog;][]{Oguri_2018}. 
The richness measure used in CAMIRA ($N_{\rm mem}$) counts the number of red-sequence galaxies
in a background-corrected circular aperture, in a similar 
manner to the $\lambda$ quantity used in redMaPPer (see \citealt{Oguri_2014} for a detailed definition).
Similarly to redMaPPer, we also ran CAMIRA in `scanning mode', using prior information
of ACT candidate positions. We find that the 5\% chance association probability corresponds to a richness
threshold of $N_{\rm mem} > 16$, by running the algorithm on a catalog of random positions drawn from 
a mock ACT DR5 cluster catalog. We use this to set the minimum $N_{\rm mem}$ threshold when considering 
cross matches against the CAMIRA catalog.

We adopted redshifts for 58 clusters from CAMIRA (only 7 of these are from the `scanning mode' run).
Fig.~\ref{fig:CAMIRA} shows some example clusters confirmed using CAMIRA.

\subsubsection{Photometric Redshifts From zCluster}
\label{sec:zCluster}
The zCluster algorithm, described in \citet{Hilton_2018}, estimates redshifts for galaxy
clusters using broadband photometry, given a priori knowledge of the cluster position. This is done using
a weighted sum of the redshift probability distributions for galaxies along the line of sight to a cluster candidate. In
addition to the redshift estimate, zCluster also provides a measure of optical density contrast,
\begin{equation}
\delta (z_{\rm c}) = \frac{n_{\rm 0.5\,Mpc}(z_{\rm c})}{A n_{\rm 3-4\,Mpc}(z_{\rm c})} - 1 \, ,
\label{eq:delta}
\end{equation}
where $z_{\rm c}$ is the estimated photometric redshift for the cluster, 
$n_{\rm 0.5\,Mpc}(z_{\rm c})$ is the number of galaxies within 0.5\,Mpc projected distance of the given
cluster position, $n_{\rm 3-4\,Mpc}(z_{\rm c})$ is a measure of the background number of galaxies
in a circular annulus 3--4\,Mpc from the cluster position, and $A$ is a factor that accounts 
for the difference in area between these two count measurements. As shown in \citet{Hilton_2018},
a $\delta$ threshold can be used to identify cluster candidates with unreliable redshift estimates.

In this work, we applied zCluster to photometric data from the Dark Energy Camera Legacy Survey 
\citep[DECaLS DR8;][]{Dey_2019}, the Kilo Degree Survey \citep[KiDS DR4;][]{Wright_2019}, and 
SDSS \citep[DR16;][]{2020ApJS..249....3A}. 
DECaLS provides optical $grz$ photometry combined with 3.4, 4.6\,\micron{} photometry 
from the Wide-field Infrared Survey Explorer mission \citep[WISE;][]{Wright_2010}, and covers
most of the ACT DR5 cluster search area footprint (\DECaLSOverlapArea{}\,deg$^2$ of overlap).
We find that zCluster is able to measure cluster redshifts out to $z \approx 1.4$ when applied
to DECaLS, due to the inclusion of the WISE data. KiDS DR4 provides \textit{ugriZYJHK$_s$} photometry over 
\KiDSOverlapArea{}\,deg$^2$ in common with ACT DR5, with near-infrared data provided by 
the VISTA Kilo degree Infrared Galaxy survey \citep[VIKING;][]{Edge_2013}.

This work benefits from several improvements that have been made to zCluster, which we briefly 
summarize here: (i) a new automated masking procedure, that constructs an area mask image using the
positions of objects in the catalog and the typical nearest-neighbour separation, resulting 
in more accurate $\delta$ estimates close to survey boundaries; (ii) bootstrap resampling is
used to estimate the uncertainty on the density contrast statistic ($\Delta \delta$) 
at all points along the redshift range, and redshifts
at which $\delta / \Delta \delta < 3$ are rejected; and (iii) we have added the ability to easily
swap the spectral template set used for the individual galaxy photometric redshift estimates.

\begin{figure}
\includegraphics[width=\columnwidth]{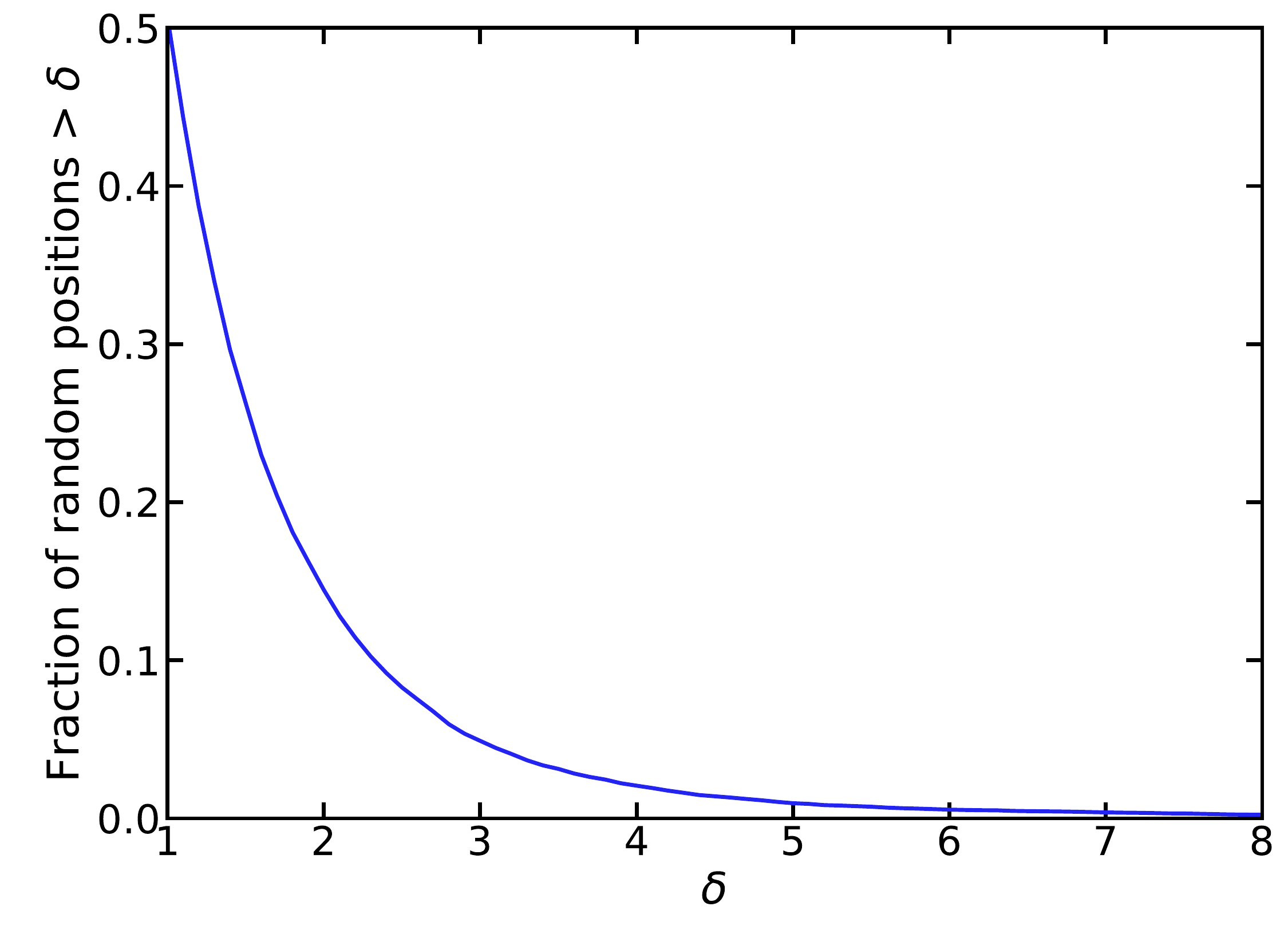}
\caption{Fraction of random positions drawn from a mock ACT DR5 cluster catalog where $\delta$,
the zCluster density contrast statistic as measured using DECaLS photometry, is greater 
than some value. We find $\delta > 3$ for 5\% of the random points.}
\label{fig:zClusterDECaLSMock}
\end{figure}

\begin{figure*}
\centering
\includegraphics[width=58mm]{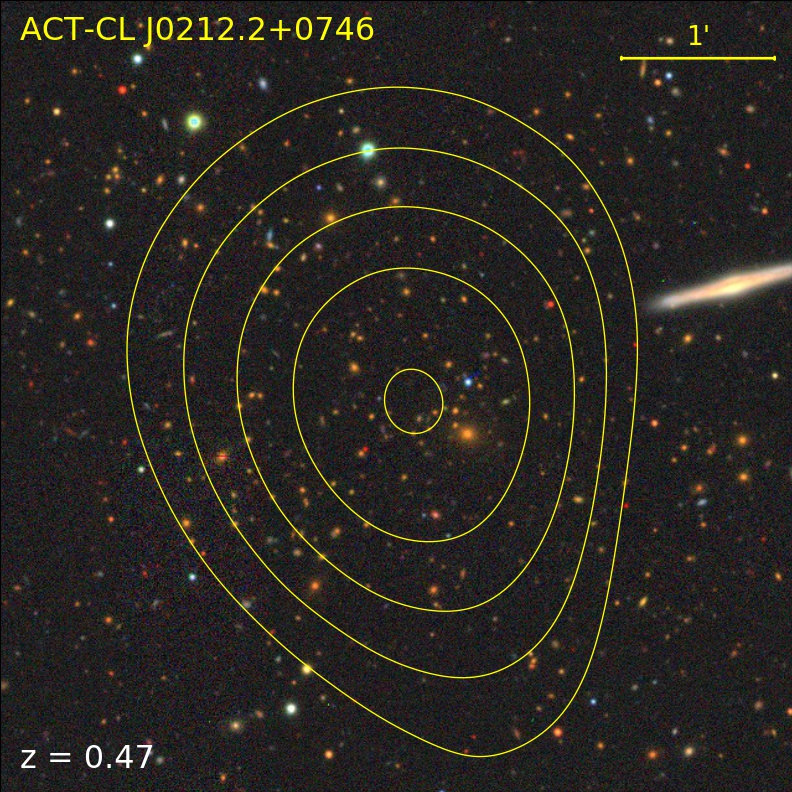}
\includegraphics[width=58mm]{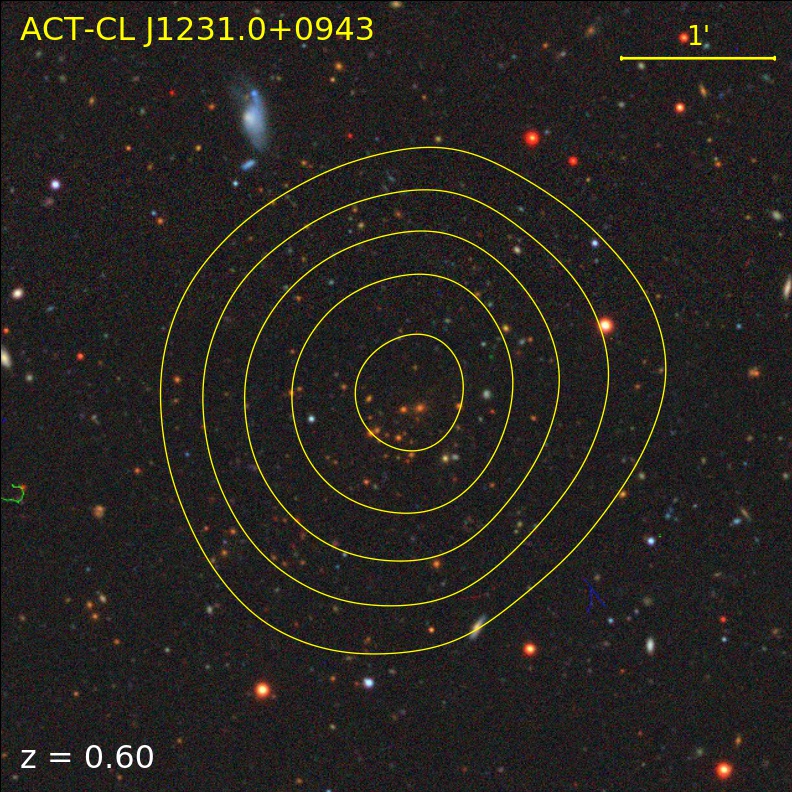}
\includegraphics[width=58mm]{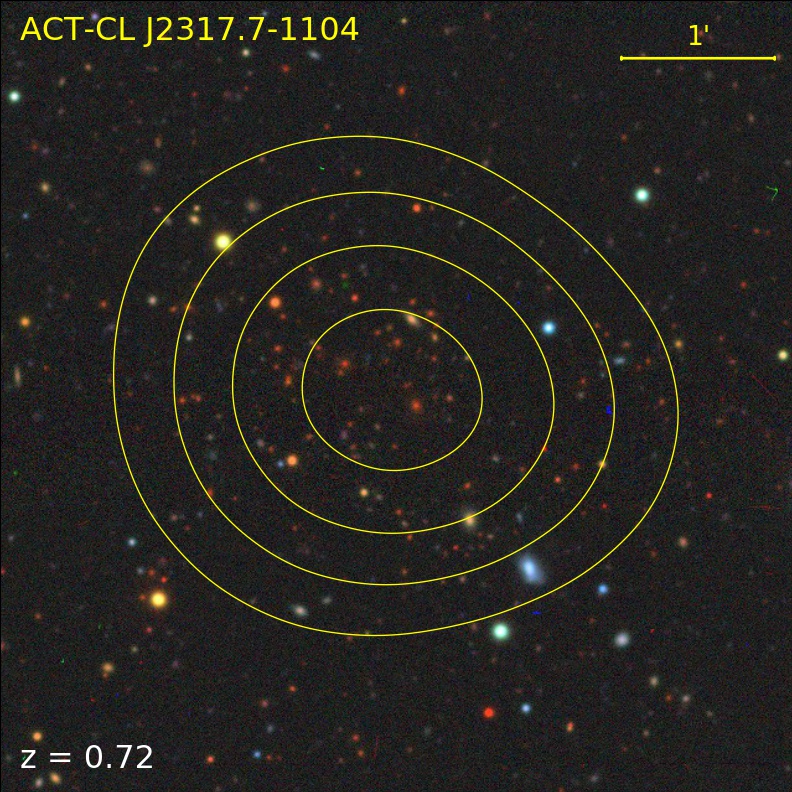}
\includegraphics[width=58mm]{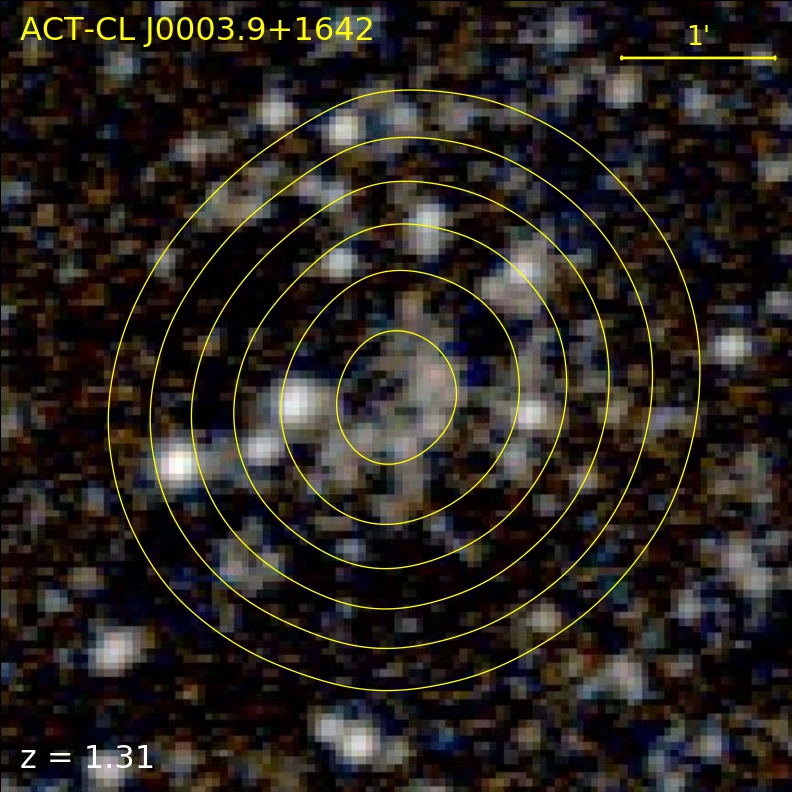}
\includegraphics[width=58mm]{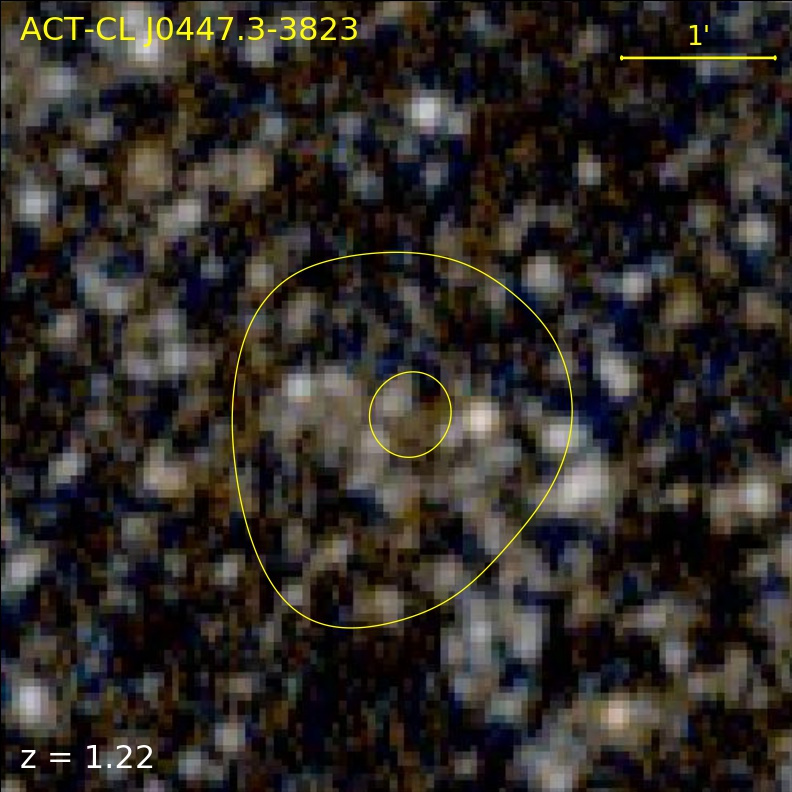}
\includegraphics[width=58mm]{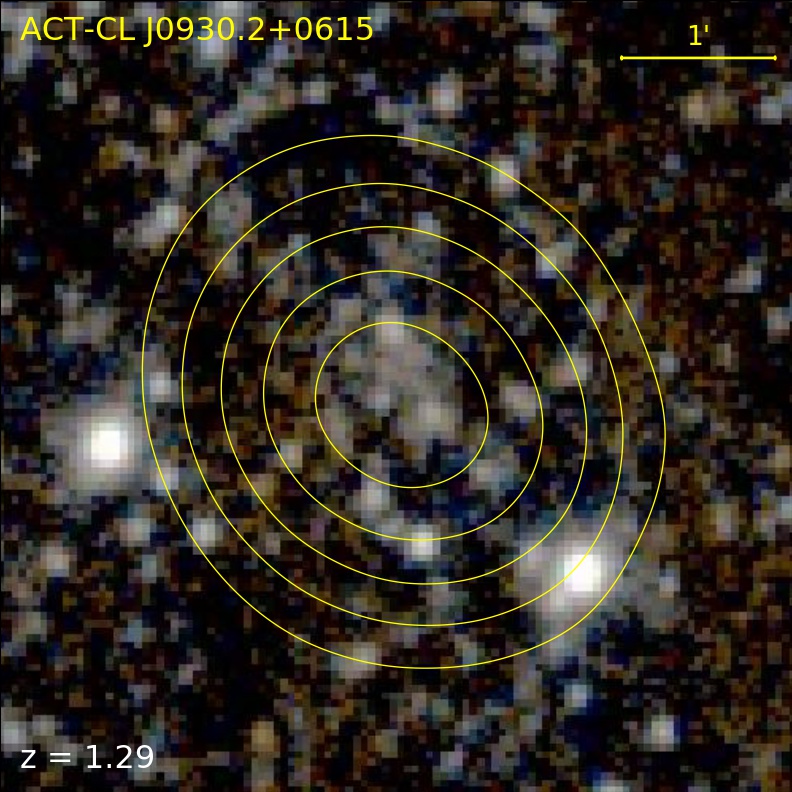}
\caption{Example DECaLS $grz$ images of ACT DR5 clusters at $z < 0.8$ (top row) and unWISE images
of $z > 1$ ACT DR5 clusters (bottom row). Each image is 5\arcmin{} on a side, with North at the top, 
East at the left. See Fig.~\ref{fig:RMDESY3} for an explanation of the contour levels.}
\label{fig:DECaLS}
\end{figure*}

While we ran zCluster on SDSS and KiDS photometry using the same set of spectral templates as used
in \citet{Hilton_2018}, i.e., the default templates from the EAZY photometric redshift code
\citep{Brammer_2008}, supplemented by the \citet[][CWW hereafter]{ColemanWuWeedman_1980} templates,
we found it necessary to switch the spectral template set in order to optimize the performance
when running on DECaLS photometry. We used a subset of the spectral templates used in the COSMOS
survey \citep{Ilbert_2009, Salvato_2011}, representing a range of normal galaxies and AGNs, 
removing all elliptical templates based on \citet{BruzualCharlot_2003} stellar population synthesis 
models (as these were found to give biased results for individual galaxies at moderate redshifts; 
we speculate that this is probably related to the extrapolation of the stellar population synthesis
models into the WISE bands), and adding in the CWW template set. 

Fig.~\ref{fig:zClusterDECaLSComparison} presents a comparison between 1168 clusters with 
spectroscopic redshifts ($z_{\rm s}$) and zCluster photometric redshift estimates ($z_{\rm c}$),
based on DECaLS photometry. Note that we have corrected the zCluster redshifts for a bias of the form 
$z = z_{\rm c}+0.02(1+z_{\rm c})$, where $z$ represents the corrected photometric redshift. We have not 
identified the source of this bias as yet, but note that this correction is sufficient to ensure that
on average the zCluster redshifts reported in this work are not biased. As shown in 
Fig.~\ref{fig:zClusterDECaLSComparison}, for clusters with $\delta > 3$, the scatter in the redshift
residuals $\Delta z/(1+z_{\rm s})$ is small ($\sigma_{\rm bw} = 0.014$; estimated using the biweight scale,
e.g., \citealt{Beers_1990}). The scatter rises to $\sigma_{\rm bw} = 0.04$ for the 16 objects beyond $z > 1$.
We find that 98\% of the redshifts are recovered within $\Delta z/(1+z_{\rm s}) < 0.05$ of the
spectroscopic redshift, so the number of catastrophic outliers is small.

To estimate the probability of a cluster candidate being associated with a random position on the sky 
where $\delta > 3$, we ran zCluster on DECaLS photometry on the same mock cluster catalog used for 
similar tests of redMaPPer and CAMIRA (Sections~\ref{sec:RM} and~\ref{sec:CAMIRA} above). 
Fig.~\ref{fig:zClusterDECaLSMock} shows the results of this exercise. We find that 5\% of random
positions have $\delta > 3$, rising to 14\% for $\delta > 2$ and 26\% for $\delta > 1.5$. 
Nevertheless, as shown in the right panel of Fig.~\ref{fig:zClusterDECaLSComparison}, the zCluster
photometric redshift estimates largely remain accurate at $\delta < 3$: we find 
$\sigma_{\rm bw} = 0.018$ for objects with $2 < \delta < 3$, with 95\% of these objects being
found within $\Delta z/(1+z_{\rm s}) < 0.05$ of the spectroscopic redshift. 

Fig.~\ref{fig:DECaLS} presents images of some example clusters confirmed using zCluster and 
DECaLS. The ACT DR5 cluster catalog contains 717 objects with redshifts provided by zCluster
(706 based on DECaLS photometry, 4 based on KiDS data, and 7 based on SDSS DR16). 
For 13 of the measurements based on DECaLS, we applied a $z > 0.6$ prior to avoid confusion with projected
lower $z$ systems that were judged not to be the source of the SZ signal following visual
inspection of the available imaging. In 96 cases where no alternative estimate is available,
we adopt zCluster redshifts with $\delta < 3$. All of these exceptions are appropriately flagged
in the \texttt{warnings} field of the cluster catalog (see Table~\ref{tab:FITSTableColumns}).

\subsubsection{Spectroscopic Redshifts From BEAMS}
\label{sec:BEAMS}
The BEAMS project (Brightest cluster galaxy Evolution with ACT, MeerKAT, and SALT) is a Large Science
Program on the Southern African Large Telescope (SALT) that is obtaining long-slit spectroscopic
observations of around 150 cluster central galaxies in a representative sample of $0.3 < z < 0.8$ ACT 
clusters. BEAMS observations began in May 2019 and at the time of writing 54 clusters have been
observed. The SALT data in hand have been processed with a modified version of the pipeline described
in \citet{Hilton_2018}. In this work, we report spectroscopic redshifts from BEAMS (labeled
SALTSpec in Table~\ref{tab:RedshiftSources}) for 15 clusters that 
would otherwise have only photometric estimates.


\subsubsection{Other Redshift Sources}
\label{sec:zOther}
We adopted a large number of redshifts used in the ACT DR5 cluster catalog from various sources in the 
literature. In particular, we used redshifts from previous published SZ surveys by 
ACT \citep{2013ApJ...765...67M, 2016MNRAS.461..248S, Hilton_2018}, \textit{Planck} 
\citep{PlanckPSZ2_2016}, and SPT \citep{, Bleem_2015, Bleem_2020, Bocquet_2019}; 
optically selected cluster catalogs based on SDSS \citep[][labelled as `WHL' in this work]{WHL_2012, WH_2015}, KiDS DR3 
\citep[][photometric redshifts based on the AMICO cluster finding algorithm]{2019MNRAS.485..498M}, and
ESO ATLAS \citep[][photometric redshifts based on the ORCA cluster finding algorithm; \citealt{Murphy_2012}]{Ansarinejad_2020}; and the 
IR-selected Massive Distant Clusters of WISE survey \citep[MaDCoWS;][]{Gonzalez_2019}, which contains more 
than 2000 high redshift ($0.7 < z < 1.5$) clusters selected from a survey area that covers most of the 
extragalactic sky.

We collected a large number of redshifts using the NASA Extragalactic Database 
(NED\footnote{\url{https://ned.ipac.caltech.edu/}}). We took care to classify such redshifts as
spectroscopic or photometric with appropriate uncertainties. References for these miscellaneous
sources can be found in the notes for Table~\ref{tab:RedshiftSources}.

\begin{deluxetable*}{p{30mm}ccp{90mm}}
\small
\tablecaption{Breakdown of redshift sources used in the ACT DR5 cluster catalog. The labels given in 
the Source column correspond to those used in the \texttt{redshiftSource} column in the 
\texttt{FITS} Table format cluster catalog (see Table~\ref{tab:FITSTableColumns}).
See Section~\ref{sec:visualInspect} for a description of how redshifts were assigned to each cluster.}
\label{tab:RedshiftSources}
\tablewidth{0pt}
\decimals
\tablehead{
\colhead{Source}       &
\colhead{Number} &
\colhead{Fraction (\%)} &
\colhead{Reference(s)} \\
}
\startdata
\raggedleft redMaPPer & 1433 & 34.2 & \citet{Rykoff_2014}; \citet{Rykoff_2016}\\
\raggedleft PublicSpec & 1184 & 28.2 & This work - based on data from 2dFLens \citep{2016MNRAS.462.4240B}; OzDES \citep{2017MNRAS.472..273C}; SDSS \citep{2020ApJS..249....3A}; and VIPERS \citep{2018AandA...609A..84S}; see Section~\ref{sec:publicSpec}\\
\raggedleft zCluster & 717 & 17.1 & This work; see Section~\ref{sec:zCluster}\\
\raggedleft WHL & 275 & 6.6 & \citet{WHL_2012}; \citet{WH_2015}\\
\raggedleft SPT & 201 & 4.8 & \citet{Bocquet_2019}; \citet{Bleem_2020}\\
\raggedleft Lit & 164 & 3.9 & See table notes\\
\raggedleft CAMIRA & 58 & 1.4 & \citet{Oguri_2018}\\
\raggedleft ACT & 52 & 1.2 & \citet{2013ApJ...765...67M}; \citet{2016MNRAS.461..248S}; \citet{Hilton_2018}\\
\raggedleft ATLAS & 51 & 1.2 & \citet{Ansarinejad_2020}\\
\raggedleft PSZ2 & 21 & 0.5 & \citet{PlanckPSZ2_2016}\\
\raggedleft SALTSpec & 15 & 0.4 & This work; see Section~\ref{sec:BEAMS}\\
\raggedleft MaDCoWS & 13 & 0.3 & \citet{Gonzalez_2019}\\
\raggedleft AMICO & 11 & 0.3 & \citet{2019MNRAS.485..498M}\\
\hline
\raggedleft Total spectroscopic & 1649 & 39.3 & \nodata \\
\raggedleft Total photometric & 2546 & 60.7 & \nodata \\
\enddata
\tablecomments{Sources for literature redshifts: \citet{1989ApJS...70....1A}; \citet{1991ApJS...76..813S}; \citet{1991ApJS...77..363S}; \citet{1994ApJS...94..583G}; \citet{1994MNRAS.269..151D}; \citet{1995ApJ...448L..93H}; \citet{1995MNRAS.274...75C}; \citet{1996ApJ...470..172S}; \citet{1998AandAS..129...31C}; \citet{1998ApJ...496L...5T}; \citet{1999ApJ...514..148D}; \citet{1999ApJS..125...35S}; \citet{2000AandAS..144..247C}; \citet{2000AN....321....1S}; \citet{2000ApJS..126..209R}; \citet{2000ApJS..129..435B}; \citet{2000MNRAS.312..663W}; \citet{2001AJ....122.2858O}; \citet{2002ApJS..140..239C}; \citet{2002MNRAS.329...87D}; \citet{2003ApJ...593...48G}; \citet{2003ApJ...594..154M}; \citet{2004AandA...423...75V}; \citet{2004AandA...425..367B}; \citet{2004AJ....128.1558S}; \citet{2004MNRAS.353..457A}; \citet{2006ApJ...638..725Z}; \citet{2006ApJ...645..955B}; \citet{2006MNRAS.366..645P}; \citet{2007ApJ...661L..33E}; \citet{2007ApJS..172..561B}; \citet{2007MNRAS.379..209S}; \citet{2008ApJ...677L..89G}; \citet{2008ApJ...682..821C}; \citet{2008MNRAS.383..879A}; \citet{2009AJ....137.2981G}; \citet{2009AJ....137.4795C}; \citet{2009AJ....138.1271D}; \citet{2010ApJ...718..876S}; \citet{2010ApJ...724.1182W}; \citet{2010MNRAS.406.1773M}; \citet{2011AandA...527A..78F}; \citet{2011ApJ...737...74G}; \citet{2011MNRAS.413.3059G}; \citet{2012AandA...538A..35C}; \citet{2012AandA...543A.102P}; \citet{2012ApJ...761...22S}; \citet{2012MNRAS.420.2120M}; \citet{2012MNRAS.423.1024M}; \citet{2013MNRAS.430..134W}; \citet{2014AandA...564A..17N}; \citet{2014ApJ...786...30C}; \citet{2014ApJ...792...76B}; \citet{2014ApJ...797...82L}; \citet{2014ApJS..213...25S}; \citet{2015AandA...582A..29P}; \citet{2015ApJ...812L..40G}; \citet{2015ApJS..216...20B}; \citet{2015MNRAS.446.2709E}; \citet{2015MNRAS.450.4248B}; \citet{2015MNRAS.452.3304P}; \citet{2019ApJ...878...66C}}
\end{deluxetable*}

\subsection{Cluster Confirmation and Redshift Assignment}
\label{sec:visualInspect}

We confirmed SZ-detected candidates as galaxy clusters using the wide variety of surveys described in
Section~\ref{sec:redshiftSources}, in combination with an extensive effort to visually
inspect the available optical/IR imaging for a large fraction of the catalog. 

To reduce the required visual classification effort, we cross-matched the cluster candidate list against
several external cluster catalogs that we deem to be reliable. The cross matching procedure makes use of the
position recovery model given in equation~(\ref{eq:posRecModel}) and shown in Fig.~\ref{fig:posRecTest}. 
We adopt the fit parameters that describe the radial distance as a function of SNR$_{2.4}$ within which 
99.7\% of the injected clusters were recovered. 
This accounts for uncertainty in the ACT cluster positions, due to noise fluctuations in the filtered maps.
However, the model does not account for position uncertainties in the external cluster catalogs.
Therefore, we add in quadrature the equivalent of an additional 0.5\,Mpc projected distance to the cross 
matching radius, evaluated at the redshift reported in the catalog being cross matched.
This serves as a conservative estimate of positional uncertainties in the external cross match catalogs.

We adopt a single redshift for each cluster in the catalog, after consideration of the various
potential redshift measurements available. Table~\ref{tab:RedshiftSources} lists the number of
redshifts used from each potential source together with the appropriate references. 
Where possible, first preference is given to a spectroscopic redshift. If this is not available, 
for clusters that have cross matches against external cluster catalogs, we select a 
photometric redshift according to the following in order of preference: 
(i) redMaPPer in DES Y3; (ii) CAMIRA; (iii) SPT; (iv) redMaPPer in SDSS; (v) WHL. The order
reflects the fact that we give preference to redshifts measured in deeper optical surveys.

We assigned redshifts from AMICO, ESO ATLAS, MaDCoWS, PSZ2, zCluster, and miscellaneous literature
sources (labeled Lit in Table~\ref{tab:RedshiftSources}) to clusters after visual inspection
of the optical/IR imaging from DECaLS, DES, KiDS, SDSS, HSC-SSP, 
Pan-STARRS \citep[PS1;][]{Flewelling_2016}, and WISE. We similarly visually inspected all objects
with redshifts derived from public spectroscopic surveys to check that the redshift assignment 
was sensible (i.e., derived from cluster member galaxies such as the BCG). 
Note, however,
that although the catalog contains clusters detected with SNR~$> 4$, visual inspection of cluster
candidates is only complete for all objects with SNR$_{2.4} > 5$. Objects with SNR$_{2.4} < 5$ have
only been visually inspected if there is some evidence from an external source that they may be 
galaxy clusters (e.g., $\delta > 3$ as measured by zCluster, or a cross match with an 
optical/IR-selected cluster catalog).

\subsection{Purity and Follow-up Completeness}
\label{sec:purity}

The fraction of optically confirmed cluster candidates above a given signal-to-noise threshold can be used to
assess the purity of a cluster sample, in the case of a complete set of follow-up observations \citep[e.g.,][]{Menanteau_2010}. Currently, we do not have all of the deep optical and IR data that would be
needed to determine the nature of all the sources in the ACT DR5 cluster candidate list in the full
\surveyArea{}\,deg$^2$ survey area. Due to the redshift-independent nature of the SZ effect, it is possible
for candidates to be located at distances that place them beyond the reach of our available imaging. 
Therefore, in the high signal-to-noise regime,
where the cluster sample is expected to be highly pure (see Fig.~\ref{fig:falsePositives}), 
the fraction of optically confirmed clusters in the ACT DR5 sample gives an indication of the completeness
of follow-up. At low signal-to-noise, this measure is instead driven by the false positive detection
rate.

\begin{figure}
\includegraphics[width=\columnwidth]{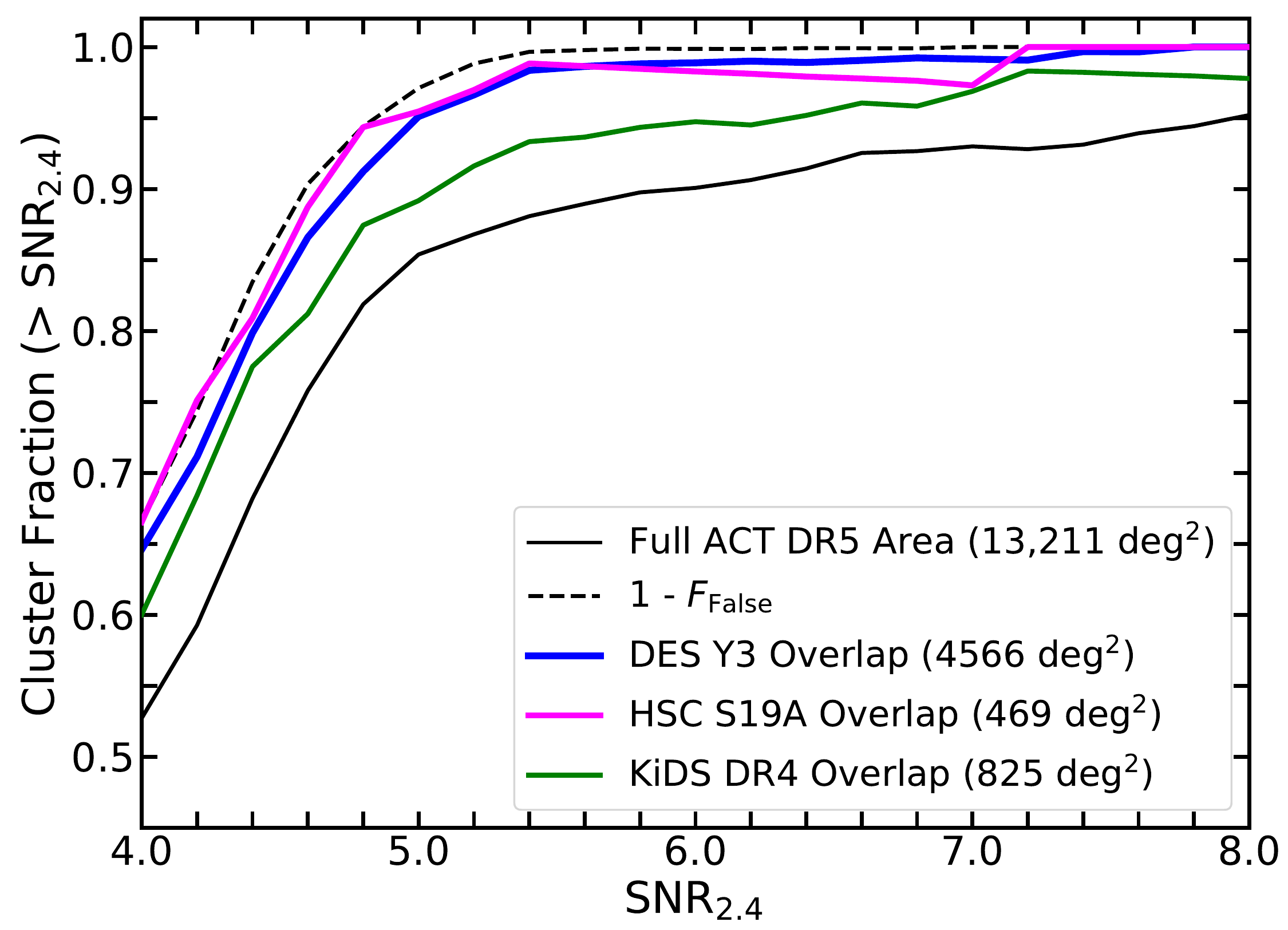}
\caption{The fraction of ACT DR5 cluster candidates that are optically confirmed as clusters (with a
redshift measurement) as a function of SNR$_{2.4}$, broken down according to overlap with
the indicated deep optical surveys (colored solid lines). Objects that are not confirmed clusters may be
contaminating false
positives (e.g., noise fluctuations in the maps) or genuine high-redshift systems that are not yet 
optically confirmed. The dashed line (labeled $1-F_{\rm False}$) shows the expected purity of the
cluster sample, based on a simulation of the survey (see Section~\ref{sec:SZDetection} and
Fig.~\ref{fig:falsePositives}). The small difference between $1-F_{\rm False}$ and the tracks for the
DES Y3 and HSC regions indicates that the optical follow-up is essentially complete for these parts of the
sky. However, the difference between $1-F_{\rm False}$ and the full ACT DR5 survey footprint indicates the 
potential for further follow-up to add up to 960 clusters to the sample.}
\label{fig:purity}
\end{figure}

Fig.~\ref{fig:purity} shows the fraction of confirmed clusters as a function of SNR$_{2.4}$ detection
threshold, broken down in terms of overlap with deep optical surveys. More than 98\% of 
the ACT DR5 candidates with SNR$_{2.4} > 5.5$ in the regions with DES Y3 and/or HSC S19A optical coverage
are confirmed as clusters and have redshift measurements. The fraction of confirmed clusters above the 
same SNR$_{2.4}$ cut is slightly lower in the region covered by KiDS DR4 (94\%), and significantly lower
when the full \surveyArea{}\,deg$^2$ ACT DR5 cluster search area is considered (89\%).
This reflects the fact that a significant fraction of the full ACT DR5 footprint
does not have complete coverage with data of similar quality to these deep optical surveys. 

As shown in Section~\ref{sec:SZDetection} and Fig.~\ref{fig:falsePositives}, we expect 34\% of 
candidates to be false positives for a selection cut of SNR$_{2.4} > 4$, based on a signal-free simulation
of the survey. We use this to predict the purity of the sample (labeled as $1 - F_{\rm False}$ in
Fig.~\ref{fig:purity}), although as noted earlier, this represents a best case scenario as the simulations
used do not fully capture all the noise sources present in the real data. We see that this traces the fraction
of candidates confirmed as clusters in the DES and HSC regions reasonably well, indicating that further optical follow-up
efforts in these areas should produce only a modest increase in the fraction of confirmed clusters.
On the other hand, only 52\% of \totalFixedSNRCandidates{} candidates with SNR$_{2.4} > 4$ detected in
the full \surveyArea{}\,deg$^2$ ACT DR5 footprint are currently optically confirmed, compared
to the 66\% expected if the estimate of the false positive rate is accurate. Therefore, further follow-up
over the full ACT DR5 area has the potential to add approximately 960 clusters to the sample.

\section{The ACT DR5 Cluster Catalog}
\label{sec:Catalog}

\subsection{Properties of the Cluster Catalog}
\label{sec:catalogProperties}

\begin{figure*}
\centering
\includegraphics[width=\columnwidth]{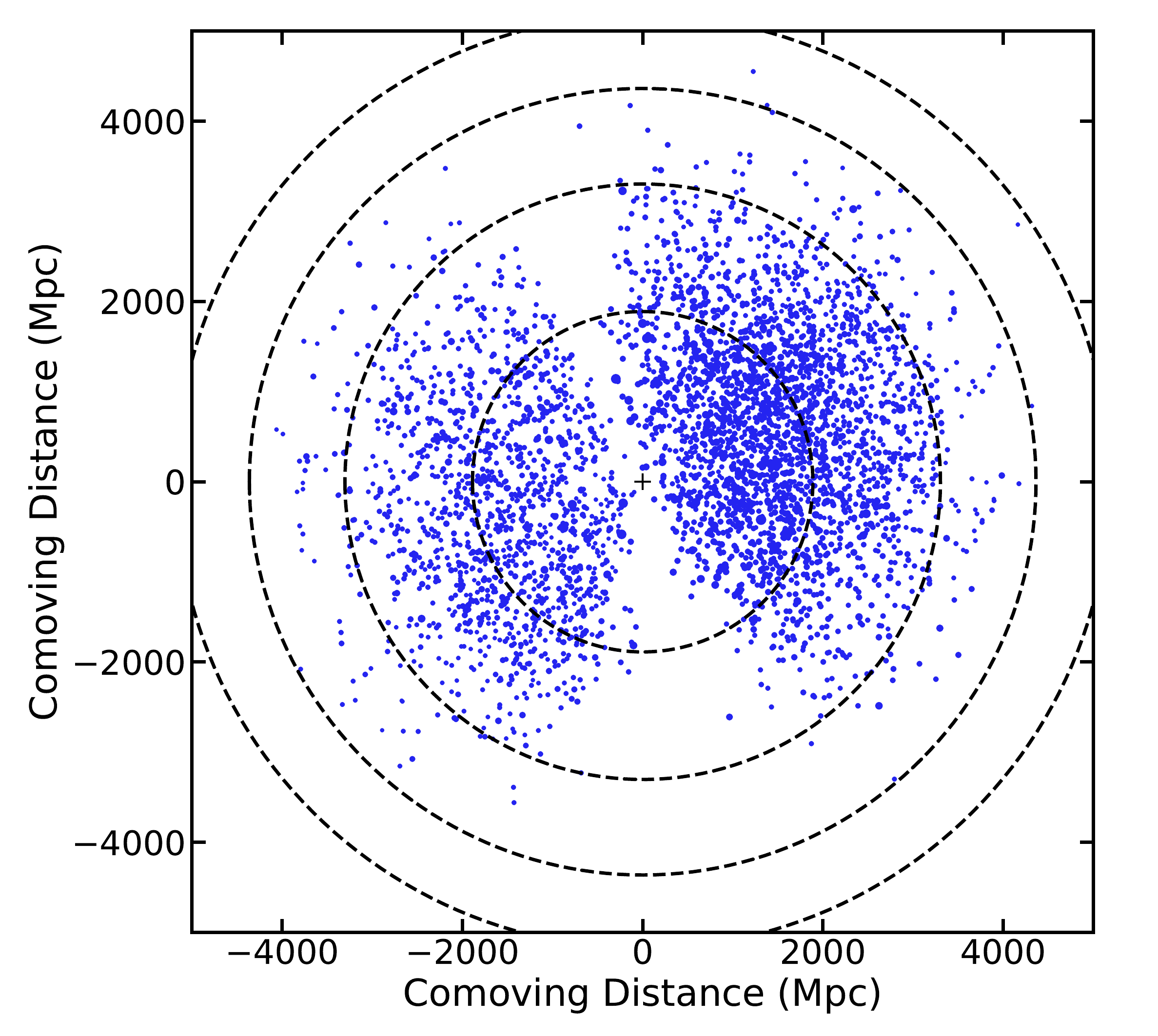}
\includegraphics[height=75mm]{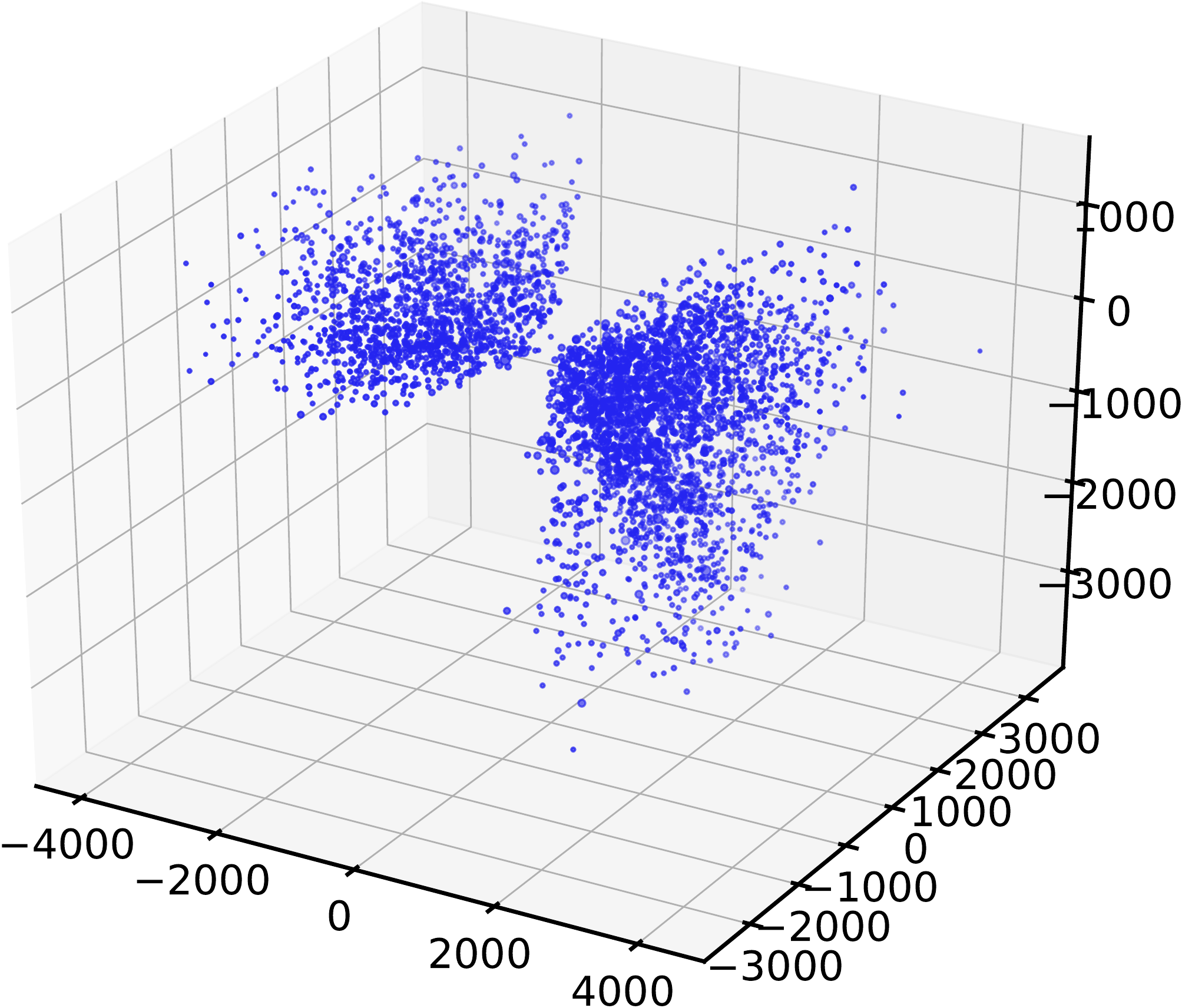}
\caption{The left panel displays a wedge plot showing the contents of the ACT DR5 cluster catalog, 
drawn in the equatorial
plane. Right Ascension is used as the angular coordinate, with 0$^\circ$ RA pointing to the right 
from the origin, and increasing anticlockwise. The radial coordinate is co-moving distance in Mpc.
Each point represents a cluster in the catalog, with the size of each point scaling
with cluster mass. The dashed circles mark the distances equivalent to redshifts 0.5, 1.0, 1.5, and
2.0, starting from the observer's location at (0, 0). The larger number of clusters seen on the right
of the plot compared to the left reflects the fact that ACT surveyed much more sky area at those
RA coordinates (see Fig.~\ref{fig:surveyArea}). The right panel shows a 3d projection of the same
information in spherical polar coordinates; here the axes are comoving distance in Mpc.}
\label{fig:wedgeplot}
\end{figure*}

This release of the ACT cluster catalog consists of \totalConfirmed{} optically confirmed
galaxy clusters detected with SNR~$> 4$ using the combination of the 98 and 150\,GHz ACT maps. 
Table~\ref{tab:FITSTableColumns}
describes the data provided in the catalog. Each cluster in the catalog has a redshift 
measurement (see Section~\ref{sec:redshiftSources}) and a set of mass estimates inferred from our
SZ observable, $\tilde{y}_0$ (see Section~\ref{sec:SZMasses}). The left panel of
Fig.~\ref{fig:wedgeplot} summarizes the contents of the catalog by showing the distribution of the 
clusters in terms of co-moving coordinates in the celestial equatorial plane (i.e., right ascension
is used as the angular coordinate). The right panel of Fig.~\ref{fig:wedgeplot} shows a similar
plot but in spherical polar coordinates.

Several studies have found evidence that cluster masses calibrated against the \citet{Arnaud_2010}
scaling relation, our fiducial mass estimate (labelled $M^{\rm UPP}_{\rm 500c}$ in this work), are 
lower than those measured from weak lensing 
\citep[e.g,][]{vonDerLinden_2014, Hoekstra_2015, Battaglia_2016, Miyatake_2019}.
For this reason, we provide a set of mass estimates that have been re-scaled according to a
richness-based weak-lensing mass calibration, following a similar
procedure to that described in \citet{Hilton_2018}. 

Using the sample of 383 SNR$_{2.4} > 6$ clusters 
(expected to be $>98$\% pure; see Section~\ref{sec:purity}) with $\lambda$ measurements from redMaPPer
in the DES Y3 footprint, and the \citet{McClintock_2019} $\lambda$--mass relation, we find that
the ratio of the average \citetalias{Arnaud_2010}--calibrated SZ-mass to the average richness-based, weak-lensing calibrated mass is
$\langle M^{\rm UPP}_{\rm 500c} \rangle / \langle M^{\rm \lambda WL}_{\rm 500c} \rangle = 0.71 \pm 0.07$.
Masses that have been rescaled according to this calibration factor are 
labelled $M^{\rm Cal}_{\rm 500c}$ throughout this work, and for convenience are provided in the cluster
catalog (see Table~\ref{tab:FITSTableColumns}). This calibration factor is in good agreement
with the value reported in \citet{Hilton_2018}, which used SDSS redMaPPer $\lambda$ measurements and
the \citet{Simet_2017} $\lambda$--mass relation. However, if we use the redMaPPer SDSS $\lambda$
measurements and the \citet{Simet_2017} relation (instead of the \citet{McClintock_2019} relation) 
with the ACT DR5 $M^{\rm UPP}_{\rm 500c}$ estimates,
then we find $\langle M^{\rm UPP}_{\rm 500c} \rangle / \langle M^{\rm \lambda WL}_{\rm 500c} \rangle = 0.66 \pm 0.08$.

\begin{figure}
\includegraphics[width=\columnwidth]{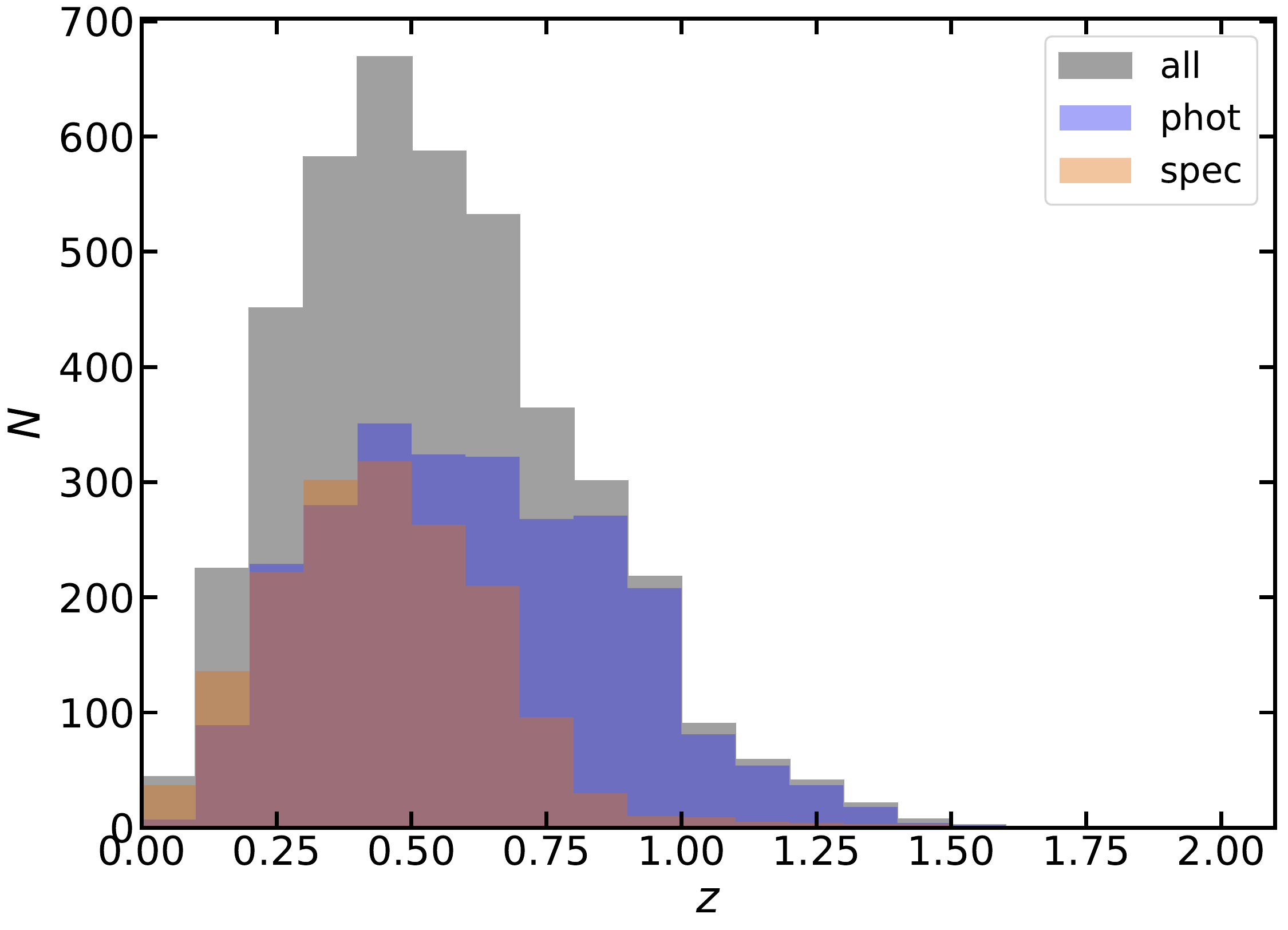}
\caption{Redshift distribution of the ACT DR5 cluster sample. The sample spans the redshift range 
$\minRedshift{} < z < \maxRedshift{}$ (median $z = \medianRedshift{}$). The distribution split
according to redshift type (spectroscopic or photometric) is also shown;
\percentageSpecRedshifts{}\% of the clusters in the sample have spectroscopic redshifts.}
\label{fig:zHist}
\end{figure}

We present the redshift distribution of the cluster sample in Fig.~\ref{fig:zHist}. The sample 
has median $z = \medianRedshift{}$, similar to other SZ-selected samples 
\citep[e.g.,][]{Hilton_2018, Bocquet_2019, Bleem_2020} and covers the redshift range 
$\minRedshift{} < z < \maxRedshift{}$. Largely due to the overlap with SDSS, a significant
fraction of the redshifts are spectroscopic (\percentageSpecRedshifts{}\%).
The highest redshift cluster in the sample, ACT-CL\,J0217.7-0345,
is detected with SNR$_{2.4} = 5.7$, and was first reported as the $z = 1.91^{+0.19}_{-0.21}$ X-ray 
selected cluster XLSSU\,J021744.1-034536 by \citet{Willis_2013}. It is also the highest redshift 
SZ detected cluster currently known \citep{Mantz_2014, Mantz_2018}. The catalog contains
\totalHighZ{} $z > 1$ clusters, which is greater than the total number of clusters reported in the
previous ACT cluster catalog \citep{Hilton_2018}. Most of the clusters
in the catalog have previously been detected in other surveys; here we report \totalNew{} new
cluster discoveries, with median $z = \medianRedshiftNew{}$. This figure excludes clusters detected
in the redMaPPer DES Y3 and CAMIRA S19A catalogs.

\begin{figure}
\includegraphics[width=\columnwidth]{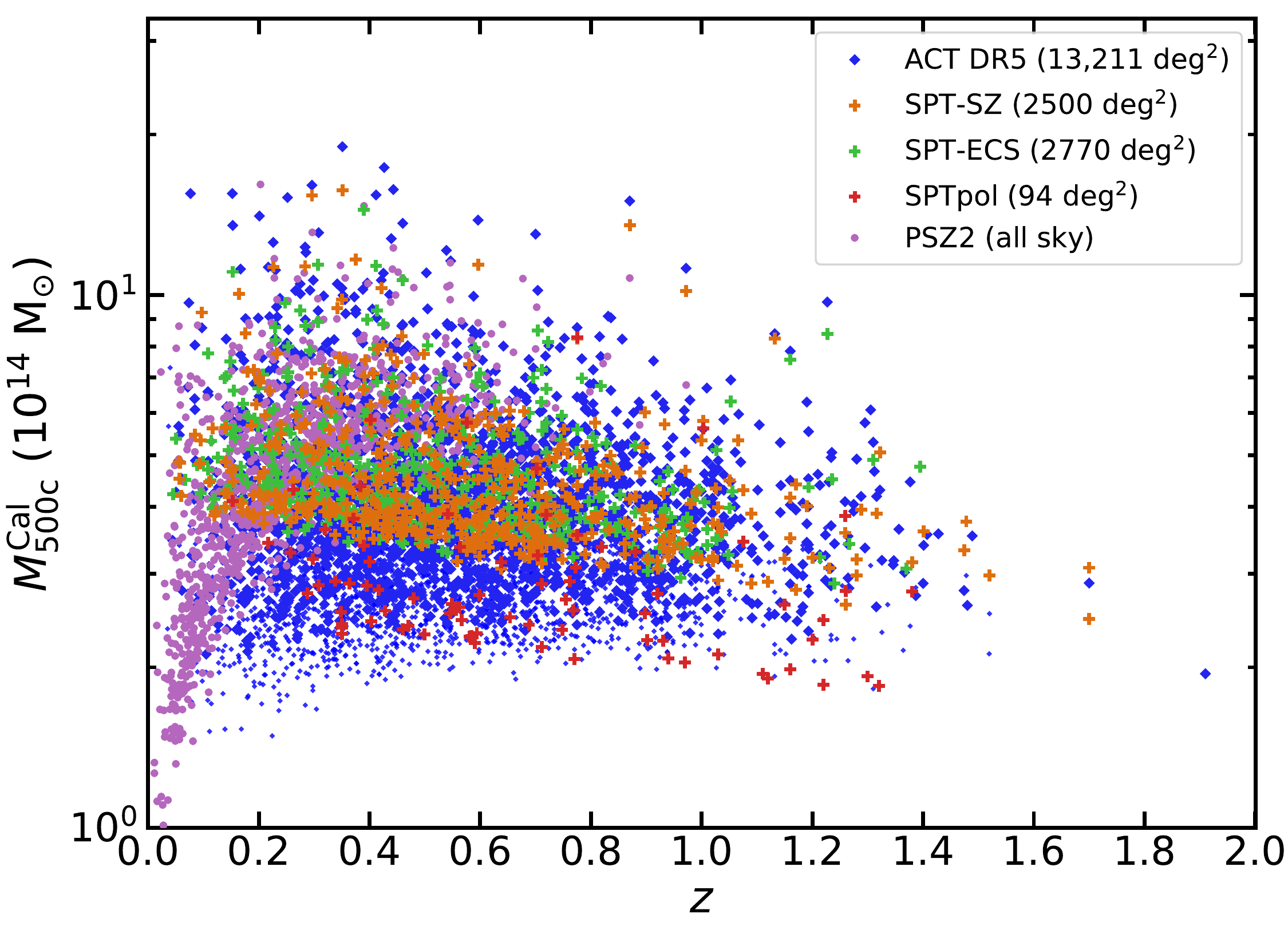}
\caption{Comparison of the ACT DR5 cluster sample in the (mass, redshift) plane with
other blind SZ surveys: PSZ2 \citep{PlanckPSZ2_2016}, SPT-SZ \citep{Bocquet_2019},
SPT-ECS \citep{Bleem_2020}, and SPTpol \citep{Huang_2020}. The large blue points show
the ACT DR5 sample selected with SNR$_{2.4} > 5$, which is similar to the detection
thresholds used in the other surveys. The small blue points extend this to 
include the full ACT DR5 sample. The ACT DR5 SZ masses displayed here have been rescaled 
according to a richness-based weak-lensing mass calibration, which is a close match
to the SPT mass scale (see Section~\ref{sec:SPTPSZ2Comparison}). Mass measurements 
from the SPT surveys and PSZ2 are as reported in the respective catalogs.
The ACT DR5 sample has been plotted behind the other surveys to aid clarity.}
\label{fig:massVz}
\end{figure}

Fig.~\ref{fig:massVz} shows the ACT DR5 sample in the (mass, redshift) plane, in comparison with 
other SZ-selected cluster samples from \textit{Planck} \citep[]{PlanckPSZ2_2016} and 
SPT \citep{Bocquet_2019, Bleem_2020, Huang_2020}. Here we show the richness-based weak-lensing
calibrated masses from ACT ($M^{\rm Cal}_{\rm 500c}$), as these are on a similar mass scale
to SPT (see Section~\ref{sec:SPTPSZ2Comparison} below). We plot both the full ACT DR5 sample down
to SNR~$> 4$ (shown as the small blue points) and a subsample with a cut of SNR$_{2.4} > 5$
applied, which is closer to the detection thresholds used in the other surveys, and more
closely resembles the sample that will be used for future cosmological analyses.
The ACT DR5 sample contains more clusters than all of the previous blind SZ cluster searches combined.
Due to the higher spatial resolution of the instruments, both the ACT DR5 and SPT samples reach to 
significantly lower mass limits than the PSZ2 catalog for $z > 0.2$. As Fig.~\ref{fig:massVz}
shows, the SPTpol sample \citep{Huang_2020} is more sensitive to lower mass clusters than 
ACT DR5 when a similar detection threshold is applied, although this survey covers only 94\,deg$^2$.

Inspection of Fig.~\ref{fig:massVz} suggests that there may be a deficit of clusters in
the redshift range $1 < z < 1.1$. This is extremely unlikely to be a real feature, and
may arise due to a bias in the photometric redshifts. 
We will investigate this further with future spectroscopic
follow-up of such high redshift systems.

\subsection{Comparison with the ACT DR3 Cluster Catalog}
\label{sec:ACTPolComp}
As discussed extensively in \citet{Choi_2020} and \citet{Aiola_2020}, there have been many changes to the ACT 
data processing pipelines at all levels of the analysis since the data release that the 
\citet{Hilton_2018} ACTPol cluster catalog (ACT DR3 hereafter) is based on. In this work, we have used maps produced
using a new co-adding procedure that incorporates data from all observing seasons and, for the first time, 
includes data taken during the day time (\citetalias{Naess_2020}). As noted in Section~\ref{sec:ACTObs},
these co-added maps include preliminary data from the 2017 and 2018 observing seasons that have
not been subjected to the full battery of tests as used in the CMB power spectrum analysis presented
in \citet{Choi_2020} and \citet{Aiola_2020}. In this work we also use a different, 
multi-frequency matched filter approach in the cluster finder compared to the algorithm described
in \citet{Hilton_2018}.

We begin by checking the recovery of ACT DR3 clusters in the ACT DR5 catalog. \citet{Hilton_2018}
reported the detection of 182 SNR~$>4$ clusters, of which 175/182 are within the ACT DR5 cluster search
area (i.e., 7 clusters fall within regions masked in DR5). Of these, 154/175 are recovered within 
2.5\arcmin{} of the position of a candidate in the ACT DR5 catalog, leaving 21 clusters that are 
not detected at SNR~$>4$. The missing 21 clusters have median
SNR$~= 4.3$ in the ACT DR3 catalog, although three SNR~$>5$ clusters (ACT-CL\,J0238.2+0245, 
ACT-CL\,J0341.9+0105, and ACT-CL J2337.6-0856) are not detected in the ACT DR5 catalog. 
Half of the missing 21 clusters were 
previously reported in other catalogs \citep{Goto_2002, Lopes_2004, Durret_2011, Menanteau_2013, Rykoff_2014, WH_2015}.
Re-running the ACT DR5 cluster search with a lower detection threshold recovers 13/21 of the missing
ACT DR3 clusters at SNR~$>3$. 

\begin{figure}
\includegraphics[width=\columnwidth]{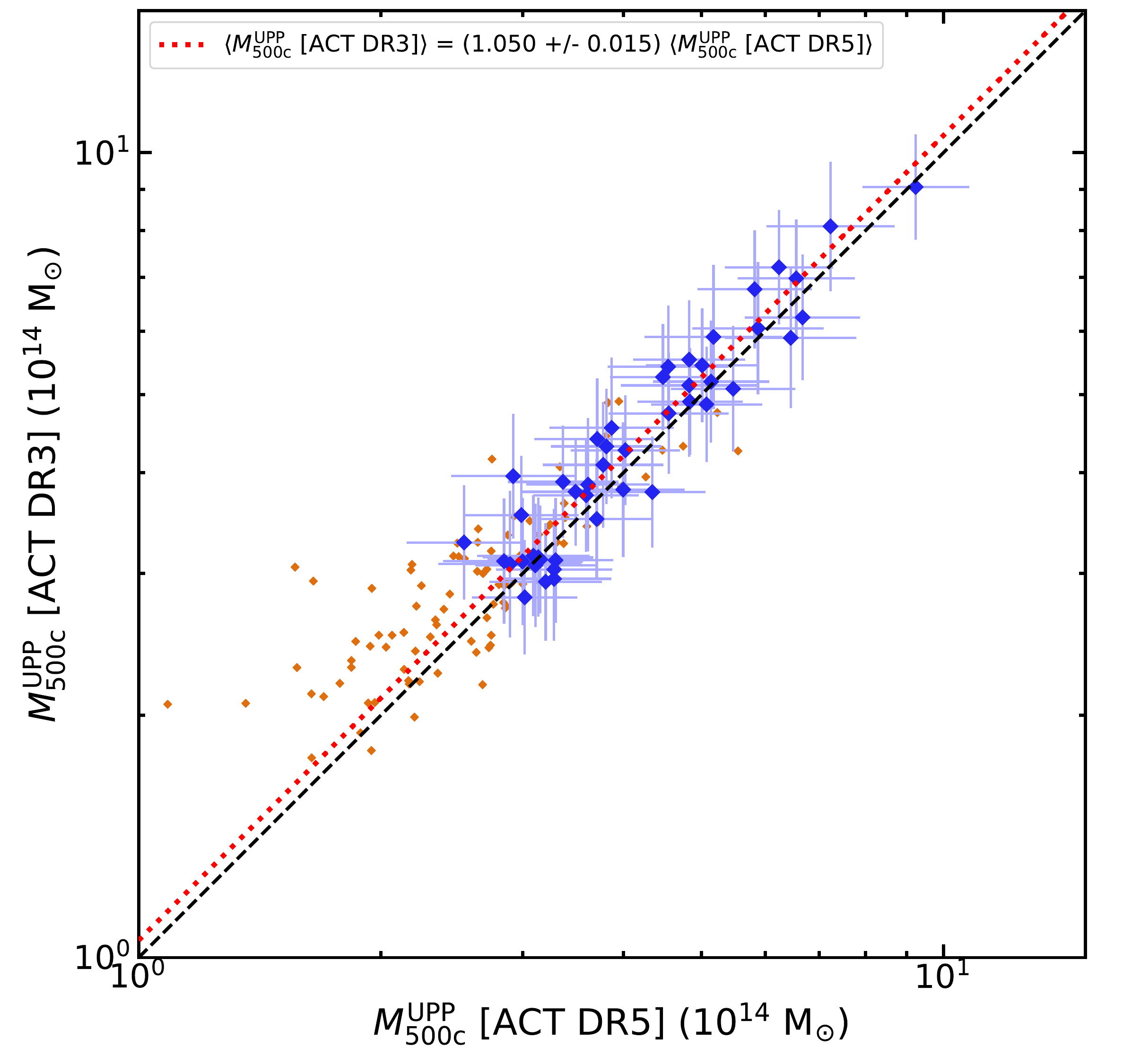}
\caption{Comparison between ACT DR5 mass estimates (this work, plotted along the horizontal
axis) and previous ACT
cluster mass estimates as reported in \citet{Hilton_2018} (labeled ACT DR3). Both analyses
assume the same scaling relation between SZ signal and mass. The large blue points are objects
with SNR$_{2.4} > 6$ in both samples (for which the unweighted mean ratio is calculated, 
shown by the dotted red line), while the small orange points (without error bars) show 
objects below this threshold. The dashed black line shows the 1:1 correlation.}
\label{fig:ACTPolMassComparison}
\end{figure}

Fig.~\ref{fig:ACTPolMassComparison} presents a comparison of the ACT DR3 mass estimates reported in
\citet{Hilton_2018} with the new measurements from ACT DR5. We highlight the objects detected
with SNR$_{2.4} > 6$ in both samples, as these should not be affected by filter noise bias
at any significant level. As expected, both sets of measurements follow a tight correlation. However, 
we see that the ACT DR5 masses are on average systematically $\approx 7$\% lower than the ACT DR3
measurements. We have verified that the difference in the filtering approach between 
\citet{Hilton_2018} and this work is not the cause (consistent $\tilde{y}_{0}$ measurements are
obtained when running either method on the same map). 
It may be the case that scale-dependent
bandpass effects \citep[see][]{Madhavacheril_2020}, which are not accounted for in this analysis,
could explain part of the offset.
While we have not yet been able to resolve 
this discrepancy, we note that gain errors at the level of a few per cent are expected in the co-added
maps analysed in this work (\citetalias{Naess_2020}). 
Therefore, we caution users of the ACT DR5 cluster catalog
that the $\tilde{y}_{0}$ measurements reported here (and in turn the $M^{\rm UPP}_{\rm 500c}$ masses) 
\textit{may} be systematically underestimated by $\approx 5-10$\%, \textit{if} the ACT DR3 catalog is taken as 
``truth''. This should be kept in mind when comparing these values against external catalogs.
However, mass calibration against external datasets can still be used to absorb any systematic
calibration error (as should be the case for the $M^{\rm Cal}_{\rm 500c}$ mass estimates).

\section{Discussion}
\label{sec:Discussion}

\subsection{The SZ Cluster Mass Scale}
\label{sec:SPTPSZ2Comparison}

\begin{figure}
\includegraphics[width=\columnwidth]{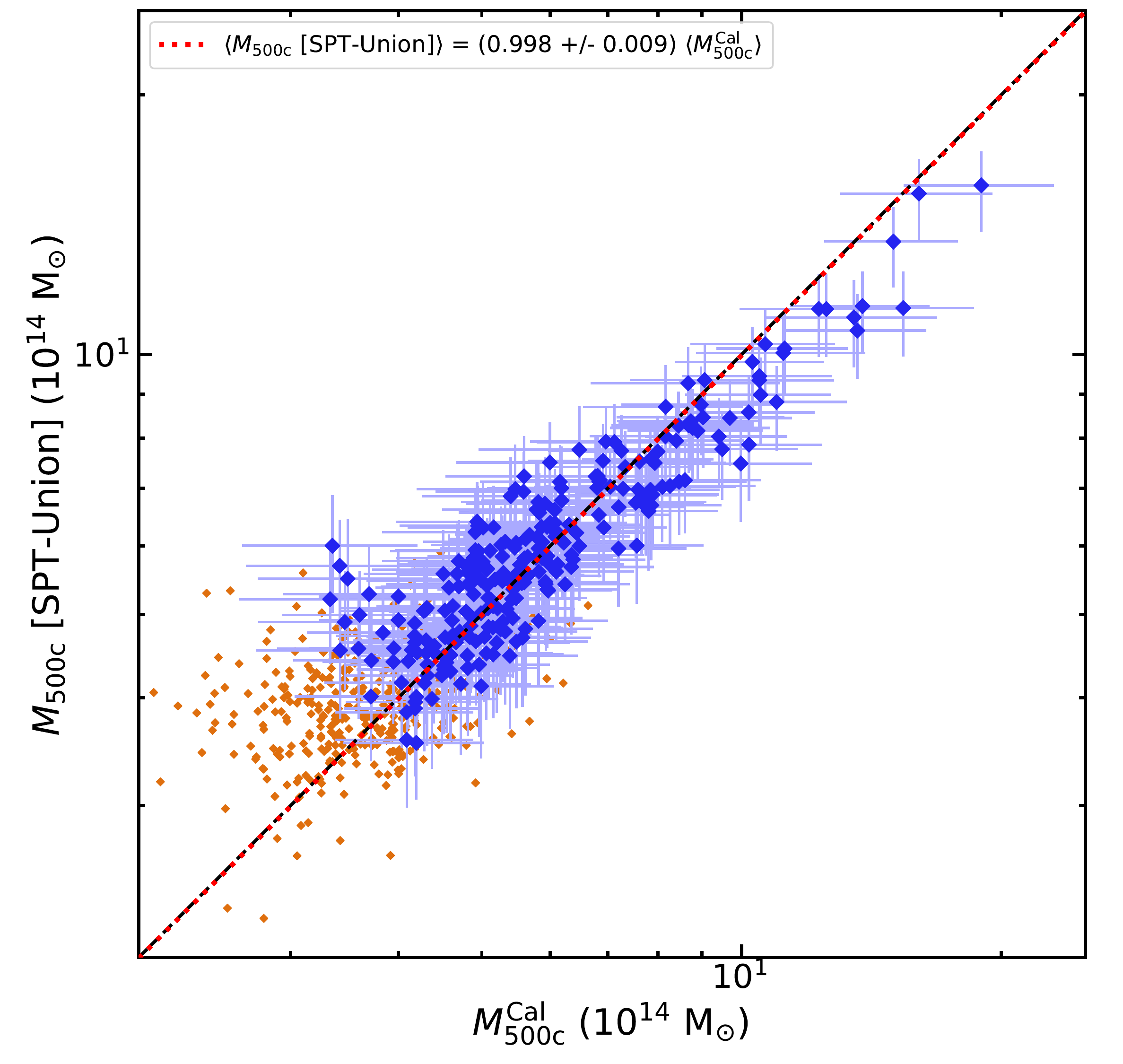}
\caption{Comparison between ACT DR5 mass estimates rescaled according to a 
richness-based weak-lensing mass calibration ($M^{\rm Cal}_{\rm 500c}$; Section~\ref{sec:Catalog})
with SPT masses reported in \citet{Bocquet_2019}, \citet{Bleem_2020}, and \citet{Huang_2020}.
The large blue points are objects with signal-to-noise~$> 6$ in both ACT and SPT
(for which the unweighted mean ratio is calculated, shown by the dotted red line), while the
small orange points (without error bars) show objects below this threshold.
The dashed black line shows the 1:1 correlation.}
\label{fig:SPTMasses}
\end{figure}

Mass calibration of cluster samples is the key systematic that limits their ability to constrain cosmological
parameters and is a topic of much debate in the literature 
\citep[e.g.,][]{vonDerLinden_2014, Hoekstra_2015, PlanckPSZ2Cosmo_2016, Battaglia_2016, Smith_2016, Medezinski_2018, Miyatake_2019}.
Cluster mass estimation based on any kind of data is dependent upon a number of assumptions.
Here we present a simple comparison of the mass estimates available in the ACT DR5 catalog with previous
SZ surveys, and a compendium of weak-lensing mass estimates \citep[CoMaLit;][]{Sereno_2015}, as
an illustration of how they may be used. Future works based on the ACT DR5 catalog will investigate this topic
in much more detail.

In \citet{Hilton_2018}, we presented comparisons between ACT cluster mass estimates and the SPT-SZ and PSZ2
catalogs. While we found excellent agreement with the \citet{Bleem_2015} SPT-SZ catalog, we noted 
an apparent mass-dependent trend when comparing the ACT masses with PSZ2 (although at low significance).
This was identified in a plot of the ACT--PSZ2 mass ratio versus the ACT mass estimate. We subsequently 
found that the apparent mass-dependent trend was an illusion driven by the combination of a regrettable 
choice in the plot axes (i.e., plotting the ACT mass instead of the PSZ2 mass as the independent coordinate), 
the very different selection of the ACT and PSZ2 cluster samples (PSZ2 detects
$z < 0.2$ clusters at high significance down to low masses while ACT does not, and the reverse is 
true at higher $z$), and that the measurements themselves are subject to a significant amount of scatter, 
especially at low signal-to-noise. We rectify this in the comparisons presented here by simply 
plotting the mass estimates in each catalog against each other.

Fig.~\ref{fig:SPTMasses} shows a comparison between the ACT DR5 masses rescaled using the 
richness-based weak-lensing mass calibration (see Section~\ref{sec:Catalog}) against a 
union of the SPT cluster catalogs 
(\citealt{Bocquet_2019, Bleem_2020, Huang_2020}; note that we make no attempt to remove duplicate objects).
This `SPT-Union' sample contains 618 clusters in common with ACT DR5 (326 from SPT-SZ, 266 from SPT-ECS,
and 26 from SPTpol), if we include all objects down to the detection thresholds used by each survey. 
The masses are clearly correlated, although there
is a tendency for the ACT mass estimates to be slightly larger than those from SPT, particularly at the high
mass end.

\begin{figure}
\includegraphics[width=\columnwidth]{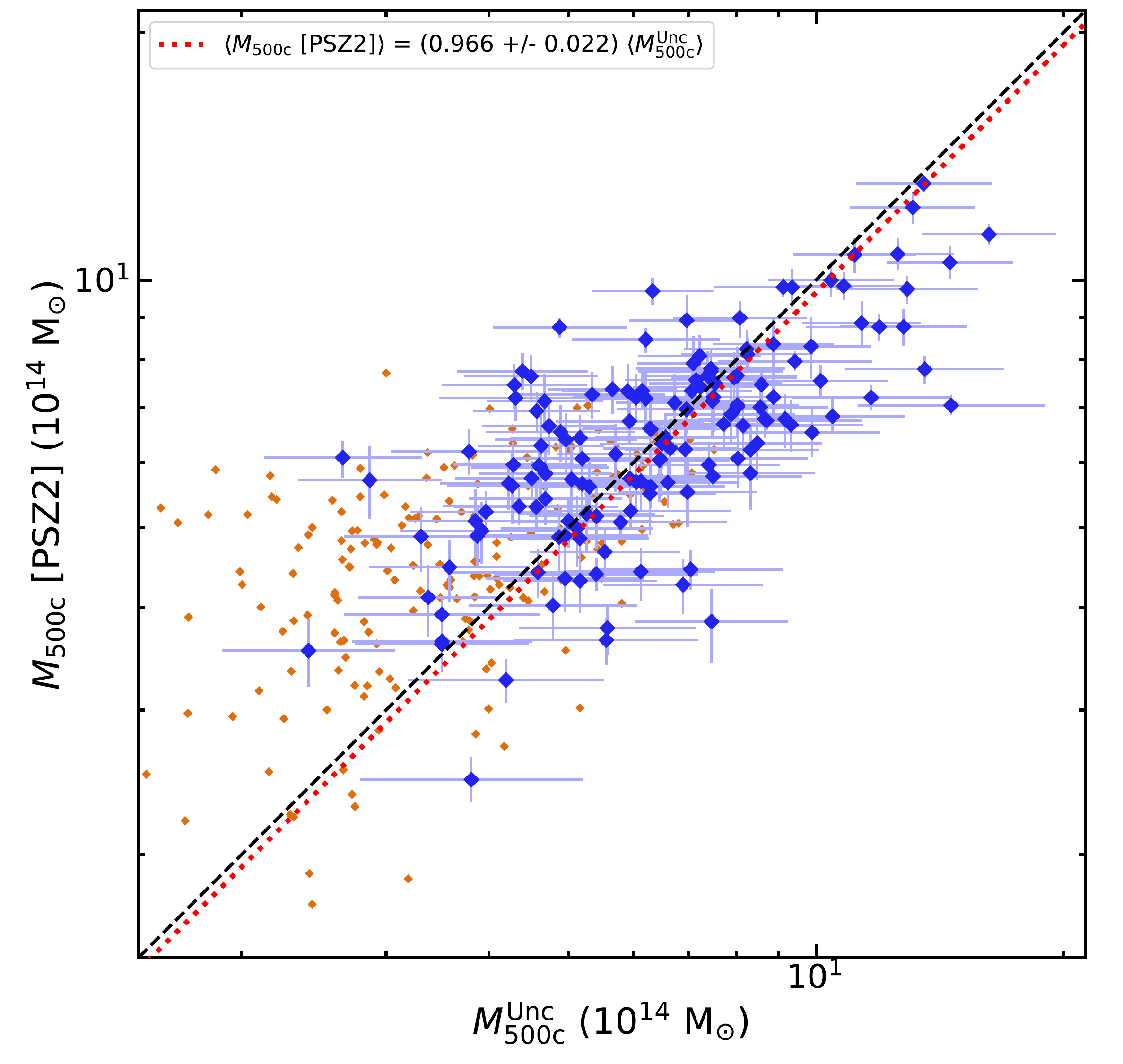}
\caption{Comparison between ACT DR5 mass estimates, uncorrected for bias due to the
steepness of the cluster mass functon ($M^{\rm Unc}_{\rm 500c}$; Section~\ref{sec:Catalog}),
with PSZ2 masses reported in \citet{PlanckPSZ2_2016}. 
The large blue points highlight objects with signal-to-noise~$> 6$ in both surveys, 
while the small orange points (without error bars) show 
objects below this threshold. The lines have the same meaning as in Fig.~\ref{fig:SPTMasses}.
}
\label{fig:PSZ2Masses}
\end{figure}

\begin{figure}
\includegraphics[width=\columnwidth]{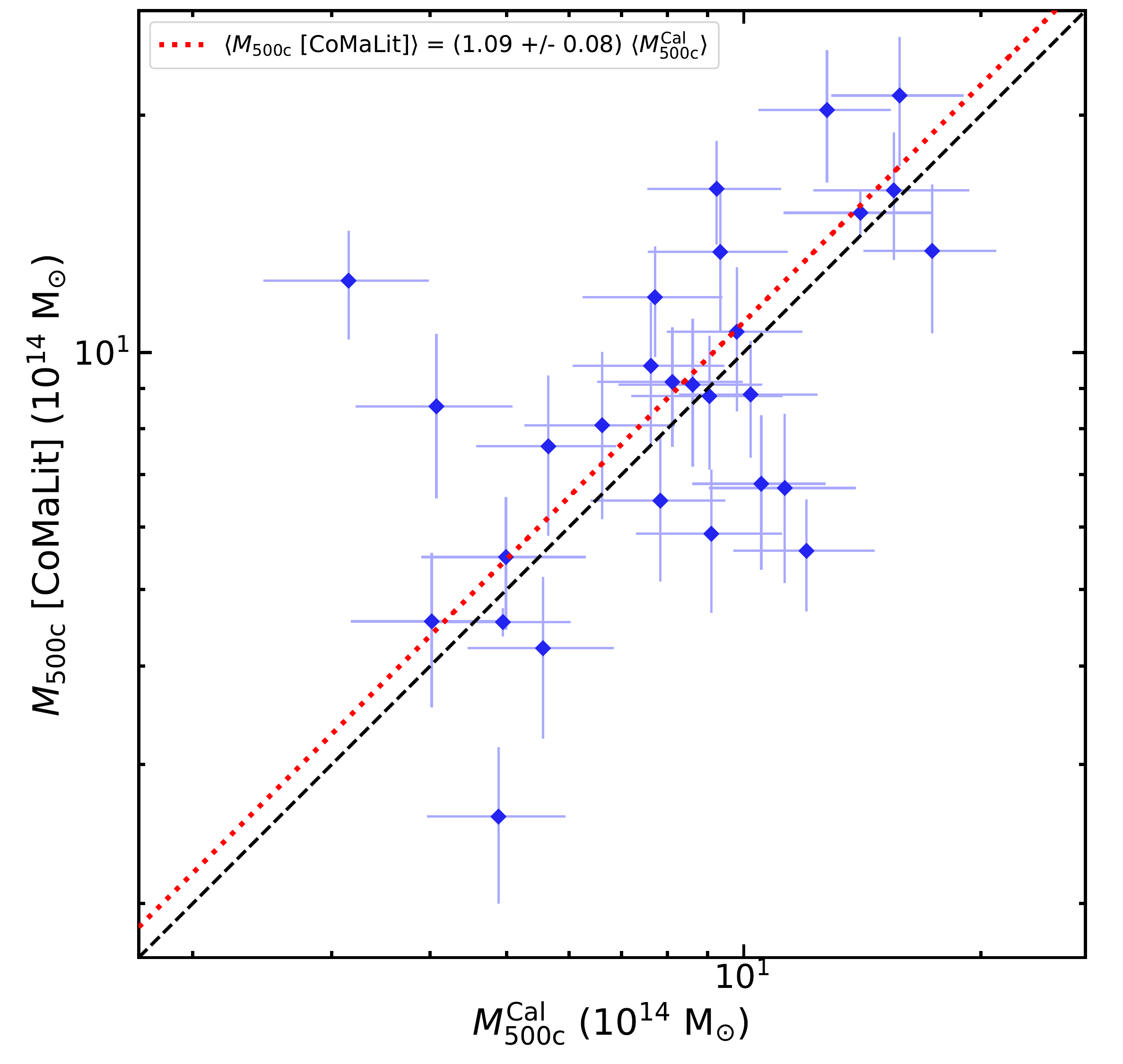}
\caption{Comparison between ACT DR5 mass estimates rescaled according to a 
richness-based weak-lensing mass calibration ($M^{\rm Cal}_{\rm 500c}$; Section~\ref{sec:Catalog})
with weak-lensing masses from the CoMaLit database \citep{Sereno_2015}.
Here we use the LC$^2$-single catalog from CoMaLit, which consists of objects modeled using a
single halo, and we restrict the selection to include only clusters with weak-lensing mass
estimates with $<25\%$ uncertainties. The lines have the same meaning as in 
Fig.~\ref{fig:SPTMasses}.}
\label{fig:CoMaLitMasses}
\end{figure}

Leaving aside any question of mass-dependent scaling for future work, we can make a simple assessment of 
the overall consistency of the mass scale between the two samples using the unweighted mean ratio of their masses \citep[e.g.,][]{2016MNRAS.461..248S, Hilton_2018}. 
Here we use the 254 objects detected at signal-to-noise$~>6$ in both samples, 
as $>98$\% of the ACT DR5 candidates with SNR$_{2.4} > 6$ were confirmed to be clusters in the DES Y3 
and HSC S19A regions (see Section~\ref{sec:purity}). Using a high signal-to-noise threshold
also mitigates the effect of the `noise floor' in the case of a significant difference in depth between
two samples (although that is not the case for the comparison here). We find 
$\langle M_{\rm 500c} \left[\rm SPT \mhyphen Union \right] \rangle = (0.998 \pm 0.009) \, \langle M^{\rm Cal}_{\rm 500c} \rangle$,
where the uncertainty is the standard error on the mean (note that this does not account for the
uncertainty on the richness-based weak-lensing mass calibration factor itself).
We find results that are consistent with this if we compare ACT DR5 against the individual SPT catalogs:
$\langle M_{\rm 500c} \left[\rm SPT \mhyphen SZ \right] \rangle = (1.027 \pm 0.012) \, \langle M^{\rm Cal}_{\rm 500c} \rangle$;
$\langle M_{\rm 500c} \left[\rm SPT \mhyphen ECS \right] \rangle = (0.961 \pm 0.014) \, \langle M^{\rm Cal}_{\rm 500c} \rangle$;
and
$\langle M_{\rm 500c} \left[\rm SPTpol \right] \rangle = (1.001 \pm 0.027) \, \langle M^{\rm Cal}_{\rm 500c} \rangle$.
Similarly to \citet{Hilton_2018}, we see that despite the differences between
how the ACT DR5 and SPT samples were constructed, and the very different method used to calibrate the
mass estimates, the richness-based weak-lensing calibrated masses are on a similar mass scale to SPT.

Fig.~\ref{fig:PSZ2Masses} presents a similar comparison between ACT DR5 and the PSZ2 catalog \citep{PlanckPSZ2_2016}.
Since we adopted the same fiducial X-ray calibrated mass scaling relation from \citetalias{Arnaud_2010} as used in the 
PSZ2 catalog, we would expect the ACT DR5 clusters to follow the same mass scale. Here the comparison is 
made against ACT DR5 mass estimates that neglect the bias correction that accounts for the steepness 
of the cluster mass function, as such a correction is not applied to the masses reported in the PSZ2 
catalog \citep[see Section~\ref{sec:SZMasses} and the discussion in][]{Battaglia_2016}.
We use a 5\arcmin{} radius to cross-match the two catalogs, resulting in a sample of
327 clusters, if we include all objects down to the detection threshold of each survey.
As shown in Fig.~\ref{fig:PSZ2Masses}, the scatter between the PSZ2 and ACT DR5 masses is large, 
but the mass scale is indeed similar: we find
$\langle M_{\rm 500c} \left[\rm PSZ2 \right] \rangle = (0.966 \pm 0.022) \, \langle M^{\rm Unc}_{\rm 500c} \rangle$
from the 148 objects detected with signal-to-noise$~> 6$ in both catalogs.

\begin{deluxetable}{cp{6cm}}
\tablecaption{Strong Lens Catalogs\label{tab:lensCodes}}
\tablewidth{0pt}
\decimals
\tablehead{
\colhead{Code} &
\colhead{Reference}
}
\startdata
M16 & \citet{More_2016}\\
D17 & \citet{Diehl_2017}\\
S18 & \citet{Sonnenfeld_2018}\\
W18 & \citet{Wong_2018}\\
P19 & \citet{Petrillo_2019}\\
J19 & \citet{JacobsJ19_2019}\\
J19a & \citet{JacobsJ19a_2019}\\
H20a & \citet{HuangH20a_2020}\\
H20b & \citet{HuangH20b_2020}\\
Jae20 & \citet{Jaelani_2020}\\
D20 & \citet{Diehl_2020}\\
\enddata
\tablecomments{Entries in the \texttt{knownLensRefCode} column of the cluster catalog
(see Table~\ref{tab:FITSTableColumns}) correspond to the Code column used here.}
\end{deluxetable}

\begin{figure*}
\centering
\includegraphics[width=58mm]{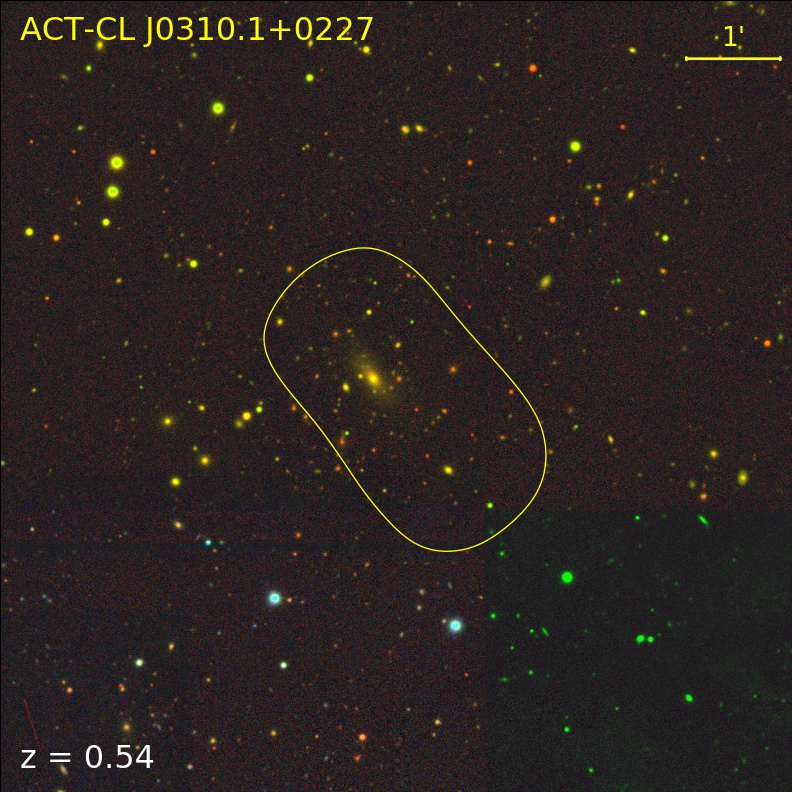}
\includegraphics[width=58mm]{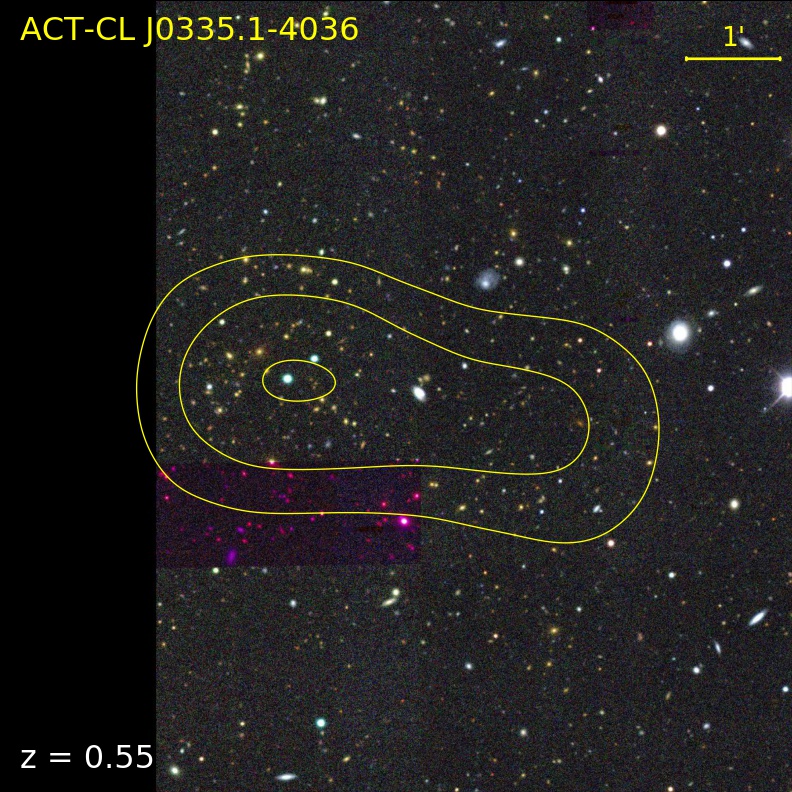}
\includegraphics[width=58mm]{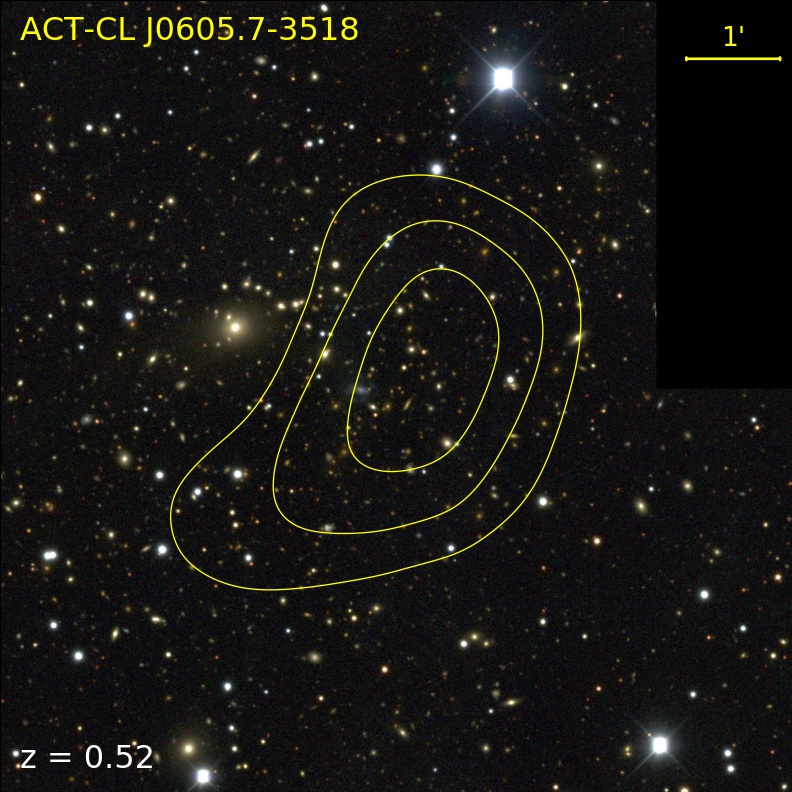}
\caption{Examples of possible projected systems (see Section~\ref{sec:projected}).
Each image is 8\arcmin{} on a side, with North at the top, East at the left. See Fig.~\ref{fig:RMDESY3}
for an explanation of the contour levels.}
\label{fig:projections}
\end{figure*}

\begin{figure*}
\centering
\includegraphics[width=58mm]{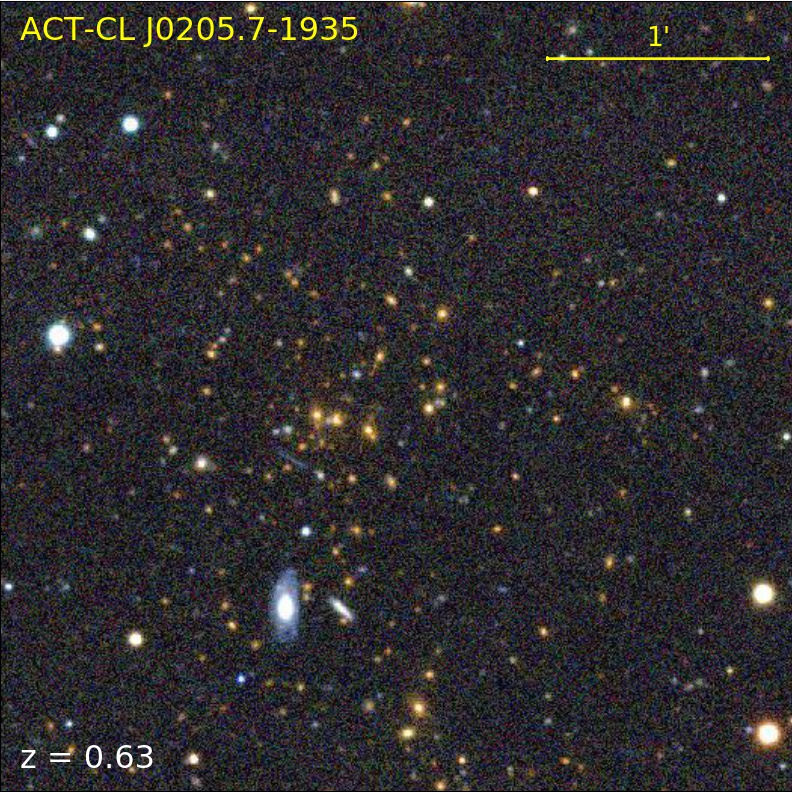}
\includegraphics[width=58mm]{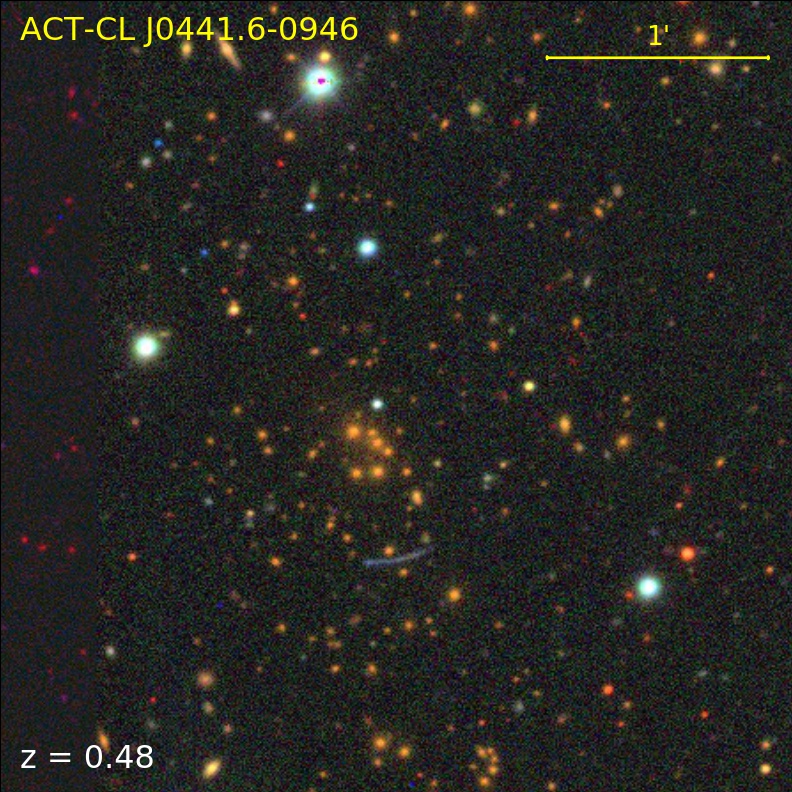}
\includegraphics[width=58mm]{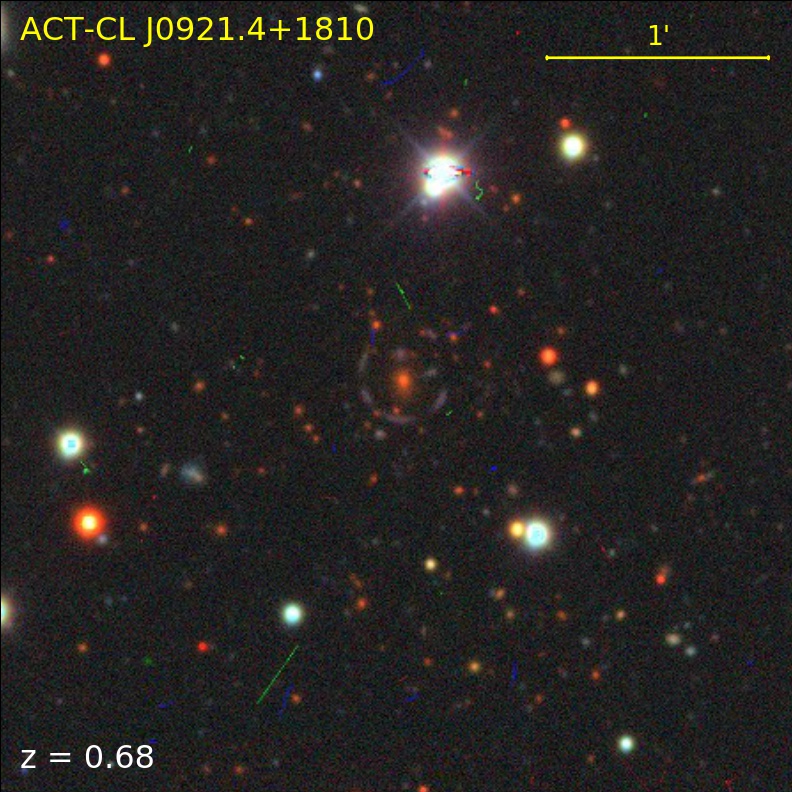}
\caption{Examples of systems that show possible strong gravitational lensing features
(see Section~\ref{sec:stronglenses}). ACT-CL\,J0205.7-1935 contains a lens candidate in the \citet{Diehl_2020}
catalog; ACT-CL\,J0441.6-0946 contains the known lens
DESI-070.4130-09.7774 \citep{HuangH20b_2020}; and the lens in ACT-CL\,J0921.4+1810 appears to be identified for the first
time in this work. Each image is 3.5\arcmin{} on a side, 
with North at the top and East at the left.}
\label{fig:stronglenses}
\end{figure*}

\begin{figure*}
\centering
\includegraphics[width=58mm]{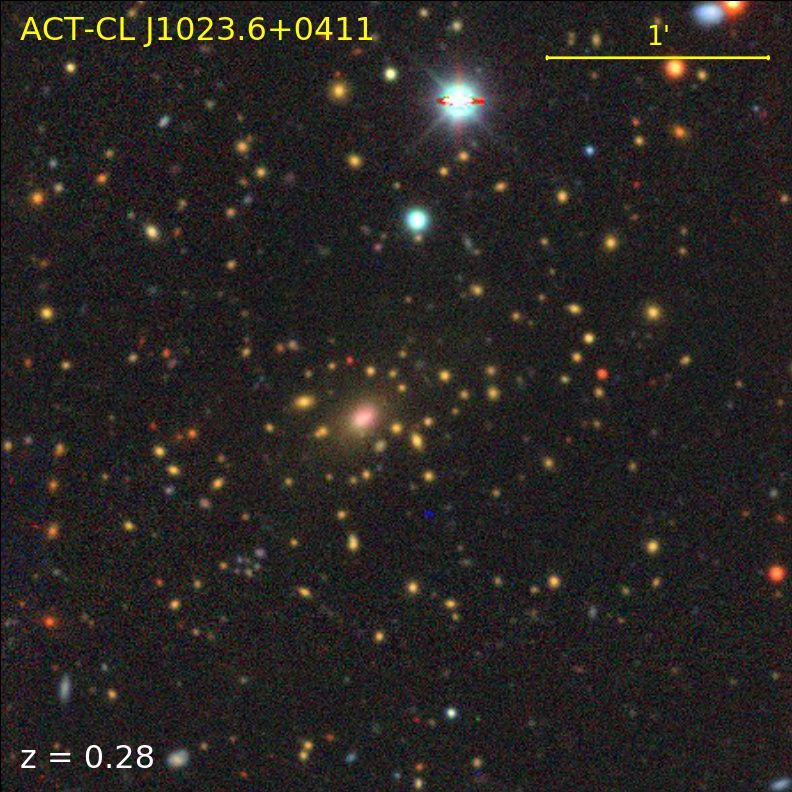}
\includegraphics[width=58mm]{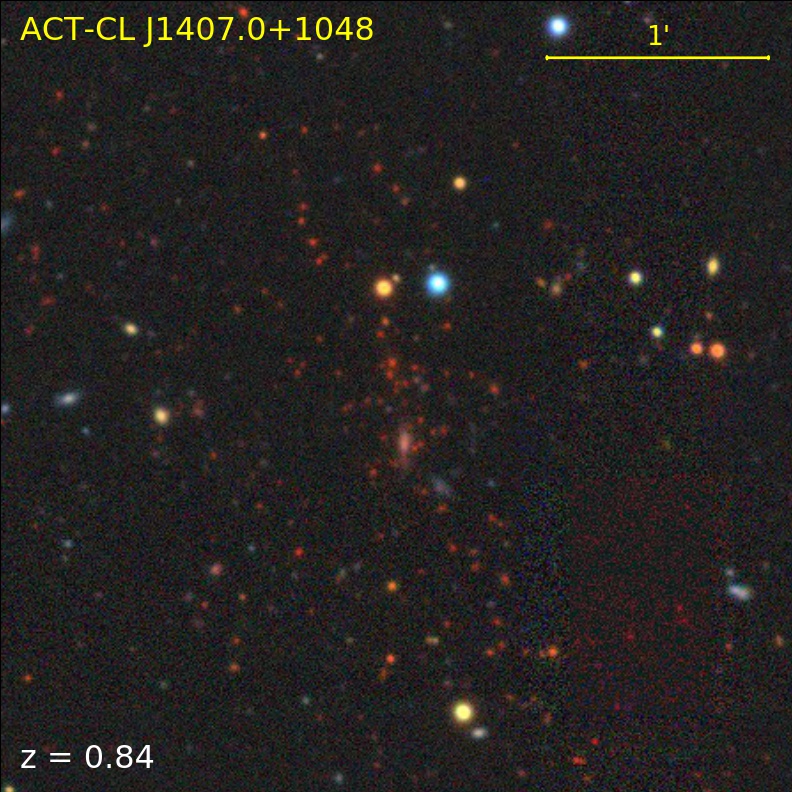}
\includegraphics[width=58mm]{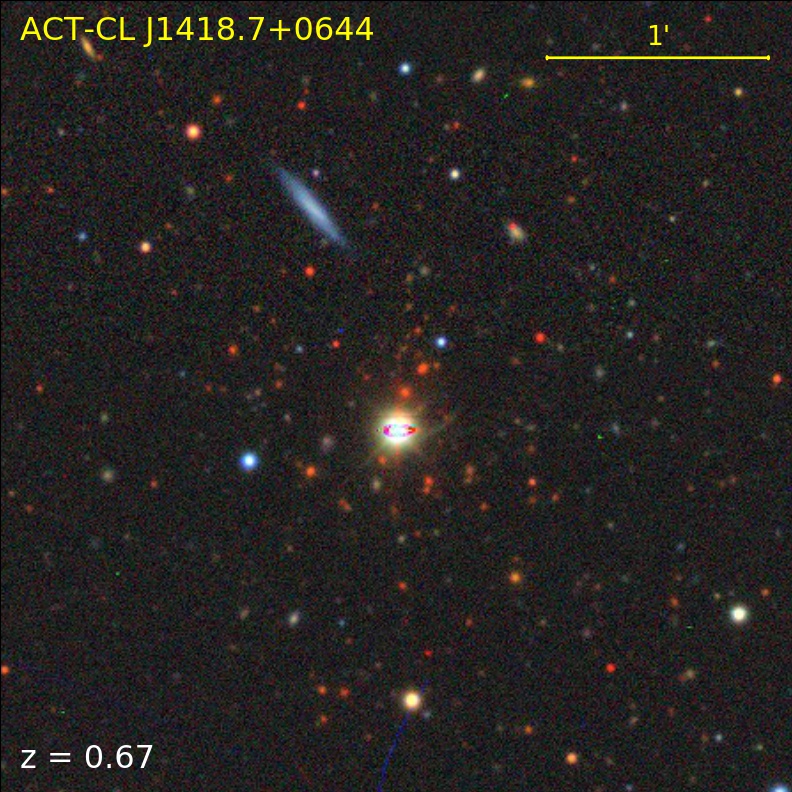}
\caption{Examples of systems with activity in the cluster core (star formation or possible AGNs;
see Section~\ref{sec:AGNs}). Each image is 3.5\arcmin{} on a side, with 
North at the top and East at the left. ACT-CL\,J1023.6+0411 is the well known $z = 0.29$
cool core ZwCl~3146 cluster; ACT-CL\,J1407.0+1048 is the second ranked cluster in
the sample at $z > 0.8$ in terms of SNR, and perhaps has a starburst galaxy in its core;
ACT-CL\,J1418.7+0644 is a high significance detection by ACT,
but was previously rejected as a false detection in the X-ray cluster catalog of 
\citet{Vikhlinin_1998}.}
\label{fig:AGNs}
\end{figure*}

\begin{figure*}
\centering
\includegraphics[width=58mm]{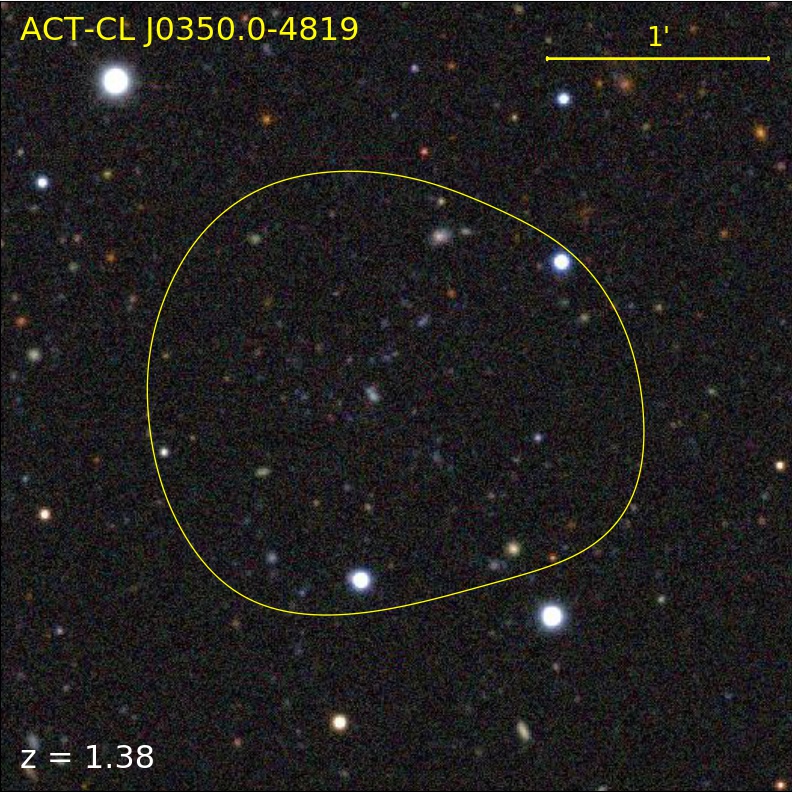}
\includegraphics[width=58mm]{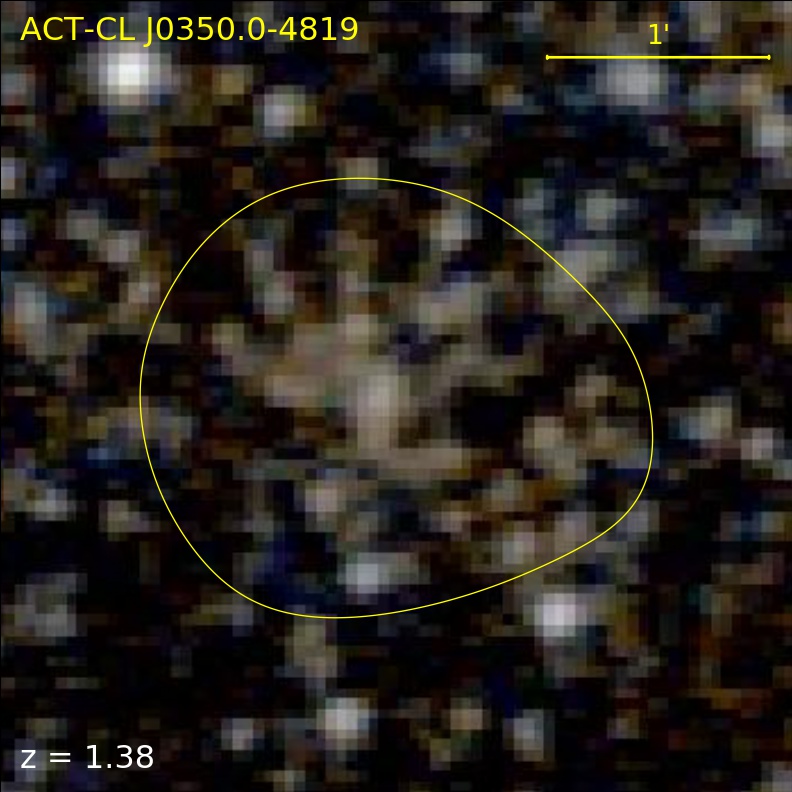}
\caption{The newly discovered high redshift cluster ACT-CL\,J0350.0-4819 at $z = 1.38$.
The left panel shows the DES Y3 $gri$ optical image, which shows a striking number of
blue galaxies within the yellow SZ signal-to-noise contour. The right panel shows the
WISE IR imaging, demonstrating that the cluster itself is a genuine high redshift system.}
\label{fig:bluecluster}
\end{figure*}

\begin{figure*}
\centering
\includegraphics[width=58mm]{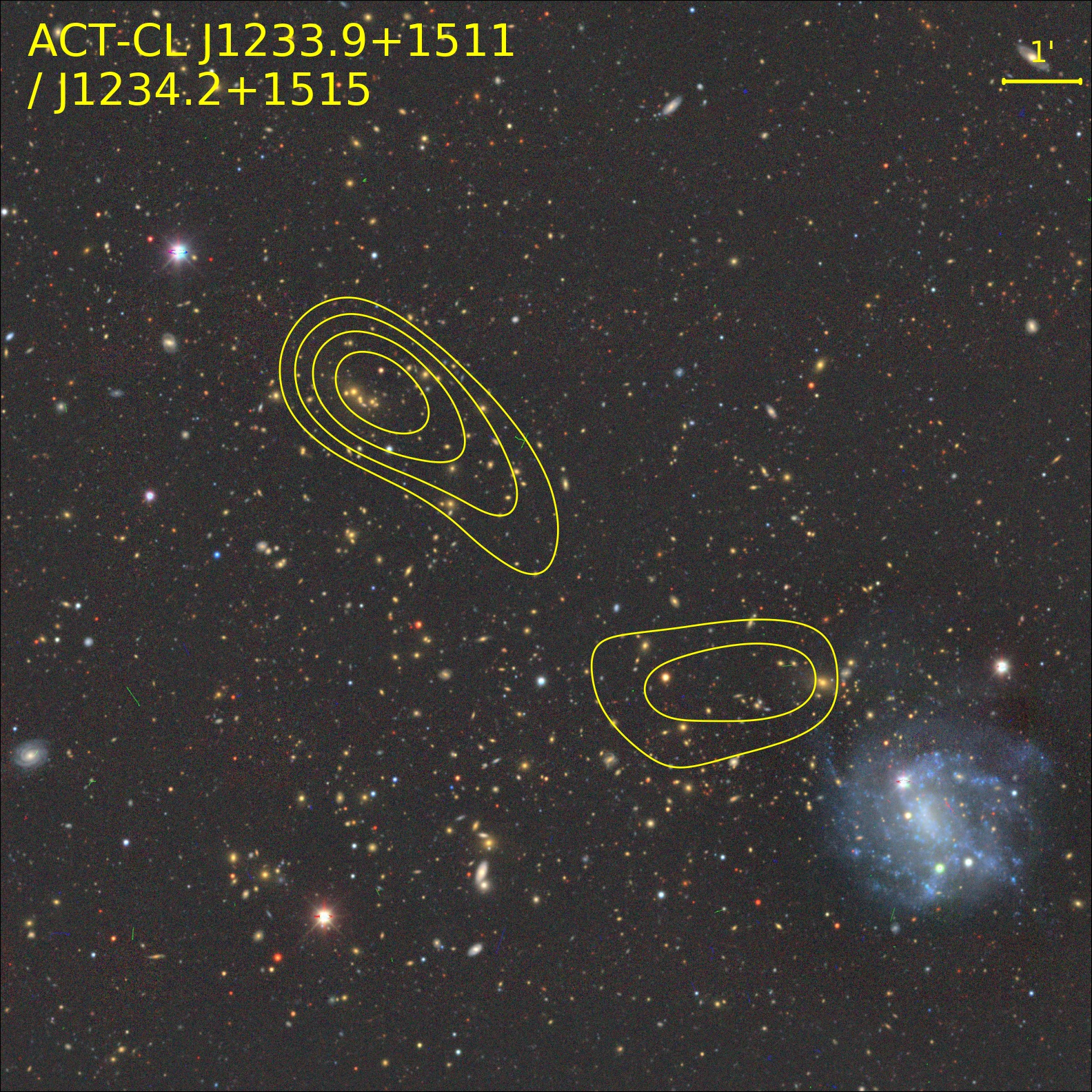}
\includegraphics[width=58mm]{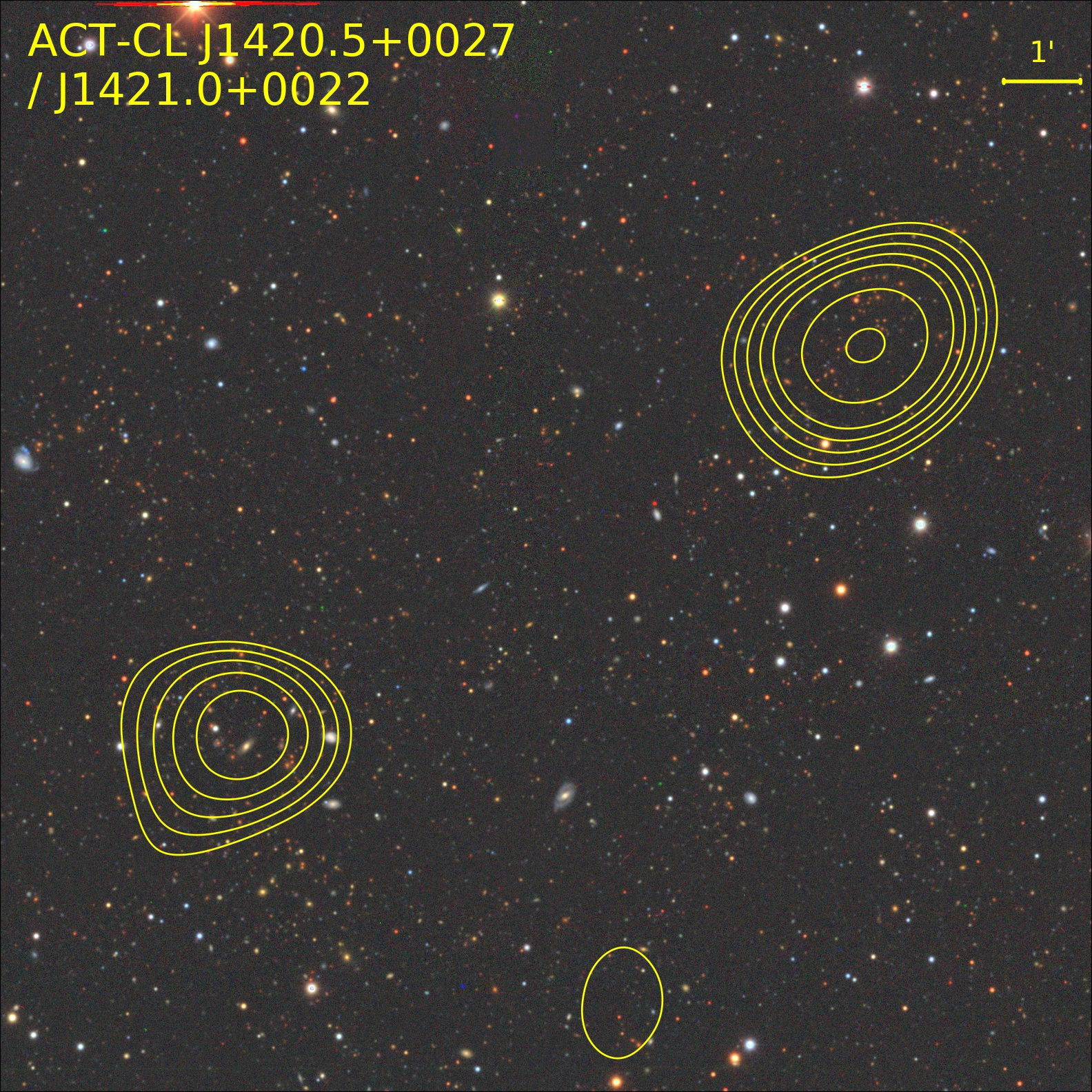}
\includegraphics[width=58mm]{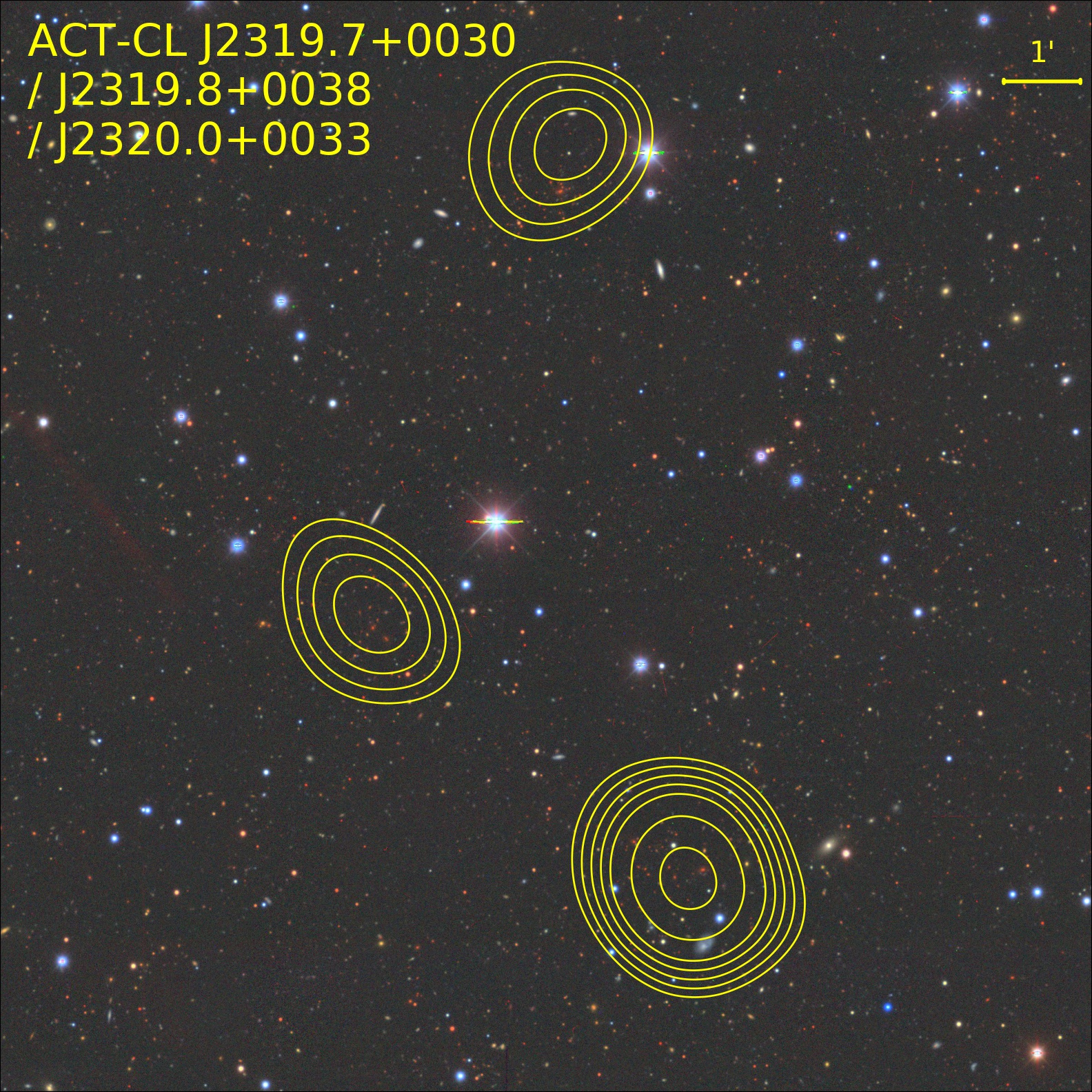}
\caption{Example DECaLS $grz$ images of multiple systems (see Section~\ref{sec:superclusters}):
ACT-CL\,J1233.9+1511/J1234.2+1515 at $z = 0.29$ (possibly a pre- or post-merger system); 
ACT-CL\,J1420.5+0027/J1421.0+0022 at $z = 0.64$ (two clusters separated by 3.6\,Mpc); and
the $z = 0.90$ triple system ACT-CL\,J2319.7+0030/J2319.8+0038/J2320.0+0033 
\citep[the RCS2 supercluster;][]{2008ApJ...677L..89G}. Each image is 13\arcmin{} on a side, with 
North at the top, East at the left. The lowest contour level shown corresponds to 
3$\sigma$ significance in the filtered ACT map. The difference between levels increases by 
0.5$\sigma$ ($3 < $~SNR~$< 5$), 1$\sigma$ ($5 < $~SNR~$< 10$), and 2$\sigma$ ($10 < $~SNR~$< 20$).}
\label{fig:superclusters}
\end{figure*}

As an independent check of the richness-based weak-lensing
mass calibration used in this work, Fig.~\ref{fig:CoMaLitMasses} presents a comparison
with a heterogeneous database of weak-lensing masses assembled from the literature
\citep{Sereno_2015}. Even though the comparison is made only with clusters that have $<25$\%
uncertainty in the weak-lensing masses, the scatter is large. Nevertheless, the overall
mass scale is consistent with the richness-based weak-lensing calibration
derived from DES observations;
$\langle M_{\rm 500c} \left[\rm CoMaLit \right] \rangle = (1.09 \pm 0.08) \, \langle M^{\rm Cal}_{\rm 500c} \rangle$.
Future work will explore the mass calibration of the ACT DR5 sample using optical weak-lensing data from
DES, HSC-SSP, and KiDS, as well as from gravitational lensing of the CMB.

\subsection{Notable Clusters}

In this Section we briefly discuss a few notable categories of systems that may be of interest for future
studies. This list is not meant to be comprehensive, and results for the most part from visual inspection 
of the cluster catalog using the available optical/IR data (see Section~\ref{sec:visualInspect}).
Further possible examples besides those mentioned here may be found by inspecting the \texttt{notes}
and \texttt{warnings} columns of the cluster catalog (see Table~\ref{tab:FITSTableColumns}).

\subsubsection{Projected Systems}
\label{sec:projected}
During visual inspection of the cluster candidates, we identified 46 systems that may be 
projections of two or more clusters at different redshifts.
These are indicated in the \texttt{warnings} field of the cluster catalog (see Table~\ref{tab:FITSTableColumns}).
Fig.~\ref{fig:projections} shows a few examples. One of these cases (ACT-CL\,J0335.1-4036) is clearly a blended
SZ detection of two systems, which the cluster finder has failed to separate because all of the pixels in
both systems are well above our detection threshold. We will seek to improve the object deblending for future
cluster catalog releases.

\subsubsection{Strong Lensing Systems}
\label{sec:stronglenses}
A search of the literature shows that there are \numKnownLenses{} known strong gravitational lenses located
within 2\,Mpc projected distance of clusters in the ACT DR5 release, as recorded in the \texttt{knownLens} field
of the cluster catalog (see Table~\ref{tab:FITSTableColumns}). Table~\ref{tab:lensCodes} lists the lens 
catalogs that were searched and the corresponding code used in the \texttt{knownLensRefCode} in the 
cluster catalog. We also identified a further \numPossNewLenses{} clusters that show possible strong
lensing features, based on visual inspection of the available optical imaging. These
are indicated in the \texttt{notes} field of the cluster catalog.
Fig.~\ref{fig:stronglenses} shows some examples of both known lenses and new candidates.

\subsubsection{Systems With Active or Star Forming Central Galaxies}
\label{sec:AGNs}
We flagged 14 systems as potentially hosting central AGNs or significant star formation, 
purely on the basis of their appearance in the available
optical/IR imaging, including the well known cool core cluster ZwCl~3146 (ACT-CL\,J1023.6+0411) 
at $z=0.29$ \citep[e.g.,][]{Romero_2020}. One of our highest significance detections at $z >0.8$ 
is a new cluster in this category, ACT-CL\,J1407.0+1048 ($z = 0.84$, SNR$_{2.4} = 33.8$,
$M^{\rm Cal}_{500c} = (9.1^{+1.7}_{-1.5}) \times 10^{14}$\,$M_{\sun}$), which has a blue
BCG as shown in the DECaLS image (Fig.~\ref{fig:AGNs}). This cluster may have similar properties
to the Phoenix cluster \citep{McDonald_2012}, but follow-up at other wavelengths is needed to
confirm this.
Some objects in this category may be ``quasars masquerading as clusters" as identified at X-ray wavelengths \citep{Somboonpanyakul_2018, Donahue_2020}, following further
investigation. For example, ACT-CL\,J1418.7+0644 (pictured in Fig.~\ref{fig:AGNs}) is detected
with SNR$_{2.4} = 12.7$ in the ACT DR5 catalog, but was rejected as a false detection in the 
X-ray cluster catalog of \citet{Vikhlinin_1998}.

\subsubsection{A Blue High Redshift Galaxy Cluster?}
\label{sec:bluecluster}
Fig.~\ref{fig:bluecluster} shows optical and WISE IR imaging of the newly discovered 
$z = 1.38$ cluster ACT-CL\,J0350.0-4819. The photometric redshift of this system was determined
using DECaLS photometry (see Section~\ref{sec:zCluster}), and we lack any spectroscopic 
information. Nevertheless, there is an apparent overdensity of galaxies with blue colors at
the cluster position, as seen in the DES $gri$ optical image. If this is not simply 
projection along the line of sight, then it may be that this system hosts an unusually
large number of star forming galaxies. We intend to obtain follow-up spectroscopy of this
system to determine if this is in fact the case.
ACT-CL\,J0350.0-4819 is also detected by the Wavelet Z Photometric optical cluster finding algorithm
(WaZP; \citealt{Aguena_2020} presents a catalog based on DES Y1), in a preliminary search of the
DES Y6 data. WaZP does not assume a red-sequence model and searches for clusters as spatial
overdensities using photometric redshifts.

\subsubsection{Multiple Systems} 
\label{sec:superclusters}

We conducted a search for pairs or groups of clusters in the catalog that may be physically associated. These objects
may be of interest for those studying cluster mergers, filaments/large scale structure around clusters, and
superclusters. We select candidates for this category as objects that have a neighbouring SZ source
within a projected distance of 10\,Mpc, and a peculiar velocity difference of $<5000$\,km\,s$^{-1}$. We find
a total of 160 such systems, consisting of 144 pairs, 15 triples, and 1 quadruple system, which are listed in
Table~\ref{tab:superclusters}. Note, however, that some clusters are part of more than one system (e.g., the 
$z = 0.49$ triple system ACT-CL\,J0059.6+1310/J0059.8+1344/J0059.9+1319 is also listed as the pairs
ACT-CL\,J0059.6+1310/J0059.9+1319 and ACT-CL\,J0059.8+1344/J0059.9+1319). We also include 
objects with photometric redshifts in this search, but flag these in the catalog, since the uncertainties 
on these redshifts are much larger than spectroscopic redshift errors.

\begin{deluxetable}{p{3.5cm}ccc}
\small
\tablecaption{Systems of Multiple SZ Sources\label{tab:superclusters}}
\tablewidth{0pt}
\decimals
\tablehead{
\colhead{Name} &
\colhead{Mean $z$}   & 
\colhead{Separation} &
\colhead{Photo-$z$?}\\
\colhead{}           &  
\colhead{}           & 
\colhead{(Mpc)} &
\colhead{}
}
\startdata
\raggedleft ACT-CL\,J0000.7$+$0225\\/J2359.5$+$0208 & 0.43 & 8.1 & \nodata \\
\raggedleft ACT-CL\,J0003.0$-$3520\\/J0003.8$-$3517 & 0.76 & 4.1 & \checkmark \\
\raggedleft ACT-CL\,J0005.0$+$0212\\/J0005.7$+$0222 & 0.84 & 7.0 & \checkmark \\
\raggedleft ACT-CL\,J0018.3$+$1618\\/J0018.5$+$1626 & 0.55 & 3.3 & \nodata \\
\raggedleft ACT-CL\,J0019.6$+$0336\\/J0020.0$+$0351 & 0.27 & 4.0 & \checkmark \\
\enddata
\tablecomments{Only a subset of the available fields in this catalog are shown here. 
Table~\ref{tab:superclusters} is published in its entirety in machine-readable format. 
A portion is shown here for guidance regarding its form and content.}
\end{deluxetable}

We find multiple systems across the redshift range $0.04 < z < 1.2$ (median $z = 0.42)$, and
the average maximum projected separation between the components of these systems is 5.8\,Mpc.
Fig.~\ref{fig:superclusters} shows a few examples.
Due to the increased depth of the ACT DR5 maps, we now detect all three components of the $z = 0.9$ RCS2 supercluster 
\citep[][recorded here as ACT-CL\,J2319.7+0030/J2319.8+0038/J2320.0+0033]{2008ApJ...677L..89G}. Some of these systems
may be pre or post-merger systems (e.g., ACT-CL\,J1233.9+1511/J1234.2+1515 at $z = 0.29$, shown in Fig.~\ref{fig:superclusters}).

\section{Summary}
\label{sec:Summary}

This work presents the first cluster catalog derived from 98 and 150\,GHz observations with the AdvACT receiver, 
covering a search area of \surveyArea{}\,deg$^2$. The catalog contains \totalConfirmed{} optically confirmed
galaxy clusters with redshift and mass estimates, making it the largest SZ-selected cluster catalog to date.
It is more than 22 times larger than the previous ACT cluster catalog \citep{Hilton_2018}, illustrating
the huge gains in sensitivity and survey speed achieved by the upgraded AdvACT receiver 
\citep{Henderson_2016}. Assuming a relation between SZ-signal and mass calibrated from X-ray 
observations \citep{Arnaud_2010}, the 90\% completeness limit of the survey for
SNR$_{2.4} > 5$ is $M_{\rm 500c} > \compLimitFull{} \times 10^{14}\,M_{\sun}$.

Thanks to the overlap with deep and wide optical surveys like DES \citep{Abbott_2018},
DECaLS \citep{Dey_2019}, HSC-SSP \citep{Aihara_2018}, KiDS \citep{Wright_2019}, and SDSS \citep{2020ApJS..249....3A},
the optical follow-up of the survey is complete over much of the survey area. The cluster sample
has median $z = \medianRedshift{}$ and covers the redshift range $\minRedshift{} < z < \maxRedshift{}$, 
with \totalHighZ{} $z > 1$ systems, and \totalNew{} newly discovered clusters. In the regions that
overlap with DES Y3, HSC S19A, and KiDS DR4, $95-98$\% of the candidates detected with SNR$_{2.4} > 6$ 
have been confirmed as clusters. 

The cluster and source detection package developed for this work is capable of analysing the next 
generation of deep, wide multi-frequency mm-wave maps that will be produced by experiments such as the 
Simons Observatory \citep{SimonsObs_2019}. It will be made publicly available at 
\url{https://github.com/simonsobs/nemo/} and on the Python Package Index (PyPI) under a 
free software license.

\acknowledgments

ACT was supported by the U.S. National Science Foundation through awards AST-0408698, AST-0965625, and AST-1440226 for the ACT project, as well as awards PHY-0355328, PHY-0855887 and PHY-1214379. Funding was also provided by Princeton University, the University of Pennsylvania, and a Canada Foundation for Innovation (CFI) award to UBC. ACT operates in the Parque Astron\'omico Atacama in northern Chile under the auspices of the Comisi\'on Nacional de Investigaci\'on (CONICYT).  The development of multichroic detectors and lenses was supported by NASA grants NNX13AE56G and NNX14AB58G. Detector research at NIST was supported by the NIST Innovations in Measurement Science program. 

CS acknowledges support from the Agencia Nacional de Investigaci\'on y Desarrollo through FONDECYT Iniciaci\'on grant no.~11191125.
SKC acknowledges support from the Cornell Presidential Postdoctoral Fellowship. RD thanks CONICYT for grant BASAL CATA AFB-170002. ZL, ES and JD are supported through NSF grant AST-1814971. KM and MHi acknowledge support from the National Research Foundation of South Africa (grant number 112132). MDN acknowledges support from NSF award AST-1454881.
DH, AM, and NS acknowledge support from NSF grant numbers AST-1513618 and AST-1907657. EC acknowledges support from the STFC Ernest Rutherford Fellowship ST/M004856/2 and STFC Consolidated Grant ST/S00033X/1, and from the Horizon 2020 ERC Starting Grant (Grant agreement No 849169). NB acknowledges support from NSF grant AST-1910021. ML was supported by a Dicke Fellowship. LP acknowledges support from the Mishrahi and Wilkinson funds. 
AJ acknowledges support from JSPS KAKENHI Grant Number JP17H02868.
JPH acknowledges funding for SZ cluster studies from NSF grant number AST-1615657.
RD thanks CONICYT for grant BASAL CATA AFB-170002.
The Flatiron Institute is supported by the Simons Foundation.
Computations were performed on Hippo at the University of KwaZulu-Natal.


Funding for the DES Projects has been provided by the DOE and NSF(USA), MEC/MICINN/MINECO(Spain), STFC(UK), HEFCE(UK). NCSA(UIUC), KICP(U. Chicago), CCAPP(Ohio State), 
MIFPA(Texas A\&M), CNPQ, FAPERJ, FINEP (Brazil), DFG(Germany) and the Collaborating Institutions in the Dark Energy Survey.

The Collaborating Institutions are Argonne Lab, UC Santa Cruz, University of Cambridge, CIEMAT-Madrid, University of Chicago, University College London, 
DES-Brazil Consortium, University of Edinburgh, ETH Z{\"u}rich, Fermilab, University of Illinois, ICE (IEEC-CSIC), IFAE Barcelona, Lawrence Berkeley Lab, 
LMU M{\"u}nchen and the associated Excellence Cluster Universe, University of Michigan, NFS's NOIRLab, University of Nottingham, Ohio State University, University of 
Pennsylvania, University of Portsmouth, SLAC National Lab, Stanford University, University of Sussex, Texas A\&M University, and the OzDES Membership Consortium.

Based in part on observations at Cerro Tololo Inter-American Observatory at NSF’s NOIRLab (NOIRLab Prop. ID 2012B-0001; PI: J. Frieman), which is managed by the Association of Universities for Research in Astronomy (AURA) under a cooperative agreement with the National Science Foundation.

The DES Data Management System is supported by the NSF under Grant Numbers AST-1138766 and AST-1536171. 
The DES participants from Spanish institutions are partially supported by MICINN under grants ESP2017-89838, PGC2018-094773, PGC2018-102021, SEV-2016-0588, SEV-2016-0597, and MDM-2015-0509, some of which include ERDF funds from the European Union. IFAE is partially funded by the CERCA program of the Generalitat de Catalunya.
Research leading to these results has received funding from the European Research
Council under the European Union's Seventh Framework Program (FP7/2007-2013) including ERC grant agreements 240672, 291329, and 306478.
We  acknowledge support from the Brazilian Instituto Nacional de Ci\^encia
e Tecnologia (INCT) do e-Universo (CNPq grant 465376/2014-2).

This manuscript has been authored by Fermi Research Alliance, LLC under Contract No. DE-AC02-07CH11359 with the U.S. Department of Energy, Office of Science, Office of High Energy Physics. 

The Hyper Suprime-Cam (HSC) collaboration includes the astronomical communities of Japan and Taiwan, and Princeton University.  The HSC instrumentation and software were developed by the National Astronomical Observatory of Japan (NAOJ), the Kavli Institute for the Physics and Mathematics of the Universe (Kavli IPMU), the University of Tokyo, the High Energy Accelerator Research Organization (KEK), the Academia Sinica Institute for Astronomy and Astrophysics in Taiwan (ASIAA), and Princeton University.  Funding was contributed by the FIRST program from the Japanese Cabinet Office, the Ministry of Education, Culture, Sports, Science and Technology (MEXT), the Japan Society for the Promotion of Science (JSPS), Japan Science and Technology Agency  (JST), the Toray Science  Foundation, NAOJ, Kavli IPMU, KEK, ASIAA, and Princeton University.

This paper makes use of software developed for the Rubin Observatory. We thank the Rubin Observatory Project for making their code available as free software at \url{http://dm.lsst.org}.

This paper is based in part on data collected at the Subaru Telescope and retrieved from the HSC data archive system, which is operated by Subaru Telescope and Astronomy Data Center (ADC) at NAOJ. Data analysis was in part carried out with the cooperation of Center for Computational Astrophysics (CfCA), NAOJ.

The Pan-STARRS1 Surveys (PS1) and the PS1 public science archive have been made possible through contributions by the Institute for Astronomy, the University of Hawaii, the Pan-STARRS Project Office, the Max Planck Society and its participating institutes, the Max Planck Institute for Astronomy, Heidelberg, and the Max Planck Institute for Extraterrestrial Physics, Garching, The Johns Hopkins University, Durham University, the University of Edinburgh, the Queen’s University Belfast, the Harvard-Smithsonian Center for Astrophysics, the Las Cumbres Observatory Global Telescope Network Incorporated, the National Central University of Taiwan, the Space Telescope Science Institute, the National Aeronautics and Space Administration under grant No. NNX08AR22G issued through the Planetary Science Division of the NASA Science Mission Directorate, the National Science Foundation grant No. AST-1238877, the University of Maryland, Eotvos Lorand University (ELTE), the Los Alamos National Laboratory, and the Gordon and Betty Moore Foundation.

The Legacy Surveys consist of three individual and complementary projects: the Dark Energy Camera Legacy Survey (DECaLS; NOAO Proposal ID \# 2014B-0404; PIs: David Schlegel and Arjun Dey), the Beijing-Arizona Sky Survey (BASS; NOAO Proposal ID \# 2015A-0801; PIs: Zhou Xu and Xiaohui Fan), and the Mayall z-band Legacy Survey (MzLS; NOAO Proposal ID \# 2016A-0453; PI: Arjun Dey). DECaLS, BASS and MzLS together include data obtained, respectively, at the Blanco telescope, Cerro Tololo Inter-American Observatory, National Optical Astronomy Observatory (NOAO); the Bok telescope, Steward Observatory, University of Arizona; and the Mayall telescope, Kitt Peak National Observatory, NOAO. The Legacy Surveys project is honored to be permitted to conduct astronomical research on Iolkam Du'ag (Kitt Peak), a mountain with particular significance to the Tohono O'odham Nation.

NOAO is operated by the Association of Universities for Research in Astronomy (AURA) under a cooperative agreement with the National Science Foundation.


The Legacy Survey team makes use of data products from the Near-Earth Object Wide-field Infrared Survey Explorer (NEOWISE), which is a project of the Jet Propulsion Laboratory/California Institute of Technology. NEOWISE is funded by the National Aeronautics and Space Administration.

The Legacy Surveys imaging of the DESI footprint is supported by the Director, Office of Science, Office of High Energy Physics of the U.S. Department of Energy under Contract No. DE-AC02-05CH1123, by the National Energy Research Scientific Computing Center, a DOE Office of Science User Facility under the same contract; and by the U.S. National Science Foundation, Division of Astronomical Sciences under Contract No. AST-0950945 to NOAO.

\software{
AstroPy \citep{Astropy_2013},
Core Cosmology Library \citep{Chisari_2019}
Pixell (\url{https://github.com/simonsobs/pixell/})
}


\clearpage

\bibliographystyle{aasjournal}
\bibliography{AdvACTClusters}

\begin{thebibliography}{}
\expandafter\ifx\csname natexlab\endcsname\relax\def\natexlab#1{#1}\fi
\providecommand{\url}[1]{\href{#1}{#1}}
\providecommand{\dodoi}[1]{doi:~\href{http://doi.org/#1}{\nolinkurl{#1}}}
\providecommand{\doeprint}[1]{\href{http://ascl.net/#1}{\nolinkurl{http://ascl.net/#1}}}
\providecommand{\doarXiv}[1]{\href{https://arxiv.org/abs/#1}{\nolinkurl{https://arxiv.org/abs/#1}}}

\bibitem[{{Abbott} {et~al.}(2018){Abbott}, {Abdalla}, {Allam}, {Amara},
  {Annis}, {Asorey}, {Avila}, {Ballester}, {Banerji}, {Barkhouse}, {Baruah},
  {Baumer}, {Bechtol}, {Becker}, {Benoit-L{\'e}vy}, {Bernstein}, {Bertin},
  {Blazek}, {Bocquet}, {Brooks}, {Brout}, {Buckley-Geer}, {Burke}, {Busti},
  {Campisano}, {Cardiel-Sas}, {Carnero Rosell}, {Carrasco Kind}, {Carretero},
  {Castander}, {Cawthon}, {Chang}, {Chen}, {Conselice}, {Costa}, {Crocce},
  {Cunha}, {D'Andrea}, {da Costa}, {Das}, {Daues}, {Davis}, {Davis}, {De
  Vicente}, {DePoy}, {DeRose}, {Desai}, {Diehl}, {Dietrich}, {Dodelson},
  {Doel}, {Drlica-Wagner}, {Eifler}, {Elliott}, {Evrard}, {Farahi}, {Fausti
  Neto}, {Fernandez}, {Finley}, {Flaugher}, {Foley}, {Fosalba}, {Friedel},
  {Frieman}, {Garc{\'\i}a-Bellido}, {Gaztanaga}, {Gerdes}, {Giannantonio},
  {Gill}, {Glazebrook}, {Goldstein}, {Gower}, {Gruen}, {Gruendl}, {Gschwend},
  {Gupta}, {Gutierrez}, {Hamilton}, {Hartley}, {Hinton}, {Hislop}, {Hollowood},
  {Honscheid}, {Hoyle}, {Huterer}, {Jain}, {James}, {Jeltema}, {Johnson},
  {Johnson}, {Kacprzak}, {Kent}, {Khullar}, {Klein}, {Kovacs}, {Koziol},
  {Krause}, {Kremin}, {Kron}, {Kuehn}, {Kuhlmann}, {Kuropatkin}, {Lahav},
  {Lasker}, {Li}, {Li}, {Liddle}, {Lima}, {Lin}, {L{\'o}pez-Reyes}, {MacCrann},
  {Maia}, {Maloney}, {Manera}, {March}, {Marriner}, {Marshall}, {Martini},
  {McClintock}, {McKay}, {McMahon}, {Melchior}, {Menanteau}, {Miller},
  {Miquel}, {Mohr}, {Morganson}, {Mould}, {Neilsen}, {Nichol}, {Nogueira},
  {Nord}, {Nugent}, {Nunes}, {Ogand o}, {Old}, {Pace}, {Palmese},
  {Paz-Chinch{\'o}n}, {Peiris}, {Percival}, {Petravick}, {Plazas}, {Poh},
  {Pond}, {Porredon}, {Pujol}, {Refregier}, {Reil}, {Ricker}, {Rollins},
  {Romer}, {Roodman}, {Rooney}, {Ross}, {Rykoff}, {Sako}, {Sanchez}, {Sanchez},
  {Santiago}, {Saro}, {Scarpine}, {Scolnic}, {Serrano}, {Sevilla-Noarbe},
  {Sheldon}, {Shipp}, {Silveira}, {Smith}, {Smith}, {Smith}, {Soares-Santos},
  {Sobreira}, {Song}, {Stebbins}, {Suchyta}, {Sullivan}, {Swanson}, {Tarle},
  {Thaler}, {Thomas}, {Thomas}, {Troxel}, {Tucker}, {Vikram}, {Vivas},
  {Walker}, {Wechsler}, {Weller}, {Wester}, {Wolf}, {Wu}, {Yanny}, {Zenteno},
  {Zhang}, {Zuntz}, {DES Collaboration}, {Juneau}, {Fitzpatrick}, {Nikutta},
  {Nidever}, {Olsen}, {Scott}, \& {NOAO Data Lab}}]{Abbott_2018}
{Abbott}, T.~M.~C., {Abdalla}, F.~B., {Allam}, S., {et~al.} 2018, \apjs, 239,
  18, \dodoi{10.3847/1538-4365/aae9f0}

\bibitem[{{Abell} {et~al.}(1989){Abell}, {Corwin}, \&
  {Olowin}}]{1989ApJS...70....1A}
{Abell}, G.~O., {Corwin}, Harold~G., J., \& {Olowin}, R.~P. 1989, \apjs, 70, 1,
  \dodoi{10.1086/191333}

\bibitem[{{Ade} {et~al.}(2019){Ade}, {Aguirre}, {Ahmed}, {Aiola}, {Ali},
  {Alonso}, {Alvarez}, {Arnold}, {Ashton}, {Austermann}, {Awan}, {Baccigalupi},
  {Baildon}, {Barron}, {Battaglia}, {Battye}, {Baxter}, {Bazarko}, {Beall},
  {Bean}, {Beck}, {Beckman}, {Beringue}, {Bianchini}, {Boada}, {Boettger},
  {Bond}, {Borrill}, {Brown}, {Bruno}, {Bryan}, {Calabrese}, {Calafut},
  {Calisse}, {Carron}, {Challinor}, {Chesmore}, {Chinone}, {Chluba}, {Cho},
  {Choi}, {Coppi}, {Cothard}, {Coughlin}, {Crichton}, {Crowley}, {Crowley},
  {Cukierman}, {D'Ewart}, {D{\"u}nner}, {de Haan}, {Devlin}, {Dicker},
  {Didier}, {Dobbs}, {Dober}, {Duell}, {Duff}, {Duivenvoorden}, {Dunkley},
  {Dusatko}, {Errard}, {Fabbian}, {Feeney}, {Ferraro}, {Flux{\`a}}, {Freese},
  {Frisch}, {Frolov}, {Fuller}, {Fuzia}, {Galitzki}, {Gallardo}, {Tomas Galvez
  Ghersi}, {Gao}, {Gawiser}, {Gerbino}, {Gluscevic}, {Goeckner-Wald}, {Golec},
  {Gordon}, {Gralla}, {Green}, {Grigorian}, {Groh}, {Groppi}, {Guan},
  {Gudmundsson}, {Han}, {Hargrave}, {Hasegawa}, {Hasselfield}, {Hattori},
  {Haynes}, {Hazumi}, {He}, {Healy}, {Henderson}, {Hervias-Caimapo}, {Hill},
  {Hill}, {Hilton}, {Hilton}, {Hincks}, {Hinshaw}, {Hlo{\v{z}}ek}, {Ho}, {Ho},
  {Howe}, {Huang}, {Hubmayr}, {Huffenberger}, {Hughes}, {Ijjas}, {Ikape},
  {Irwin}, {Jaffe}, {Jain}, {Jeong}, {Kaneko}, {Karpel}, {Katayama}, {Keating},
  {Kernasovskiy}, {Keskitalo}, {Kisner}, {Kiuchi}, {Klein}, {Knowles},
  {Koopman}, {Kosowsky}, {Krachmalnicoff}, {Kuenstner}, {Kuo}, {Kusaka},
  {Lashner}, {Lee}, {Lee}, {Leon}, {Leung}, {Lewis}, {Li}, {Li}, {Limon},
  {Linder}, {Lopez-Caraballo}, {Louis}, {Lowry}, {Lungu}, {Madhavacheril},
  {Mak}, {Maldonado}, {Mani}, {Mates}, {Matsuda}, {Maurin}, {Mauskopf}, {May},
  {McCallum}, {McKenney}, {McMahon}, {Meerburg}, {Meyers}, {Miller},
  {Mirmelstein}, {Moodley}, {Munchmeyer}, {Munson}, {Naess}, {Nati},
  {Navaroli}, {Newburgh}, {Nguyen}, {Niemack}, {Nishino}, {Orlowski-Scherer},
  {Page}, {Partridge}, {Peloton}, {Perrotta}, {Piccirillo}, {Pisano},
  {Poletti}, {Puddu}, {Puglisi}, {Raum}, {Reichardt}, {Remazeilles},
  {Rephaeli}, {Riechers}, {Rojas}, {Roy}, {Sadeh}, {Sakurai}, {Salatino},
  {Sathyanarayana Rao}, {Schaan}, {Schmittfull}, {Sehgal}, {Seibert}, {Seljak},
  {Sherwin}, {Shimon}, {Sierra}, {Sievers}, {Sikhosana}, {Silva-Feaver},
  {Simon}, {Sinclair}, {Siritanasak}, {Smith}, {Smith}, {Spergel}, {Staggs},
  {Stein}, {Stevens}, {Stompor}, {Suzuki}, {Tajima}, {Takakura}, {Teply},
  {Thomas}, {Thorne}, {Thornton}, {Trac}, {Tsai}, {Tucker}, {Ullom},
  {Vagnozzi}, {van Engelen}, {Van Lanen}, {Van Winkle}, {Vavagiakis},
  {Verg{\`e}s}, {Vissers}, {Wagoner}, {Walker}, {Ward}, {Westbrook},
  {Whitehorn}, {Williams}, {Williams}, {Wollack}, {Xu}, {Yu}, {Yu}, {Zago},
  {Zhang}, {Zhu}, \& {Simons Observatory Collaboration}}]{SimonsObs_2019}
{Ade}, P., {Aguirre}, J., {Ahmed}, Z., {et~al.} 2019, \jcap, 2019, 056,
  \dodoi{10.1088/1475-7516/2019/02/056}

\bibitem[{{Aguena} {et~al.}(2020)}]{Aguena_2020}
{Aguena}, M., {et~al.} 2020, in preparation, 000.
\newblock \doarXiv{TBC}

\bibitem[{{Ahumada} {et~al.}(2020){Ahumada}, {Allende Prieto}, {Almeida},
  {Anders}, {Anderson}, {Andrews}, {Anguiano}, {Arcodia}, {Armengaud},
  {Aubert}, \& et~al.}]{2020ApJS..249....3A}
{Ahumada}, R., {Allende Prieto}, C., {Almeida}, A., {et~al.} 2020, \apjs, 249,
  3, \dodoi{10.3847/1538-4365/ab929e}

\bibitem[{{Aihara} {et~al.}(2018){Aihara}, {Arimoto}, {Armstrong}, {Arnouts},
  {Bahcall}, {Bickerton}, {Bosch}, {Bundy}, {Capak}, {Chan}, {Chiba}, {Coupon},
  {Egami}, {Enoki}, {Finet}, {Fujimori}, {Fujimoto}, {Furusawa}, {Furusawa},
  {Goto}, {Goulding}, {Greco}, {Greene}, {Gunn}, {Hamana}, {Harikane},
  {Hashimoto}, {Hattori}, {Hayashi}, {Hayashi}, {He{\l}miniak}, {Higuchi},
  {Hikage}, {Ho}, {Hsieh}, {Huang}, {Huang}, {Ikeda}, {Imanishi}, {Inoue},
  {Iwasawa}, {Iwata}, {Jaelani}, {Jian}, {Kamata}, {Karoji}, {Kashikawa},
  {Katayama}, {Kawanomoto}, {Kayo}, {Koda}, {Koike}, {Kojima}, {Komiyama},
  {Konno}, {Koshida}, {Koyama}, {Kusakabe}, {Leauthaud}, {Lee}, {Lin}, {Lin},
  {Lupton}, {Mand elbaum}, {Matsuoka}, {Medezinski}, {Mineo}, {Miyama},
  {Miyatake}, {Miyazaki}, {Momose}, {More}, {More}, {Moritani}, {Moriya},
  {Morokuma}, {Mukae}, {Murata}, {Murayama}, {Nagao}, {Nakata}, {Niida},
  {Niikura}, {Nishizawa}, {Obuchi}, {Oguri}, {Oishi}, {Okabe}, {Okamoto},
  {Okura}, {Ono}, {Onodera}, {Onoue}, {Osato}, {Ouchi}, {Price}, {Pyo}, {Sako},
  {Sawicki}, {Shibuya}, {Shimasaku}, {Shimono}, {Shirasaki}, {Silverman},
  {Simet}, {Speagle}, {Spergel}, {Strauss}, {Sugahara}, {Sugiyama}, {Suto},
  {Suyu}, {Suzuki}, {Tait}, {Takada}, {Takata}, {Tamura}, {Tanaka}, {Tanaka},
  {Tanaka}, {Tanaka}, {Terai}, {Terashima}, {Toba}, {Tominaga}, {Toshikawa},
  {Turner}, {Uchida}, {Uchiyama}, {Umetsu}, {Uraguchi}, {Urata}, {Usuda},
  {Utsumi}, {Wang}, {Wang}, {Wong}, {Yabe}, {Yamada}, {Yamanoi}, {Yasuda},
  {Yeh}, {Yonehara}, \& {Yuma}}]{Aihara_2018}
{Aihara}, H., {Arimoto}, N., {Armstrong}, R., {et~al.} 2018, \pasj, 70, S4,
  \dodoi{10.1093/pasj/psx066}

\bibitem[{{Aiola} {et~al.}(2020){Aiola}, {Calabrese}, {Maurin}, {Naess},
  {Schmitt}, {Abitbol}, {Addison}, {Ade}, {Alonso}, {Amiri}, {Amodeo},
  {Angile}, {Austermann}, {Baildon}, {Battaglia}, {Beall}, {Bean}, {Becker},
  {Bond}, {Bruno}, {Calafut}, {Campusano}, {Carrero}, {Chesmore}, {Cho.},
  {Choi}, {Clark}, {Cothard}, {Crichton}, {Crowley}, {Darwish}, {Datta},
  {Denison}, {Devlin}, {Duell}, {Duff}, {Duivenvoorden}, {Dunkley},
  {D{\"u}nner}, {Essinger-Hileman}, {Fankhanel}, {Ferraro}, {Fox}, {Fuzia},
  {Gallardo}, {Gluscevic}, {Golec}, {Grace}, {Gralla}, {Guan}, {Hall},
  {Halpern}, {Han}, {Hargrave}, {Hasselfield}, {Helton}, {Henderson},
  {Hensley}, {Hill}, {Hilton}, {Hilton}, {Hincks}, {Hlo{\v{z}}ek}, {Ho},
  {Hubmayr}, {Huffenberger}, {Hughes}, {Infante}, {Irwin}, {Jackson}, {Klein},
  {Knowles}, {Koopman}, {Kosowsky}, {Lakey}, {Li}, {Li}, {Li}, {Lokken},
  {Louis}, {Lungu}, {MacInnis}, {Madhavacheril}, {Maldonado}, {Mallaby-Kay},
  {Marsden}, {McMahon}, {Menanteau}, {Moodley}, {Morton}, {Namikawa}, {Nati},
  {Newburgh}, {Nibarger}, {Nicola}, {Niemack}, {Nolta}, {Orlowski-Sherer},
  {Page}, {Pappas}, {Partridge}, {Phakathi}, {Prince}, {Puddu}, {Qu}, {Rivera},
  {Robertson}, {Rojas}, {Salatino}, {Schaan}, {Schillaci}, {Sehgal}, {Sherwin},
  {Sierra}, {Sievers}, {Sifon}, {Sikhosana}, {Simon}, {Spergel}, {Staggs},
  {Stevens}, {Storer}, {Sunder}, {Switzer}, {Thorne}, {Thornton}, {Trac},
  {Treu}, {Tucker}, {Vale}, {Van Engelen}, {Van Lanen}, {Vavagiakis},
  {Wagoner}, {Wang}, {Ward}, {Wollack}, {Xu}, {Zago}, \& {Zhu}}]{Aiola_2020}
{Aiola}, S., {Calabrese}, E., {Maurin}, L., {et~al.} 2020, arXiv e-prints,
  arXiv:2007.07288.
\newblock \doarXiv{2007.07288}

\bibitem[{{Allen} {et~al.}(2008){Allen}, {Rapetti}, {Schmidt}, {Ebeling},
  {Morris}, \& {Fabian}}]{2008MNRAS.383..879A}
{Allen}, S.~W., {Rapetti}, D.~A., {Schmidt}, R.~W., {et~al.} 2008, \mnras, 383,
  879, \dodoi{10.1111/j.1365-2966.2007.12610.x}

\bibitem[{{Allen} {et~al.}(2004){Allen}, {Schmidt}, {Ebeling}, {Fabian}, \&
  {van Speybroeck}}]{2004MNRAS.353..457A}
{Allen}, S.~W., {Schmidt}, R.~W., {Ebeling}, H., {Fabian}, A.~C., \& {van
  Speybroeck}, L. 2004, \mnras, 353, 457,
  \dodoi{10.1111/j.1365-2966.2004.08080.x}

\bibitem[{{Ansarinejad} {et~al.}(2020)}]{Ansarinejad_2020}
{Ansarinejad}, B., {et~al.} 2020, in preparation, 000.
\newblock \doarXiv{TBC}

\bibitem[{{Arnaud} {et~al.}(2005){Arnaud}, {Pointecouteau}, \&
  {Pratt}}]{Arnaud_2005}
{Arnaud}, M., {Pointecouteau}, E., \& {Pratt}, G.~W. 2005, A\&A, 441, 893,
  \dodoi{10.1051/0004-6361:20052856}

\bibitem[{{Arnaud} {et~al.}(2010){Arnaud}, {Pratt}, {Piffaretti},
  {B{\"o}hringer}, {Croston}, \& {Pointecouteau}}]{Arnaud_2010}
{Arnaud}, M., {Pratt}, G.~W., {Piffaretti}, R., {et~al.} 2010, A\&A, 517, A92,
  \dodoi{10.1051/0004-6361/200913416}

\bibitem[{{Astropy Collaboration} {et~al.}(2013){Astropy Collaboration},
  {Robitaille}, {Tollerud}, {Greenfield}, {Droettboom}, {Bray}, {Aldcroft},
  {Davis}, {Ginsburg}, {Price-Whelan}, {Kerzendorf}, {Conley}, {Crighton},
  {Barbary}, {Muna}, {Ferguson}, {Grollier}, {Parikh}, {Nair}, {Unther},
  {Deil}, {Woillez}, {Conseil}, {Kramer}, {Turner}, {Singer}, {Fox}, {Weaver},
  {Zabalza}, {Edwards}, {Azalee Bostroem}, {Burke}, {Casey}, {Crawford},
  {Dencheva}, {Ely}, {Jenness}, {Labrie}, {Lim}, {Pierfederici}, {Pontzen},
  {Ptak}, {Refsdal}, {Servillat}, \& {Streicher}}]{Astropy_2013}
{Astropy Collaboration}, {Robitaille}, T.~P., {Tollerud}, E.~J., {et~al.} 2013,
  \aap, 558, A33, \dodoi{10.1051/0004-6361/201322068}

\bibitem[{{Barkhouse} {et~al.}(2006){Barkhouse}, {Green}, {Vikhlinin}, {Kim},
  {Perley}, {Cameron}, {Silverman}, {Mossman}, {Burenin}, {Jannuzi}, {Kim},
  {Smith}, {Smith}, {Tananbaum}, \& {Wilkes}}]{2006ApJ...645..955B}
{Barkhouse}, W.~A., {Green}, P.~J., {Vikhlinin}, A., {et~al.} 2006, \apj, 645,
  955, \dodoi{10.1086/504457}

\bibitem[{{Battaglia} {et~al.}(2016){Battaglia}, {Leauthaud}, {Miyatake},
  {Hasselfield}, {Gralla}, {Allison}, {Bond}, {Calabrese}, {Crichton},
  {Devlin}, {Dunkley}, {D{\"u}nner}, {Erben}, {Ferrara}, {Halpern}, {Hilton},
  {Hill}, {Hincks}, {Hlo{\v z}ek}, {Huffenberger}, {Hughes}, {Kneib},
  {Kosowsky}, {Makler}, {Marriage}, {Menanteau}, {Miller}, {Moodley}, {Moraes},
  {Niemack}, {Page}, {Shan}, {Sehgal}, {Sherwin}, {Sievers}, {Sif{\'o}n},
  {Spergel}, {Staggs}, {Taylor}, {Thornton}, {van Waerbeke}, \&
  {Wollack}}]{Battaglia_2016}
{Battaglia}, N., {Leauthaud}, A., {Miyatake}, H., {et~al.} 2016, \jcap, 8, 013,
  \dodoi{10.1088/1475-7516/2016/08/013}

\bibitem[{{Beers} {et~al.}(1990){Beers}, {Flynn}, \& {Gebhardt}}]{Beers_1990}
{Beers}, T.~C., {Flynn}, K., \& {Gebhardt}, K. 1990, AJ, 100, 32

\bibitem[{{Bhattacharya} {et~al.}(2013){Bhattacharya}, {Habib}, {Heitmann}, \&
  {Vikhlinin}}]{Bhattacharya_2013}
{Bhattacharya}, S., {Habib}, S., {Heitmann}, K., \& {Vikhlinin}, A. 2013, \apj,
  766, 32, \dodoi{10.1088/0004-637X/766/1/32}

\bibitem[{{Birkinshaw}(1999)}]{Birkinshaw_1999}
{Birkinshaw}, M. 1999, Physics Reports, 310, 97,
  \dodoi{10.1016/S0370-1573(98)00080-5}

\bibitem[{{Blake} {et~al.}(2016){Blake}, {Amon}, {Childress}, {Erben},
  {Glazebrook}, {Harnois-Deraps}, {Heymans}, {Hildebrandt}, {Hinton},
  {Janssens}, {Johnson}, {Joudaki}, {Klaes}, {Kuijken}, {Lidman}, {Marin},
  {Parkinson}, {Poole}, \& {Wolf}}]{2016MNRAS.462.4240B}
{Blake}, C., {Amon}, A., {Childress}, M., {et~al.} 2016, \mnras, 462, 4240,
  \dodoi{10.1093/mnras/stw1990}

\bibitem[{{Bleem} {et~al.}(2015{\natexlab{a}}){Bleem}, {Stalder}, {Brodwin},
  {Busha}, {Gladders}, {High}, {Rest}, \& {Wechsler}}]{2015ApJS..216...20B}
{Bleem}, L.~E., {Stalder}, B., {Brodwin}, M., {et~al.} 2015{\natexlab{a}},
  \apjs, 216, 20, \dodoi{10.1088/0067-0049/216/1/20}

\bibitem[{{Bleem} {et~al.}(2015{\natexlab{b}}){Bleem}, {Stalder}, {de Haan},
  {Aird}, {Allen}, {Applegate}, {Ashby}, {Bautz}, {Bayliss}, {Benson},
  {Bocquet}, {Brodwin}, {Carlstrom}, {Chang}, {Chiu}, {Cho}, {Clocchiatti},
  {Crawford}, {Crites}, {Desai}, {Dietrich}, {Dobbs}, {Foley}, {Forman},
  {George}, {Gladders}, {Gonzalez}, {Halverson}, {Hennig}, {Hoekstra},
  {Holder}, {Holzapfel}, {Hrubes}, {Jones}, {Keisler}, {Knox}, {Lee}, {Leitch},
  {Liu}, {Lueker}, {Luong-Van}, {Mantz}, {Marrone}, {McDonald}, {McMahon},
  {Meyer}, {Mocanu}, {Mohr}, {Murray}, {Padin}, {Pryke}, {Reichardt}, {Rest},
  {Ruel}, {Ruhl}, {Saliwanchik}, {Saro}, {Sayre}, {Schaffer}, {Schrabback},
  {Shirokoff}, {Song}, {Spieler}, {Stanford}, {Staniszewski}, {Stark}, {Story},
  {Stubbs}, {Vanderlinde}, {Vieira}, {Vikhlinin}, {Williamson}, {Zahn}, \&
  {Zenteno}}]{Bleem_2015}
{Bleem}, L.~E., {Stalder}, B., {de Haan}, T., {et~al.} 2015{\natexlab{b}},
  \apjs, 216, 27, \dodoi{10.1088/0067-0049/216/2/27}

\bibitem[{{Bleem} {et~al.}(2020){Bleem}, {Bocquet}, {Stalder}, {Gladders},
  {Ade}, {Allen}, {Anderson}, {Annis}, {Ashby}, {Austermann}, {Avila}, {Avva},
  {Bayliss}, {Beall}, {Bechtol}, {Bender}, {Benson}, {Bertin}, {Bianchini},
  {Blake}, {Brodwin}, {Brooks}, {Buckley-Geer}, {Burke}, {Carlstrom}, {Rosell},
  {Carrasco Kind}, {Carretero}, {Chang}, {Chiang}, {Citron}, {Moran},
  {Costanzi}, {Crawford}, {Crites}, {da Costa}, {de Haan}, {De Vicente},
  {Desai}, {Diehl}, {Dietrich}, {Dobbs}, {Eifler}, {Everett}, {Flaugher},
  {Floyd}, {Frieman}, {Gallicchio}, {Garc{\'\i}a-Bellido}, {George}, {Gerdes},
  {Gilbert}, {Gruen}, {Gruendl}, {Gschwend}, {Gupta}, {Gutierrez}, {Halverson},
  {Harrington}, {Henning}, {Heymans}, {Holder}, {Hollowood}, {Holzapfel},
  {Honscheid}, {Hrubes}, {Huang}, {Hubmayr}, {Irwin}, {James}, {Jeltema},
  {Joudaki}, {Khullar}, {Klein}, {Knox}, {Kuropatkin}, {Lee}, {Li}, {Lidman},
  {Lowitz}, {MacCrann}, {Mahler}, {Maia}, {Marshall}, {McDonald}, {McMahon},
  {Melchior}, {Menanteau}, {Meyer}, {Miquel}, {Mocanu}, {Mohr}, {Montgomery},
  {Nadolski}, {Natoli}, {Nibarger}, {Noble}, {Novosad}, {Padin}, {Palmese},
  {Parkinson}, {Patil}, {Paz-Chinch{\'o}n}, {Plazas}, {Pryke}, {Ramachandra},
  {Reichardt}, {Remolina Gonz{\'a}lez}, {Romer}, {Roodman}, {Ruhl}, {Rykoff},
  {Saliwanchik}, {Sanchez}, {Saro}, {Sayre}, {Schaffer}, {Schrabback},
  {Serrano}, {Sharon}, {Sievers}, {Smecher}, {Smith}, {Soares-Santos}, {Stark},
  {Story}, {Suchyta}, {Tarle}, {Tucker}, {Vanderlinde}, {Veach}, {Vieira},
  {Wang}, {Weller}, {Whitehorn}, {Wu}, {Yefremenko}, \& {Zhang}}]{Bleem_2020}
{Bleem}, L.~E., {Bocquet}, S., {Stalder}, B., {et~al.} 2020, \apjs, 247, 25,
  \dodoi{10.3847/1538-4365/ab6993}

\bibitem[{{Bocquet} {et~al.}(2019){Bocquet}, {Dietrich}, {Schrabback}, {Bleem},
  {Klein}, {Allen}, {Applegate}, {Ashby}, {Bautz}, {Bayliss}, {Benson},
  {Brodwin}, {Bulbul}, {Canning}, {Capasso}, {Carlstrom}, {Chang}, {Chiu},
  {Cho}, {Clocchiatti}, {Crawford}, {Crites}, {de Haan}, {Desai}, {Dobbs},
  {Foley}, {Forman}, {Garmire}, {George}, {Gladders}, {Gonzalez}, {Grandis},
  {Gupta}, {Halverson}, {Hlavacek-Larrondo}, {Hoekstra}, {Holder}, {Holzapfel},
  {Hou}, {Hrubes}, {Huang}, {Jones}, {Khullar}, {Knox}, {Kraft}, {Lee}, {von
  der Linden}, {Luong-Van}, {Mantz}, {Marrone}, {McDonald}, {McMahon}, {Meyer},
  {Mocanu}, {Mohr}, {Morris}, {Padin}, {Patil}, {Pryke}, {Rapetti},
  {Reichardt}, {Rest}, {Ruhl}, {Saliwanchik}, {Saro}, {Sayre}, {Schaffer},
  {Shirokoff}, {Stalder}, {Stanford}, {Staniszewski}, {Stark}, {Story},
  {Strazzullo}, {Stubbs}, {Vanderlinde}, {Vieira}, {Vikhlinin}, {Williamson},
  \& {Zenteno}}]{Bocquet_2019}
{Bocquet}, S., {Dietrich}, J.~P., {Schrabback}, T., {et~al.} 2019, \apj, 878,
  55, \dodoi{10.3847/1538-4357/ab1f10}

\bibitem[{{B{\"o}hringer} {et~al.}(2000){B{\"o}hringer}, {Voges}, {Huchra},
  {McLean}, {Giacconi}, {Rosati}, {Burg}, {Mader}, {Schuecker}, {Simi{\c{c}}},
  {Komossa}, {Reiprich}, {Retzlaff}, \& {Tr{\"u}mper}}]{2000ApJS..129..435B}
{B{\"o}hringer}, H., {Voges}, W., {Huchra}, J.~P., {et~al.} 2000, \apjs, 129,
  435, \dodoi{10.1086/313427}

\bibitem[{{B{\"o}hringer} {et~al.}(2004){B{\"o}hringer}, {Schuecker}, {Guzzo},
  {Collins}, {Voges}, {Cruddace}, {Ortiz-Gil}, {Chincarini}, {De Grandi},
  {Edge}, {MacGillivray}, {Neumann}, {Schindler}, \&
  {Shaver}}]{2004AandA...425..367B}
{B{\"o}hringer}, H., {Schuecker}, P., {Guzzo}, L., {et~al.} 2004, \aap, 425,
  367, \dodoi{10.1051/0004-6361:20034484}

\bibitem[{{Bradley} {et~al.}(2014){Bradley}, {Zitrin}, {Coe}, {Bouwens},
  {Postman}, {Balestra}, {Grillo}, {Monna}, {Rosati}, {Seitz}, {Host}, {Lemze},
  {Moustakas}, {Moustakas}, {Shu}, {Zheng}, {Broadhurst}, {Carrasco}, {Jouvel},
  {Koekemoer}, {Medezinski}, {Meneghetti}, {Nonino}, {Smit}, {Umetsu},
  {Bartelmann}, {Ben{\'\i}tez}, {Donahue}, {Ford}, {Infante}, {Jimenez-Teja},
  {Kelson}, {Lahav}, {Maoz}, {Melchior}, {Merten}, \&
  {Molino}}]{2014ApJ...792...76B}
{Bradley}, L.~D., {Zitrin}, A., {Coe}, D., {et~al.} 2014, \apj, 792, 76,
  \dodoi{10.1088/0004-637X/792/1/76}

\bibitem[{{Brammer} {et~al.}(2008){Brammer}, {van Dokkum}, \&
  {Coppi}}]{Brammer_2008}
{Brammer}, G.~B., {van Dokkum}, P.~G., \& {Coppi}, P. 2008, \apj, 686, 1503,
  \dodoi{10.1086/591786}

\bibitem[{{Bruzual} \& {Charlot}(2003)}]{BruzualCharlot_2003}
{Bruzual}, G., \& {Charlot}, S. 2003, MNRAS, 344, 1000

\bibitem[{{Buddendiek} {et~al.}(2015){Buddendiek}, {Schrabback}, {Greer},
  {Hoekstra}, {Sommer}, {Eifler}, {Erben}, {Erler}, {Hicks}, {High},
  {Hildebrandt}, {Marrone}, {Morris}, {Muzzin}, {Reiprich}, {Schirmer},
  {Schneider}, \& {von der Linden}}]{2015MNRAS.450.4248B}
{Buddendiek}, A., {Schrabback}, T., {Greer}, C.~H., {et~al.} 2015, \mnras, 450,
  4248, \dodoi{10.1093/mnras/stv783}

\bibitem[{{Burenin} {et~al.}(2007){Burenin}, {Vikhlinin}, {Hornstrup},
  {Ebeling}, {Quintana}, \& {Mescheryakov}}]{2007ApJS..172..561B}
{Burenin}, R.~A., {Vikhlinin}, A., {Hornstrup}, A., {et~al.} 2007, \apjs, 172,
  561, \dodoi{10.1086/519457}

\bibitem[{{Caccianiga} {et~al.}(2000){Caccianiga}, {Maccacaro}, {Wolter},
  {Della Ceca}, \& {Gioia}}]{2000AandAS..144..247C}
{Caccianiga}, A., {Maccacaro}, T., {Wolter}, A., {Della Ceca}, R., \& {Gioia},
  I.~M. 2000, \aaps, 144, 247, \dodoi{10.1051/aas:2000344}

\bibitem[{{Calabretta} \& {Greisen}(2002)}]{Calabretta_2002}
{Calabretta}, M.~R., \& {Greisen}, E.~W. 2002, \aap, 395, 1077,
  \dodoi{10.1051/0004-6361:20021327}

\bibitem[{{Cappi} {et~al.}(1998){Cappi}, {Held}, \&
  {Marano}}]{1998AandAS..129...31C}
{Cappi}, A., {Held}, E.~V., \& {Marano}, B. 1998, \aaps, 129, 31,
  \dodoi{10.1051/aas:1998172}

\bibitem[{{Carlstrom} {et~al.}(2002){Carlstrom}, {Holder}, \&
  {Reese}}]{Carlstrom_2002}
{Carlstrom}, J.~E., {Holder}, G.~P., \& {Reese}, E.~D. 2002, ARA\&A, 40, 643,
  \dodoi{10.1146/annurev.astro.40.060401.093803}

\bibitem[{{Cavagnolo} {et~al.}(2008){Cavagnolo}, {Donahue}, {Voit}, \&
  {Sun}}]{2008ApJ...682..821C}
{Cavagnolo}, K.~W., {Donahue}, M., {Voit}, G.~M., \& {Sun}, M. 2008, \apj, 682,
  821, \dodoi{10.1086/588630}

\bibitem[{{Childress} {et~al.}(2017){Childress}, {Lidman}, {Davis}, {Tucker},
  {Asorey}, {Yuan}, {Abbott}, {Abdalla}, {Allam}, {Annis}, {Banerji},
  {Benoit-L{\'e}vy}, {Bernard}, {Bertin}, {Brooks}, {Buckley-Geer}, {Burke},
  {Carnero Rosell}, {Carollo}, {Carrasco Kind}, {Carretero}, {Castander},
  {Cunha}, {da Costa}, {D'Andrea}, {Doel}, {Eifler}, {Evrard}, {Flaugher},
  {Foley}, {Fosalba}, {Frieman}, {Garc{\'\i}a-Bellido}, {Glazebrook},
  {Goldstein}, {Gruen}, {Gruendl}, {Gschwend}, {Gupta}, {Gutierrez}, {Hinton},
  {Hoormann}, {James}, {Kessler}, {Kim}, {King}, {Kovacs}, {Kuehn}, {Kuhlmann},
  {Kuropatkin}, {Lagattuta}, {Lewis}, {Li}, {Lima}, {Lin}, {Macaulay}, {Maia},
  {Marriner}, {March}, {Marshall}, {Martini}, {McMahon}, {Menanteau}, {Miquel},
  {Moller}, {Morganson}, {Mould}, {Mudd}, {Muthukrishna}, {Nichol}, {Nord},
  {Ogando}, {Ostrovski}, {Parkinson}, {Plazas}, {Reed}, {Reil}, {Romer},
  {Rykoff}, {Sako}, {Sanchez}, {Scarpine}, {Schindler}, {Schubnell}, {Scolnic},
  {Sevilla-Noarbe}, {Seymour}, {Sharp}, {Smith}, {Soares-Santos}, {Sobreira},
  {Sommer}, {Spinka}, {Suchyta}, {Sullivan}, {Swanson}, {Tarle}, {Uddin},
  {Walker}, {Wester}, \& {Zhang}}]{2017MNRAS.472..273C}
{Childress}, M.~J., {Lidman}, C., {Davis}, T.~M., {et~al.} 2017, \mnras, 472,
  273, \dodoi{10.1093/mnras/stx1872}

\bibitem[{{Chisari} {et~al.}(2019){Chisari}, {Alonso}, {Krause}, {Leonard},
  {Bull}, {Neveu}, {Villarreal}, {Singh}, {McClintock}, {Ellison}, {Du},
  {Zuntz}, {Mead}, {Joudaki}, {Lorenz}, {Tr{\"o}ster}, {Sanchez}, {Lanusse},
  {Ishak}, {Hlozek}, {Blazek}, {Campagne}, {Almoubayyed}, {Eifler}, {Kirby},
  {Kirkby}, {Plaszczynski}, {Slosar}, {Vrastil}, {Wagoner}, \& {LSST Dark
  Energy Science Collaboration}}]{Chisari_2019}
{Chisari}, N.~E., {Alonso}, D., {Krause}, E., {et~al.} 2019, \apjs, 242, 2,
  \dodoi{10.3847/1538-4365/ab1658}

\bibitem[{{Choi} {et~al.}(2018){Choi}, {Austermann}, {Beall}, {Crowley},
  {Datta}, {Duff}, {Gallardo}, {Ho}, {Hubmayr}, {Koopman}, {Li}, {Nati},
  {Niemack}, {Page}, {Salatino}, {Simon}, {Staggs}, {Stevens}, {Ullom}, \&
  {Wollack}}]{Choi_2018}
{Choi}, S.~K., {Austermann}, J., {Beall}, J.~A., {et~al.} 2018, Journal of Low
  Temperature Physics, 193, 267, \dodoi{10.1007/s10909-018-1982-4}

\bibitem[{{Choi} {et~al.}(2020){Choi}, {Hasselfield}, {Ho}, {Koopman}, {Lungu},
  {Abitbol}, {Addison}, {Ade}, {Aiola}, {Alonso}, {Amiri}, {Amodeo}, {Angile},
  {Austermann}, {Baildon}, {Battaglia}, {Beall}, {Bean}, {Becker}, {Bond},
  {Bruno}, {Calabrese}, {Calafut}, {Campusano}, {Carrero}, {Chesmore}, {Cho.},
  {Clark}, {Cothard}, {Crichton}, {Crowley}, {Darwish}, {Datta}, {Denison},
  {Devlin}, {Duell}, {Duff}, {Duivenvoorden}, {Dunkley}, {D{\"u}nner},
  {Essinger-Hileman}, {Fankhanel}, {Ferraro}, {Fox}, {Fuzia}, {Gallardo},
  {Gluscevic}, {Golec}, {Grace}, {Gralla}, {Guan}, {Hall}, {Halpern}, {Han},
  {Hargrave}, {Henderson}, {Hensley}, {Hill}, {Hilton}, {Hilton}, {Hincks},
  {Hlo{\v{z}}ek}, {Hubmayr}, {Huffenberger}, {Hughes}, {Infante}, {Irwin},
  {Jackson}, {Klein}, {Knowles}, {Kosowsky}, {Lakey}, {Li}, {Li}, {Li},
  {Lokken}, {Louis}, {MacInnis}, {Madhavacheril}, {Maldonado}, {Mallaby-Kay},
  {Marsden}, {Maurin}, {McMahon}, {Menanteau}, {Moodley}, {Morton}, {Naess},
  {Namikawa}, {Nati}, {Newburgh}, {Nibarger}, {Nicola}, {Niemack}, {Nolta},
  {Orlowski-Sherer}, {Page}, {Pappas}, {Partridge}, {Phakathi}, {Prince},
  {Puddu}, {Qu}, {Rivera}, {Robertson}, {Rojas}, {Salatino}, {Schaan},
  {Schillaci}, {Schmitt}, {Sehgal}, {Sherwin}, {Sierra}, {Sievers}, {Sifon},
  {Sikhosana}, {Simon}, {Spergel}, {Staggs}, {Stevens}, {Storer}, {Sunder},
  {Switzer}, {Thorne}, {Thornton}, {Trac}, {Treu}, {Tucker}, {Vale}, {Van
  Engelen}, {Van Lanen}, {Vavagiakis}, {Wagoner}, {Wang}, {Ward}, {Wollack},
  {Xu}, {Zago}, \& {Zhu}}]{Choi_2020}
{Choi}, S.~K., {Hasselfield}, M., {Ho}, S.-P.~P., {et~al.} 2020, arXiv
  e-prints, arXiv:2007.07289.
\newblock \doarXiv{2007.07289}

\bibitem[{{Chon} \& {B{\"o}hringer}(2012)}]{2012AandA...538A..35C}
{Chon}, G., \& {B{\"o}hringer}, H. 2012, \aap, 538, A35,
  \dodoi{10.1051/0004-6361/201117996}

\bibitem[{{Coleman} {et~al.}(1980){Coleman}, {Wu}, \&
  {Weedman}}]{ColemanWuWeedman_1980}
{Coleman}, G.~D., {Wu}, C.-C., \& {Weedman}, D.~W. 1980, \apjs, 43, 393,
  \dodoi{10.1086/190674}

\bibitem[{{Connor} {et~al.}(2019){Connor}, {Kelson}, {Blanc}, \&
  {Boutsia}}]{2019ApJ...878...66C}
{Connor}, T., {Kelson}, D.~D., {Blanc}, G.~A., \& {Boutsia}, K. 2019, \apj,
  878, 66, \dodoi{10.3847/1538-4357/ab1f7a}

\bibitem[{{Coziol} {et~al.}(2009){Coziol}, {Andernach}, {Caretta},
  {Alamo-Mart{\'\i}nez}, \& {Tago}}]{2009AJ....137.4795C}
{Coziol}, R., {Andernach}, H., {Caretta}, C.~A., {Alamo-Mart{\'\i}nez}, K.~A.,
  \& {Tago}, E. 2009, \aj, 137, 4795, \dodoi{10.1088/0004-6256/137/6/4795}

\bibitem[{{Crawford} {et~al.}(1995){Crawford}, {Edge}, {Fabian}, {Allen},
  {Bohringer}, {Ebeling}, {McMahon}, \& {Voges}}]{1995MNRAS.274...75C}
{Crawford}, C.~S., {Edge}, A.~C., {Fabian}, A.~C., {et~al.} 1995, \mnras, 274,
  75, \dodoi{10.1093/mnras/274.1.75}

\bibitem[{{Crawford} {et~al.}(2014){Crawford}, {Wirth}, \&
  {Bershady}}]{2014ApJ...786...30C}
{Crawford}, S.~M., {Wirth}, G.~D., \& {Bershady}, M.~A. 2014, \apj, 786, 30,
  \dodoi{10.1088/0004-637X/786/1/30}

\bibitem[{{Cruddace} {et~al.}(2002){Cruddace}, {Voges}, {B{\"o}hringer},
  {Collins}, {Romer}, {MacGillivray}, {Yentis}, {Schuecker}, {Ebeling}, \& {De
  Grandi}}]{2002ApJS..140..239C}
{Cruddace}, R., {Voges}, W., {B{\"o}hringer}, H., {et~al.} 2002, \apjs, 140,
  239, \dodoi{10.1086/324519}

\bibitem[{{Dalton} {et~al.}(1994){Dalton}, {Efstathiou}, {Maddox}, \&
  {Sutherland }}]{1994MNRAS.269..151D}
{Dalton}, G.~B., {Efstathiou}, G., {Maddox}, S.~J., \& {Sutherland }, W.~J.
  1994, \mnras, 269, 151, \dodoi{10.1093/mnras/269.1.151}

\bibitem[{{Dawson} {et~al.}(2009){Dawson}, {Aldering}, {Amanullah}, {Barbary},
  {Barrientos}, {Brodwin}, {Connolly}, {Dey}, {Doi}, {Donahue}, {Eisenhardt},
  {Ellingson}, {Faccioli}, {Fadeyev}, {Fakhouri}, {Fruchter}, {Gilbank},
  {Gladders}, {Goldhaber}, {Gonzalez}, {Goobar}, {Gude}, {Hattori}, {Hoekstra},
  {Huang}, {Ihara}, {Jannuzi}, {Johnston}, {Kashikawa}, {Koester}, {Konishi},
  {Kowalski}, {Lidman}, {Linder}, {Lubin}, {Meyers}, {Morokuma}, {Munshi},
  {Mullis}, {Oda}, {Panagia}, {Perlmutter}, {Postman}, {Pritchard}, {Rhodes},
  {Rosati}, {Rubin}, {Schlegel}, {Spadafora}, {Stanford}, {Stanishev}, {Stern},
  {Strovink}, {Suzuki}, {Takanashi}, {Tokita}, {Wagner}, {Wang}, {Yasuda},
  {Yee}, \& {Supernova Cosmology Project}}]{2009AJ....138.1271D}
{Dawson}, K.~S., {Aldering}, G., {Amanullah}, R., {et~al.} 2009, \aj, 138,
  1271, \dodoi{10.1088/0004-6256/138/5/1271}

\bibitem[{{De Grandi} {et~al.}(1999){De Grandi}, {B{\"o}hringer}, {Guzzo},
  {Molendi}, {Chincarini}, {Collins}, {Cruddace}, {Neumann}, {Schindler},
  {Schuecker}, \& {Voges}}]{1999ApJ...514..148D}
{De Grandi}, S., {B{\"o}hringer}, H., {Guzzo}, L., {et~al.} 1999, \apj, 514,
  148, \dodoi{10.1086/306939}

\bibitem[{{De Propris} {et~al.}(2002){De Propris}, {Couch}, {Colless},
  {Dalton}, {Collins}, {Baugh}, {Bland -Hawthorn}, {Bridges}, {Cannon}, {Cole},
  {Cross}, {Deeley}, {Driver}, {Efstathiou}, {Ellis}, {Frenk}, {Glazebrook},
  {Jackson}, {Lahav}, {Lewis}, {Lumsden}, {Maddox}, {Madgwick}, {Moody},
  {Norberg}, {Peacock}, {Percival}, {Peterson}, {Sutherland}, \&
  {Taylor}}]{2002MNRAS.329...87D}
{De Propris}, R., {Couch}, W.~J., {Colless}, M., {et~al.} 2002, \mnras, 329,
  87, \dodoi{10.1046/j.1365-8711.2002.04958.x}

\bibitem[{{Dey} {et~al.}(2019){Dey}, {Schlegel}, {Lang}, {Blum}, {Burleigh},
  {Fan}, {Findlay}, {Finkbeiner}, {Herrera}, {Juneau}, {Landriau}, {Levi},
  {McGreer}, {Meisner}, {Myers}, {Moustakas}, {Nugent}, {Patej}, {Schlafly},
  {Walker}, {Valdes}, {Weaver}, {Y{\`e}che}, {Zou}, {Zhou}, {Abareshi},
  {Abbott}, {Abolfathi}, {Aguilera}, {Alam}, {Allen}, {Alvarez}, {Annis},
  {Ansarinejad}, {Aubert}, {Beechert}, {Bell}, {BenZvi}, {Beutler}, {Bielby},
  {Bolton}, {Brice{\~n}o}, {Buckley-Geer}, {Butler}, {Calamida}, {Carlberg},
  {Carter}, {Casas}, {Castander}, {Choi}, {Comparat}, {Cukanovaite}, {Delubac},
  {DeVries}, {Dey}, {Dhungana}, {Dickinson}, {Ding}, {Donaldson}, {Duan},
  {Duckworth}, {Eftekharzadeh}, {Eisenstein}, {Etourneau}, {Fagrelius},
  {Farihi}, {Fitzpatrick}, {Font-Ribera}, {Fulmer}, {G{\"a}nsicke},
  {Gaztanaga}, {George}, {Gerdes}, {Gontcho}, {Gorgoni}, {Green}, {Guy},
  {Harmer}, {Hernand ez}, {Honscheid}, {Huang}, {James}, {Jannuzi}, {Jiang},
  {Joyce}, {Karcher}, {Karkar}, {Kehoe}, {Kneib}, {Kueter-Young}, {Lan},
  {Lauer}, {Le Guillou}, {Le Van Suu}, {Lee}, {Lesser}, {Perreault Levasseur},
  {Li}, {Mann}, {Marshall}, {Mart{\'\i}nez-V{\'a}zquez}, {Martini}, {du Mas des
  Bourboux}, {McManus}, {Meier}, {M{\'e}nard}, {Metcalfe},
  {Mu{\~n}oz-Guti{\'e}rrez}, {Najita}, {Napier}, {Narayan}, {Newman}, {Nie},
  {Nord}, {Norman}, {Olsen}, {Paat}, {Palanque-Delabrouille}, {Peng},
  {Poppett}, {Poremba}, {Prakash}, {Rabinowitz}, {Raichoor}, {Rezaie},
  {Robertson}, {Roe}, {Ross}, {Ross}, {Rudnick}, {Safonova}, {Saha},
  {S{\'a}nchez}, {Savary}, {Schweiker}, {Scott}, {Seo}, {Shan}, {Silva},
  {Slepian}, {Soto}, {Sprayberry}, {Staten}, {Stillman}, {Stupak}, {Summers},
  {Sien Tie}, {Tirado}, {Vargas-Maga{\~n}a}, {Vivas}, {Wechsler}, {Williams},
  {Yang}, {Yang}, {Yapici}, {Zaritsky}, {Zenteno}, {Zhang}, {Zhang}, {Zhou}, \&
  {Zhou}}]{Dey_2019}
{Dey}, A., {Schlegel}, D.~J., {Lang}, D., {et~al.} 2019, \aj, 157, 168,
  \dodoi{10.3847/1538-3881/ab089d}

\bibitem[{{Diehl} {et~al.}(2017){Diehl}, {Buckley-Geer}, {Lindgren}, {Nord},
  {Gaitsch}, {Gaitsch}, {Lin}, {Allam}, {Collett}, {Furlanetto}, {Gill},
  {More}, {Nightingale}, {Odden}, {Pellico}, {Tucker}, {da Costa}, {Fausti
  Neto}, {Kuropatkin}, {Soares-Santos}, {Welch}, {Zhang}, {Frieman}, {Abdalla},
  {Annis}, {Benoit-L{\'e}vy}, {Bertin}, {Brooks}, {Burke}, {Carnero Rosell},
  {Carrasco Kind}, {Carretero}, {Cunha}, {D'Andrea}, {Desai}, {Dietrich},
  {Drlica-Wagner}, {Evrard}, {Finley}, {Flaugher}, {Garc{\'\i}a-Bellido},
  {Gerdes}, {Goldstein}, {Gruen}, {Gruendl}, {Gschwend}, {Gutierrez}, {James},
  {Kuehn}, {Kuhlmann}, {Lahav}, {Li}, {Lima}, {Maia}, {Marshall}, {Menanteau},
  {Miquel}, {Nichol}, {Nugent}, {Ogand o}, {Plazas}, {Reil}, {Romer}, {Sako},
  {Sanchez}, {Santiago}, {Scarpine}, {Schindler}, {Schubnell},
  {Sevilla-Noarbe}, {Sheldon}, {Smith}, {Sobreira}, {Suchyta}, {Swanson},
  {Tarle}, {Thomas}, {Walker}, \& {DES Collaboration}}]{Diehl_2017}
{Diehl}, H.~T., {Buckley-Geer}, E.~J., {Lindgren}, K.~A., {et~al.} 2017, \apjs,
  232, 15, \dodoi{10.3847/1538-4365/aa8667}

\bibitem[{{Diehl} {et~al.}(2020)}]{Diehl_2020}
{Diehl}, H.~T., {et~al.} 2020, in preparation, 000.
\newblock \doarXiv{TBC}

\bibitem[{{Donahue} {et~al.}(2020){Donahue}, {Funkhouser}, {Koeppe}, {Frisbie},
  \& {Voit}}]{Donahue_2020}
{Donahue}, M., {Funkhouser}, K., {Koeppe}, D., {Frisbie}, R. L.~S., \& {Voit},
  G.~M. 2020, \apj, 889, 121, \dodoi{10.3847/1538-4357/ab64da}

\bibitem[{{D{\"u}nner} {et~al.}(2013)}]{Dunner_2013}
{D{\"u}nner}, R., {et~al.} 2013, ApJ, 762, 10,
  \dodoi{10.1088/0004-637X/762/1/10}

\bibitem[{{Durret} {et~al.}(2011){Durret}, {Adami}, {Cappi}, {Maurogordato},
  {M{\'a}rquez}, {Ilbert}, {Coupon}, {Arnouts}, {Benoist}, {Blaizot}, {Edorh},
  {Garilli}, {Guennou}, {Le Brun}, {Le F{\`e}vre}, {Mazure}, {McCracken},
  {Mellier}, {Mezrag}, {Slezak}, {Tresse}, \& {Ulmer}}]{Durret_2011}
{Durret}, F., {Adami}, C., {Cappi}, A., {et~al.} 2011, \aap, 535, A65,
  \dodoi{10.1051/0004-6361/201116985}

\bibitem[{{Ebeling} {et~al.}(2007){Ebeling}, {Barrett}, {Donovan}, {Ma},
  {Edge}, \& {van Speybroeck}}]{2007ApJ...661L..33E}
{Ebeling}, H., {Barrett}, E., {Donovan}, D., {et~al.} 2007, \apjl, 661, L33,
  \dodoi{10.1086/518603}

\bibitem[{{Edge} {et~al.}(2013){Edge}, {Sutherland}, {Kuijken}, {Driver},
  {McMahon}, {Eales}, \& {Emerson}}]{Edge_2013}
{Edge}, A., {Sutherland}, W., {Kuijken}, K., {et~al.} 2013, The Messenger, 154,
  32

\bibitem[{{Ehlert} {et~al.}(2015){Ehlert}, {Allen}, {Brandt}, {Canning}, {Luo},
  {Mantz}, {Morris}, {von der Linden}, \& {Xue}}]{2015MNRAS.446.2709E}
{Ehlert}, S., {Allen}, S.~W., {Brandt}, W.~N., {et~al.} 2015, \mnras, 446,
  2709, \dodoi{10.1093/mnras/stu2091}

\bibitem[{{Fassbender} {et~al.}(2011){Fassbender}, {B{\"o}hringer}, {Santos},
  {Pratt}, {{\v{S}}uhada}, {Kohnert}, {Lerchster}, {Rovilos}, {Pierini},
  {Chon}, {Schwope}, {Lamer}, {M{\"u}hlegger}, {Rosati}, {Quintana}, {Nastasi},
  {de Hoon}, {Seitz}, \& {Mohr}}]{2011AandA...527A..78F}
{Fassbender}, R., {B{\"o}hringer}, H., {Santos}, J.~S., {et~al.} 2011, \aap,
  527, A78, \dodoi{10.1051/0004-6361/201015204}

\bibitem[{{Flewelling} {et~al.}(2016){Flewelling}, {Magnier}, {Chambers},
  {Heasley}, {Holmberg}, {Huber}, {Sweeney}, {Waters}, {Chen}, {Farrow},
  {Hasinger}, {Henderson}, {Long}, {Metcalfe}, {Nieto-Santisteban}, {Norberg},
  {Saglia}, {Szalay}, {Rest}, {Thakar}, {Tonry}, {Valenti}, {Werner}, {White},
  {Denneau}, {Draper}, {Hodapp}, {Jedicke}, {Kaiser}, {Kudritzki}, {Price},
  {Wainscoat}, {Chastel}, {McClean}, {Postman}, \& {Shiao}}]{Flewelling_2016}
{Flewelling}, H.~A., {Magnier}, E.~A., {Chambers}, K.~C., {et~al.} 2016, ArXiv
  e-prints.
\newblock \doarXiv{1612.05243}

\bibitem[{{Fowler} {et~al.}(2007){Fowler}, {Niemack}, {Dicker}, {Aboobaker},
  {Ade}, {Battistelli}, {Devlin}, {Fisher}, {Halpern}, {Hargrave}, {Hincks},
  {Kaul}, {Klein}, {Lau}, {Limon}, {Marriage}, {Mauskopf}, {Page}, {Staggs},
  {Swetz}, {Switzer}, {Thornton}, \& {Tucker}}]{Fowler_2007}
{Fowler}, J.~W., {Niemack}, M.~D., {Dicker}, S.~R., {et~al.} 2007, \ao, 46,
  3444, \dodoi{10.1364/AO.46.003444}

\bibitem[{{Gal} {et~al.}(2009){Gal}, {Lopes}, {de Carvalho}, {Kohl-Moreira},
  {Capelato}, \& {Djorgovski}}]{2009AJ....137.2981G}
{Gal}, R.~R., {Lopes}, P.~A.~A., {de Carvalho}, R.~R., {et~al.} 2009, \aj, 137,
  2981, \dodoi{10.1088/0004-6256/137/2/2981}

\bibitem[{{Geach} {et~al.}(2011){Geach}, {Murphy}, \&
  {Bower}}]{2011MNRAS.413.3059G}
{Geach}, J.~E., {Murphy}, D. N.~A., \& {Bower}, R.~G. 2011, \mnras, 413, 3059,
  \dodoi{10.1111/j.1365-2966.2011.18380.x}

\bibitem[{{Gilbank} {et~al.}(2008){Gilbank}, {Yee}, {Ellingson}, {Hicks},
  {Gladders}, {Barrientos}, \& {Keeney}}]{2008ApJ...677L..89G}
{Gilbank}, D.~G., {Yee}, H.~K.~C., {Ellingson}, E., {et~al.} 2008, \apjl, 677,
  L89, \dodoi{10.1086/588138}

\bibitem[{{Gioia} \& {Luppino}(1994)}]{1994ApJS...94..583G}
{Gioia}, I.~M., \& {Luppino}, G.~A. 1994, \apjs, 94, 583,
  \dodoi{10.1086/192083}

\bibitem[{{Gladders} {et~al.}(2003){Gladders}, {Hoekstra}, {Yee}, {Hall}, \&
  {Barrientos}}]{2003ApJ...593...48G}
{Gladders}, M.~D., {Hoekstra}, H., {Yee}, H.~K.~C., {Hall}, P.~B., \&
  {Barrientos}, L.~F. 2003, \apj, 593, 48, \dodoi{10.1086/376518}

\bibitem[{{Gonzalez} {et~al.}(2015){Gonzalez}, {Decker}, {Brodwin},
  {Eisenhardt}, {Marrone}, {Stanford}, {Stern}, {Wylezalek}, {Aldering},
  {Abdulla}, {Boone}, {Carlstrom}, {Fagrelius}, {Gettings}, {Greer}, {Hayden},
  {Leitch}, {Lin}, {Mantz}, {Muchovej}, {Perlmutter}, \&
  {Zeimann}}]{2015ApJ...812L..40G}
{Gonzalez}, A.~H., {Decker}, B., {Brodwin}, M., {et~al.} 2015, \apjl, 812, L40,
  \dodoi{10.1088/2041-8205/812/2/L40}

\bibitem[{{Gonzalez} {et~al.}(2019){Gonzalez}, {Gettings}, {Brodwin},
  {Eisenhardt}, {Stanford}, {Wylezalek}, {Decker}, {Marrone}, {Moravec},
  {O'Donnell}, {Stalder}, {Stern}, {Abdulla}, {Brown}, {Carlstrom}, {Chambers},
  {Hayden}, {Lin}, {Magnier}, {Masci}, {Mantz}, {McDonald}, {Mo}, {Perlmutter},
  {Wright}, \& {Zeimann}}]{Gonzalez_2019}
{Gonzalez}, A.~H., {Gettings}, D.~P., {Brodwin}, M., {et~al.} 2019, \apjs, 240,
  33, \dodoi{10.3847/1538-4365/aafad2}

\bibitem[{{Goto} {et~al.}(2002){Goto}, {Sekiguchi}, {Nichol}, {Bahcall}, {Kim},
  {Annis}, {Ivezi{\'c}}, {Brinkmann}, {Hennessy}, {Szokoly}, \&
  {Tucker}}]{Goto_2002}
{Goto}, T., {Sekiguchi}, M., {Nichol}, R.~C., {et~al.} 2002, \aj, 123, 1807,
  \dodoi{10.1086/339303}

\bibitem[{{Gralla} {et~al.}(2011){Gralla}, {Sharon}, {Gladders}, {Marrone},
  {Barrientos}, {Bayliss}, {Bonamente}, {Bulbul}, {Carlstrom}, {Culverhouse},
  {Gilbank}, {Greer}, {Hasler}, {Hawkins}, {Hennessy}, {Joy}, {Koester},
  {Lamb}, {Leitch}, {Miller}, {Mroczkowski}, {Muchovej}, {Oguri}, {Plagge},
  {Pryke}, \& {Woody}}]{2011ApJ...737...74G}
{Gralla}, M.~B., {Sharon}, K., {Gladders}, M.~D., {et~al.} 2011, \apj, 737, 74,
  \dodoi{10.1088/0004-637X/737/2/74}

\bibitem[{{Hasselfield} {et~al.}(2013)}]{Hasselfield_2013}
{Hasselfield}, M., {et~al.} 2013, JCAP, 7, 8,
  \dodoi{10.1088/1475-7516/2013/07/008}

\bibitem[{{Henderson} {et~al.}(2016){Henderson}, {Allison}, {Austermann},
  {Baildon}, {Battaglia}, {Beall}, {Becker}, {De Bernardis}, {Bond},
  {Calabrese}, {Choi}, {Coughlin}, {Crowley}, {Datta}, {Devlin}, {Duff},
  {Dunkley}, {D{\"u}nner}, {van Engelen}, {Gallardo}, {Grace}, {Hasselfield},
  {Hills}, {Hilton}, {Hincks}, {Hloẑek}, {Ho}, {Hubmayr}, {Huffenberger},
  {Hughes}, {Irwin}, {Koopman}, {Kosowsky}, {Li}, {McMahon}, {Munson}, {Nati},
  {Newburgh}, {Niemack}, {Niraula}, {Page}, {Pappas}, {Salatino}, {Schillaci},
  {Schmitt}, {Sehgal}, {Sherwin}, {Sievers}, {Simon}, {Spergel}, {Staggs},
  {Stevens}, {Thornton}, {Van Lanen}, {Vavagiakis}, {Ward}, \&
  {Wollack}}]{Henderson_2016}
{Henderson}, S.~W., {Allison}, R., {Austermann}, J., {et~al.} 2016, Journal of
  Low Temperature Physics, 184, 772, \dodoi{10.1007/s10909-016-1575-z}

\bibitem[{{Hilton} {et~al.}(2018){Hilton}, {Hasselfield}, {Sif{\'o}n},
  {Battaglia}, {Aiola}, {Bharadwaj}, {Bond}, {Choi}, {Crichton}, {Datta},
  {Devlin}, {Dunkley}, {D{\"u}nner}, {Gallardo}, {Gralla}, {Hincks}, {Ho},
  {Hubmayr}, {Huffenberger}, {Hughes}, {Koopman}, {Kosowsky}, {Louis},
  {Madhavacheril}, {Marriage}, {Maurin}, {McMahon}, {Miyatake}, {Moodley},
  {N{\ae}ss}, {Nati}, {Newburgh}, {Niemack}, {Oguri}, {Page}, {Partridge},
  {Schmitt}, {Sievers}, {Spergel}, {Staggs}, {Trac}, {van Engelen},
  {Vavagiakis}, \& {Wollack}}]{Hilton_2018}
{Hilton}, M., {Hasselfield}, M., {Sif{\'o}n}, C., {et~al.} 2018, \apjs, 235,
  20, \dodoi{10.3847/1538-4365/aaa6cb}

\bibitem[{Ho {et~al.}(2017)Ho, Austermann, Beall, Choi, Cothard, Crowley,
  Datta, Devlin, Duff, Gallardo, Hasselfield, Henderson, Hilton, Hubmayr,
  Koopman, Li, McMahon, Niemack, Salatino, Simon, Staggs, Ward, Ullom,
  Vavagiakis, \& Wollack}]{Ho_2017}
Ho, S.-P.~P., Austermann, J., Beall, J.~A., {et~al.} 2017, in Millimeter,
  Submillimeter, and Far-Infrared Detectors and Instrumentation for Astronomy
  VIII, ed. W.~S. Holland \& J.~Zmuidzinas, Vol. 9914, International Society
  for Optics and Photonics (SPIE), 301 -- 315, \dodoi{10.1117/12.2233113}

\bibitem[{{Hoekstra} {et~al.}(2015){Hoekstra}, {Herbonnet}, {Muzzin}, {Babul},
  {Mahdavi}, {Viola}, \& {Cacciato}}]{Hoekstra_2015}
{Hoekstra}, H., {Herbonnet}, R., {Muzzin}, A., {et~al.} 2015, \mnras, 449, 685,
  \dodoi{10.1093/mnras/stv275}

\bibitem[{{Huang} {et~al.}(2020{\natexlab{a}}){Huang}, {Bleem}, {Stalder},
  {Ade}, {Allen}, {Anderson}, {Austermann}, {Avva}, {Beall}, {Bender},
  {Benson}, {Bianchini}, {Bocquet}, {Brodwin}, {Carlstrom}, {Chang}, {Chiang},
  {Citron}, {Moran}, {Crawford}, {Crites}, {Haan}, {Dobbs}, {Everett}, {Floyd},
  {Gallicchio}, {George}, {Gilbert}, {Gladders}, {Guns}, {Gupta}, {Halverson},
  {Harrington}, {Henning}, {Hilton}, {Holder}, {Holzapfel}, {Hrubes},
  {Hubmayr}, {Irwin}, {Khullar}, {Knox}, {Lee}, {Li}, {Lowitz}, {McDonald},
  {McMahon}, {Meyer}, {Mocanu}, {Montgomery}, {Nadolski}, {Natoli}, {Nibarger},
  {Noble}, {Novosad}, {Padin}, {Patil}, {Pryke}, {Reichardt}, {Ruhl},
  {Saliwanchik}, {Saro}, {Sayre}, {Schaffer}, {Sharon}, {Sievers}, {Smecher},
  {Stark}, {Story}, {Tucker}, {Vanderlinde}, {Veach}, {Vieira}, {Wang},
  {Whitehorn}, {Wu}, \& {Yefremenko}}]{Huang_2020}
{Huang}, N., {Bleem}, L.~E., {Stalder}, B., {et~al.} 2020{\natexlab{a}}, \aj,
  159, 110, \dodoi{10.3847/1538-3881/ab6a96}

\bibitem[{{Huang} {et~al.}(2020{\natexlab{b}}){Huang}, {Storfer}, {Ravi},
  {Pilon}, {Domingo}, {Schlegel}, {Bailey}, {Dey}, {Gupta}, {Herrera},
  {Juneau}, {Landriau}, {Lang}, {Meisner}, {Moustakas}, {Myers}, {Schlafly},
  {Valdes}, {Weaver}, {Yang}, \& {Y{\`e}che}}]{HuangH20a_2020}
{Huang}, X., {Storfer}, C., {Ravi}, V., {et~al.} 2020{\natexlab{b}}, \apj, 894,
  78, \dodoi{10.3847/1538-4357/ab7ffb}

\bibitem[{{Huang} {et~al.}(2020{\natexlab{c}}){Huang}, {Storfer}, {Gu}, {Ravi},
  {Pilon}, {Sheu}, {Venguswamy}, {Bankda}, {Dey}, {Landriau}, {Lang},
  {Meisner}, {Moustakas}, {Myers}, {Sajith}, {Schlafly}, \&
  {Schlegel}}]{HuangH20b_2020}
{Huang}, X., {Storfer}, C., {Gu}, A., {et~al.} 2020{\natexlab{c}}, arXiv
  e-prints, arXiv:2005.04730.
\newblock \doarXiv{2005.04730}

\bibitem[{{Hughes} {et~al.}(1995){Hughes}, {Birkinshaw}, \&
  {Huchra}}]{1995ApJ...448L..93H}
{Hughes}, J.~P., {Birkinshaw}, M., \& {Huchra}, J.~P. 1995, \apjl, 448, L93,
  \dodoi{10.1086/309609}

\bibitem[{{Ilbert} {et~al.}(2009){Ilbert}, {Capak}, {Salvato}, {Aussel},
  {McCracken}, {Sanders}, {Scoville}, {Kartaltepe}, {Arnouts}, {Le Floc'h},
  {Mobasher}, {Taniguchi}, {Lamareille}, {Leauthaud}, {Sasaki}, {Thompson},
  {Zamojski}, {Zamorani}, {Bardelli}, {Bolzonella}, {Bongiorno}, {Brusa},
  {Caputi}, {Carollo}, {Contini}, {Cook}, {Coppa}, {Cucciati}, {de la Torre},
  {de Ravel}, {Franzetti}, {Garilli}, {Hasinger}, {Iovino}, {Kampczyk},
  {Kneib}, {Knobel}, {Kovac}, {Le Borgne}, {Le Brun}, {Le F{\`e}vre}, {Lilly},
  {Looper}, {Maier}, {Mainieri}, {Mellier}, {Mignoli}, {Murayama}, {Pell{\`o}},
  {Peng}, {P{\'e}rez-Montero}, {Renzini}, {Ricciardelli}, {Schiminovich},
  {Scodeggio}, {Shioya}, {Silverman}, {Surace}, {Tanaka}, {Tasca}, {Tresse},
  {Vergani}, \& {Zucca}}]{Ilbert_2009}
{Ilbert}, O., {Capak}, P., {Salvato}, M., {et~al.} 2009, \apj, 690, 1236,
  \dodoi{10.1088/0004-637X/690/2/1236}

\bibitem[{{Itoh} {et~al.}(1998){Itoh}, {Kohyama}, \& {Nozawa}}]{Itoh_1998}
{Itoh}, N., {Kohyama}, Y., \& {Nozawa}, S. 1998, \apj, 502, 7,
  \dodoi{10.1086/305876}

\bibitem[{{Jacobs} {et~al.}(2019{\natexlab{a}}){Jacobs}, {Collett},
  {Glazebrook}, {McCarthy}, {Qin}, {Abbott}, {Abdalla}, {Annis}, {Avila},
  {Bechtol}, {Bertin}, {Brooks}, {Buckley-Geer}, {Burke}, {Carnero Rosell},
  {Carrasco Kind}, {Carretero}, {da Costa}, {Davis}, {De Vicente}, {Desai},
  {Diehl}, {Doel}, {Eifler}, {Flaugher}, {Frieman}, {Garc{\'\i}a-Bellido},
  {Gaztanaga}, {Gerdes}, {Goldstein}, {Gruen}, {Gruendl}, {Gschwend},
  {Gutierrez}, {Hartley}, {Hollowood}, {Honscheid}, {Hoyle}, {James}, {Kuehn},
  {Kuropatkin}, {Lahav}, {Li}, {Lima}, {Lin}, {Maia}, {Martini}, {Miller},
  {Miquel}, {Nord}, {Plazas}, {Sanchez}, {Scarpine}, {Schubnell}, {Serrano},
  {Sevilla-Noarbe}, {Smith}, {Soares-Santos}, {Sobreira}, {Suchyta}, {Swanson},
  {Tarle}, {Vikram}, {Walker}, {Zhang}, {Zuntz}, \& {DES
  Collaboration}}]{JacobsJ19_2019}
{Jacobs}, C., {Collett}, T., {Glazebrook}, K., {et~al.} 2019{\natexlab{a}},
  \mnras, 484, 5330, \dodoi{10.1093/mnras/stz272}

\bibitem[{{Jacobs} {et~al.}(2019{\natexlab{b}}){Jacobs}, {Collett},
  {Glazebrook}, {Buckley-Geer}, {Diehl}, {Lin}, {McCarthy}, {Qin}, {Odden},
  {Caso Escudero}, {Dial}, {Yung}, {Gaitsch}, {Pellico}, {Lindgren}, {Abbott},
  {Annis}, {Avila}, {Brooks}, {Burke}, {Carnero Rosell}, {Carrasco Kind},
  {Carretero}, {da Costa}, {De Vicente}, {Fosalba}, {Frieman},
  {Garc{\'\i}a-Bellido}, {Gaztanaga}, {Goldstein}, {Gruen}, {Gruendl},
  {Gschwend}, {Hollowood}, {Honscheid}, {Hoyle}, {James}, {Krause},
  {Kuropatkin}, {Lahav}, {Lima}, {Maia}, {Marshall}, {Miquel}, {Plazas},
  {Roodman}, {Sanchez}, {Scarpine}, {Serrano}, {Sevilla-Noarbe}, {Smith},
  {Sobreira}, {Suchyta}, {Swanson}, {Tarle}, {Vikram}, {Walker}, {Zhang}, \&
  {DES Collaboration}}]{JacobsJ19a_2019}
---. 2019{\natexlab{b}}, \apjs, 243, 17, \dodoi{10.3847/1538-4365/ab26b6}

\bibitem[{{Jaelani} {et~al.}(2020){Jaelani}, {More}, {Oguri}, {Sonnenfeld},
  {Suyu}, {Rusu}, {Wong}, {Chan}, {Kayo}, {Lee}, {Chao}, {Coupon}, {Inoue}, \&
  {Futamase}}]{Jaelani_2020}
{Jaelani}, A.~T., {More}, A., {Oguri}, M., {et~al.} 2020, \mnras, 495, 1291,
  \dodoi{10.1093/mnras/staa1062}

\bibitem[{{Klein} {et~al.}(2019){Klein}, {Grandis}, {Mohr}, {Paulus}, {Abbott},
  {Annis}, {Avila}, {Bertin}, {Brooks}, {Buckley-Geer}, {Rosell}, {Kind},
  {Carretero}, {Castander}, {Cunha}, {D'Andrea}, {da Costa}, {De Vicente},
  {Desai}, {Diehl}, {Dietrich}, {Doel}, {Evrard}, {Flaugher}, {Fosalba},
  {Frieman}, {Garc{\'\i}a-Bellido}, {Gaztanaga}, {Giles}, {Gruen}, {Gruendl},
  {Gschwend}, {Gutierrez}, {Hartley}, {Hollowood}, {Honscheid}, {Hoyle},
  {James}, {Jeltema}, {Kuehn}, {Kuropatkin}, {Lima}, {Maia}, {March},
  {Marshall}, {Menanteau}, {Miquel}, {Ogando}, {Plazas}, {Romer}, {Roodman},
  {Sanchez}, {Scarpine}, {Schindler}, {Serrano}, {Sevilla-Noarbe}, {Smith},
  {Smith}, {Soares-Santos}, {Sobreira}, {Suchyta}, {Swanson}, {Tarle},
  {Thomas}, {Vikram}, \& {DES Collaboration}}]{Klein_2019}
{Klein}, M., {Grandis}, S., {Mohr}, J.~J., {et~al.} 2019, \mnras, 488, 739,
  \dodoi{10.1093/mnras/stz1463}

\bibitem[{{Lauer} {et~al.}(2014){Lauer}, {Postman}, {Strauss}, {Graves}, \&
  {Chisari}}]{2014ApJ...797...82L}
{Lauer}, T.~R., {Postman}, M., {Strauss}, M.~A., {Graves}, G.~J., \& {Chisari},
  N.~E. 2014, \apj, 797, 82, \dodoi{10.1088/0004-637X/797/2/82}

\bibitem[{{Lopes} {et~al.}(2004){Lopes}, {de Carvalho}, {Gal}, {Djorgovski},
  {Odewahn}, {Mahabal}, \& {Brunner}}]{Lopes_2004}
{Lopes}, P.~A.~A., {de Carvalho}, R.~R., {Gal}, R.~R., {et~al.} 2004, \aj, 128,
  1017, \dodoi{10.1086/423038}

\bibitem[{{Madhavacheril} {et~al.}(2020){Madhavacheril}, {Hill}, {N{\ae}ss},
  {Addison}, {Aiola}, {Baildon}, {Battaglia}, {Bean}, {Bond}, {Calabrese},
  {Calafut}, {Choi}, {Darwish}, {Datta}, {Devlin}, {Dunkley}, {D{\"u}nner},
  {Ferraro}, {Gallardo}, {Gluscevic}, {Halpern}, {Han}, {Hasselfield},
  {Hilton}, {Hincks}, {Hlo{\v{z}}ek}, {Ho}, {Huffenberger}, {Hughes},
  {Koopman}, {Kosowsky}, {Lokken}, {Louis}, {Lungu}, {MacInnis}, {Maurin},
  {McMahon}, {Moodley}, {Nati}, {Niemack}, {Page}, {Partridge}, {Robertson},
  {Sehgal}, {Schaan}, {Schillaci}, {Sherwin}, {Sif{\'o}n}, {Simon}, {Spergel},
  {Staggs}, {Storer}, {van Engelen}, {Vavagiakis}, {Wollack}, \&
  {Xu}}]{Madhavacheril_2020}
{Madhavacheril}, M.~S., {Hill}, J.~C., {N{\ae}ss}, S., {et~al.} 2020, \prd,
  102, 023534, \dodoi{10.1103/PhysRevD.102.023534}

\bibitem[{{Mann} \& {Ebeling}(2012)}]{2012MNRAS.420.2120M}
{Mann}, A.~W., \& {Ebeling}, H. 2012, \mnras, 420, 2120,
  \dodoi{10.1111/j.1365-2966.2011.20170.x}

\bibitem[{{Mantz} {et~al.}(2010){Mantz}, {Allen}, {Ebeling}, {Rapetti}, \&
  {Drlica-Wagner}}]{2010MNRAS.406.1773M}
{Mantz}, A., {Allen}, S.~W., {Ebeling}, H., {Rapetti}, D., \& {Drlica-Wagner},
  A. 2010, \mnras, 406, 1773, \dodoi{10.1111/j.1365-2966.2010.16993.x}

\bibitem[{{Mantz} {et~al.}(2014){Mantz}, {Abdulla}, {Carlstrom}, {Greer},
  {Leitch}, {Marrone}, {Muchovej}, {Adami}, {Birkinshaw}, {Bremer}, {Clerc},
  {Giles}, {Horellou}, {Maughan}, {Pacaud}, {Pierre}, \& {Willis}}]{Mantz_2014}
{Mantz}, A.~B., {Abdulla}, Z., {Carlstrom}, J.~E., {et~al.} 2014, \apj, 794,
  157, \dodoi{10.1088/0004-637X/794/2/157}

\bibitem[{{Mantz} {et~al.}(2018){Mantz}, {Abdulla}, {Allen}, {Carlstrom},
  {Logan}, {Marrone}, {Maughan}, {Willis}, {Pacaud}, \& {Pierre}}]{Mantz_2018}
{Mantz}, A.~B., {Abdulla}, Z., {Allen}, S.~W., {et~al.} 2018, \aap, 620, A2,
  \dodoi{10.1051/0004-6361/201630096}

\bibitem[{{Marriage} {et~al.}(2011)}]{Marriage_2011}
{Marriage}, T.~A., {et~al.} 2011, ApJ, 737, 61,
  \dodoi{10.1088/0004-637X/737/2/61}

\bibitem[{{Maturi} {et~al.}(2019){Maturi}, {Bellagamba}, {Radovich},
  {Roncarelli}, {Sereno}, {Moscardini}, {Bardelli}, \&
  {Puddu}}]{2019MNRAS.485..498M}
{Maturi}, M., {Bellagamba}, F., {Radovich}, M., {et~al.} 2019, \mnras, 485,
  498, \dodoi{10.1093/mnras/stz294}

\bibitem[{{McClintock} {et~al.}(2019){McClintock}, {Varga}, {Gruen}, {Rozo},
  {Rykoff}, {Shin}, {Melchior}, {DeRose}, {Seitz}, {Dietrich}, {Sheldon},
  {Zhang}, {von der Linden}, {Jeltema}, {Mantz}, {Romer}, {Allen}, {Becker},
  {Bermeo}, {Bhargava}, {Costanzi}, {Everett}, {Farahi}, {Hamaus}, {Hartley},
  {Hollowood}, {Hoyle}, {Israel}, {Li}, {MacCrann}, {Morris}, {Palmese},
  {Plazas}, {Pollina}, {Rau}, {Simet}, {Soares-Santos}, {Troxel}, {Vergara
  Cervantes}, {Wechsler}, {Zuntz}, {Abbott}, {Abdalla}, {Allam}, {Annis},
  {Avila}, {Bridle}, {Brooks}, {Burke}, {Carnero Rosell}, {Carrasco Kind},
  {Carretero}, {Castander}, {Crocce}, {Cunha}, {D'Andrea}, {da Costa}, {Davis},
  {De Vicente}, {Diehl}, {Doel}, {Drlica-Wagner}, {Evrard}, {Flaugher},
  {Fosalba}, {Frieman}, {Garc{\'\i}a-Bellido}, {Gaztanaga}, {Gerdes},
  {Giannantonio}, {Gruendl}, {Gutierrez}, {Honscheid}, {James}, {Kirk},
  {Krause}, {Kuehn}, {Lahav}, {Li}, {Lima}, {March}, {Marshall}, {Menanteau},
  {Miquel}, {Mohr}, {Nord}, {Ogando}, {Roodman}, {Sanchez}, {Scarpine},
  {Schindler}, {Sevilla-Noarbe}, {Smith}, {Smith}, {Sobreira}, {Suchyta},
  {Swanson}, {Tarle}, {Tucker}, {Vikram}, {Walker}, {Weller}, \& {DES
  Collaboration}}]{McClintock_2019}
{McClintock}, T., {Varga}, T.~N., {Gruen}, D., {et~al.} 2019, \mnras, 482,
  1352, \dodoi{10.1093/mnras/sty2711}

\bibitem[{{McDonald} {et~al.}(2012)}]{McDonald_2012}
{McDonald}, M., {et~al.} 2012, Nature, 488, 349, \dodoi{10.1038/nature11379}

\bibitem[{{Medezinski} {et~al.}(2018){Medezinski}, {Battaglia}, {Umetsu},
  {Oguri}, {Miyatake}, {Nishizawa}, {Sif{\'o}n}, {Spergel}, {Chiu}, {Lin},
  {Bahcall}, \& {Komiyama}}]{Medezinski_2018}
{Medezinski}, E., {Battaglia}, N., {Umetsu}, K., {et~al.} 2018, \pasj, 70, S28,
  \dodoi{10.1093/pasj/psx128}

\bibitem[{{Mehrtens} {et~al.}(2012){Mehrtens}, {Romer}, {Hilton},
  {Lloyd-Davies}, {Miller}, {Stanford}, {Hosmer}, {Hoyle}, {Collins}, {Liddle},
  {Viana}, {Nichol}, {Stott}, {Dubois}, {Kay}, {Sahl{\'e}n}, {Young}, {Short},
  {Christodoulou}, {Watson}, {Davidson}, {Harrison}, {Baruah}, {Smith},
  {Burke}, {Mayers}, {Deadman}, {Rooney}, {Edmondson}, {West}, {Campbell},
  {Edge}, {Mann}, {Sabirli}, {Wake}, {Benoist}, {da Costa}, {Maia}, \&
  {Ogando}}]{2012MNRAS.423.1024M}
{Mehrtens}, N., {Romer}, A.~K., {Hilton}, M., {et~al.} 2012, \mnras, 423, 1024,
  \dodoi{10.1111/j.1365-2966.2012.20931.x}

\bibitem[{{Melin} {et~al.}(2006){Melin}, {Bartlett}, \&
  {Delabrouille}}]{Melin_2006}
{Melin}, J.~B., {Bartlett}, J.~G., \& {Delabrouille}, J. 2006, \aap, 459, 341,
  \dodoi{10.1051/0004-6361:20065034}

\bibitem[{{Menanteau} {et~al.}(2010)}]{Menanteau_2010}
{Menanteau}, F., {et~al.} 2010, ApJ, 723, 1523,
  \dodoi{10.1088/0004-637X/723/2/1523}

\bibitem[{{Menanteau} {et~al.}(2013{\natexlab{a}}){Menanteau}, {Sif{\'o}n},
  {Barrientos}, {Battaglia}, {Bond}, {Crichton}, {Das}, {Devlin}, {Dicker},
  {D{\"u}nner}, {Gralla}, {Hajian}, {Hasselfield}, {Hilton}, {Hincks},
  {Hughes}, {Infante}, {Kosowsky}, {Marriage}, {Marsden}, {Moodley}, {Niemack},
  {Nolta}, {Page}, {Partridge}, {Reese}, {Schmitt}, {Sievers}, {Spergel},
  {Staggs}, {Switzer}, \& {Wollack}}]{2013ApJ...765...67M}
{Menanteau}, F., {Sif{\'o}n}, C., {Barrientos}, L.~F., {et~al.}
  2013{\natexlab{a}}, \apj, 765, 67, \dodoi{10.1088/0004-637X/765/1/67}

\bibitem[{{Menanteau} {et~al.}(2013{\natexlab{b}})}]{Menanteau_2013}
{Menanteau}, F., {et~al.} 2013{\natexlab{b}}, ApJ, 765, 67,
  \dodoi{10.1088/0004-637X/765/1/67}

\bibitem[{{Miyatake} {et~al.}(2019){Miyatake}, {Battaglia}, {Hilton},
  {Medezinski}, {Nishizawa}, {More}, {Aiola}, {Bahcall}, {Bond}, {Calabrese},
  {Choi}, {Devlin}, {Dunkley}, {Dunner}, {Fuzia}, {Gallardo}, {Gralla},
  {Hasselfield}, {Halpern}, {Hikage}, {Hill}, {Hincks}, {Hlo{\v{z}}ek},
  {Huffenberger}, {Hughes}, {Koopman}, {Kosowsky}, {Louis}, {Madhavacheril},
  {McMahon}, {Mandelbaum}, {Marriage}, {Maurin}, {Miyazaki}, {Moodley},
  {Murata}, {Naess}, {Newburgh}, {Niemack}, {Nishimichi}, {Okabe}, {Oguri},
  {Osato}, {Page}, {Partridge}, {Robertson}, {Sehgal}, {Sherwin}, {Shirasaki},
  {Sievers}, {Sif{\'o}n}, {Simon}, {Spergel}, {Staggs}, {Stein}, {Takada},
  {Trac}, {Umetsu}, {van Engelen}, \& {Wollack}}]{Miyatake_2019}
{Miyatake}, H., {Battaglia}, N., {Hilton}, M., {et~al.} 2019, \apj, 875, 63,
  \dodoi{10.3847/1538-4357/ab0af0}

\bibitem[{{Miyazaki} {et~al.}(2018){Miyazaki}, {Komiyama}, {Kawanomoto}, {Doi},
  {Furusawa}, {Hamana}, {Hayashi}, {Ikeda}, {Kamata}, {Karoji}, {Koike},
  {Kurakami}, {Miyama}, {Morokuma}, {Nakata}, {Namikawa}, {Nakaya}, {Nariai},
  {Obuchi}, {Oishi}, {Okada}, {Okura}, {Tait}, {Takata}, {Tanaka}, {Tanaka},
  {Terai}, {Tomono}, {Uraguchi}, {Usuda}, {Utsumi}, {Yamada}, {Yamanoi},
  {Aihara}, {Fujimori}, {Mineo}, {Miyatake}, {Oguri}, {Uchida}, {Tanaka},
  {Yasuda}, {Takada}, {Murayama}, {Nishizawa}, {Sugiyama}, {Chiba}, {Futamase},
  {Wang}, {Chen}, {Ho}, {Liaw}, {Chiu}, {Ho}, {Lai}, {Lee}, {Jeng}, {Iwamura},
  {Armstrong}, {Bickerton}, {Bosch}, {Gunn}, {Lupton}, {Loomis}, {Price},
  {Smith}, {Strauss}, {Turner}, {Suzuki}, {Miyazaki}, {Muramatsu}, {Yamamoto},
  {Endo}, {Ezaki}, {Ito}, {Kawaguchi}, {Sofuku}, {Taniike}, {Akutsu}, {Dojo},
  {Kasumi}, {Matsuda}, {Imoto}, {Miwa}, {Suzuki}, {Takeshi}, \&
  {Yokota}}]{Miyazaki_2018}
{Miyazaki}, S., {Komiyama}, Y., {Kawanomoto}, S., {et~al.} 2018, \pasj, 70, S1,
  \dodoi{10.1093/pasj/psx063}

\bibitem[{{More} {et~al.}(2016){More}, {Verma}, {Marshall}, {More}, {Baeten},
  {Wilcox}, {Macmillan}, {Cornen}, {Kapadia}, {Parrish}, {Snyder}, {Davis},
  {Gavazzi}, {Lintott}, {Simpson}, {Miller}, {Smith}, {Paget}, {Saha},
  {K{\"u}ng}, \& {Collett}}]{More_2016}
{More}, A., {Verma}, A., {Marshall}, P.~J., {et~al.} 2016, \mnras, 455, 1191,
  \dodoi{10.1093/mnras/stv1965}

\bibitem[{{Mroczkowski} {et~al.}(2019){Mroczkowski}, {Nagai}, {Basu}, {Chluba},
  {Sayers}, {Adam}, {Churazov}, {Crites}, {Di Mascolo}, {Eckert},
  {Macias-Perez}, {Mayet}, {Perotto}, {Pointecouteau}, {Romero}, {Ruppin},
  {Scannapieco}, \& {ZuHone}}]{Mroczkowski_2019}
{Mroczkowski}, T., {Nagai}, D., {Basu}, K., {et~al.} 2019, \ssr, 215, 17,
  \dodoi{10.1007/s11214-019-0581-2}

\bibitem[{{Mullis} {et~al.}(2003){Mullis}, {McNamara}, {Quintana}, {Vikhlinin},
  {Henry}, {Gioia}, {Hornstrup}, {Forman}, \& {Jones}}]{2003ApJ...594..154M}
{Mullis}, C.~R., {McNamara}, B.~R., {Quintana}, H., {et~al.} 2003, \apj, 594,
  154, \dodoi{10.1086/376866}

\bibitem[{{Murphy} {et~al.}(2012){Murphy}, {Geach}, \& {Bower}}]{Murphy_2012}
{Murphy}, D.~N.~A., {Geach}, J.~E., \& {Bower}, R.~G. 2012, \mnras, 420, 1861,
  \dodoi{10.1111/j.1365-2966.2011.19782.x}

\bibitem[{{Naess} {et~al.}(2020){Naess}, {Aiola}, {Austermann}, {Battaglia},
  {Beall}, {Becker}, {Bond}, {Calabrese}, {Choi}, {Cothard}, {Crowley},
  {Darwish}, {Datta}, {Denison}, {Devlin}, {Duell}, {Duff}, {Duivenvoorden},
  {Dunkley}, {D{\"u}nner}, {Fox}, {Gallardo}, {Halpern}, {Han}, {Hasselfield},
  {Hill}, {Hilton}, {Hilton}, {Hincks}, {Hlo{\v{z}}ek}, {Ho}, {Hubmayr},
  {Huffenberger}, {Hughes}, {Kosowsky}, {Louis}, {Madhavacheril}, {McMahon},
  {Moodley}, {Nati}, {Nibarger}, {Niemack}, {Page}, {Partridge}, {Salatino},
  {Schaan}, {Schillaci}, {Schmitt}, {Sherwin}, {Sehgal}, {Sif{\'o}n},
  {Spergel}, {Staggs}, {Stevens}, {Storer}, {Ullom}, {Vale}, {Van Engelen},
  {Van Lanen}, {Vavagiakis}, {Wollack}, \& {Xu}}]{Naess_2020}
{Naess}, S., {Aiola}, S., {Austermann}, J.~E., {et~al.} 2020, arXiv e-prints,
  arXiv:2007.07290.
\newblock \doarXiv{2007.07290}

\bibitem[{{Nastasi} {et~al.}(2014){Nastasi}, {B{\"o}hringer}, {Fassbender}, {de
  Hoon}, {Lamer}, {Mohr}, {Padilla}, {Pratt}, {Quintana}, {Rosati}, {Santos},
  {Schwope}, {{\v{S}}uhada}, \& {Verdugo}}]{2014AandA...564A..17N}
{Nastasi}, A., {B{\"o}hringer}, H., {Fassbender}, R., {et~al.} 2014, \aap, 564,
  A17, \dodoi{10.1051/0004-6361/201322321}

\bibitem[{{Oegerle} \& {Hill}(2001)}]{2001AJ....122.2858O}
{Oegerle}, W.~R., \& {Hill}, J.~M. 2001, \aj, 122, 2858, \dodoi{10.1086/323536}

\bibitem[{{Oguri}(2014)}]{Oguri_2014}
{Oguri}, M. 2014, \mnras, 444, 147, \dodoi{10.1093/mnras/stu1446}

\bibitem[{{Oguri} {et~al.}(2018){Oguri}, {Lin}, {Lin}, {Nishizawa}, {More},
  {More}, {Hsieh}, {Medezinski}, {Miyatake}, {Jian}, {Lin}, {Takada}, {Okabe},
  {Speagle}, {Coupon}, {Leauthaud}, {Lupton}, {Miyazaki}, {Price}, {Tanaka},
  {Chiu}, {Komiyama}, {Okura}, {Tanaka}, \& {Usuda}}]{Oguri_2018}
{Oguri}, M., {Lin}, Y.-T., {Lin}, S.-C., {et~al.} 2018, \pasj, 70, S20,
  \dodoi{10.1093/pasj/psx042}

\bibitem[{{Oke}(1974)}]{Oke_1974}
{Oke}, J.~B. 1974, ApJS, 27, 21

\bibitem[{{Petrillo} {et~al.}(2019){Petrillo}, {Tortora}, {Vernardos},
  {Koopmans}, {Verdoes Kleijn}, {Bilicki}, {Napolitano}, {Chatterjee},
  {Covone}, {Dvornik}, {Erben}, {Getman}, {Giblin}, {Heymans}, {de Jong},
  {Kuijken}, {Schneider}, {Shan}, {Spiniello}, \& {Wright}}]{Petrillo_2019}
{Petrillo}, C.~E., {Tortora}, C., {Vernardos}, G., {et~al.} 2019, \mnras, 484,
  3879, \dodoi{10.1093/mnras/stz189}

\bibitem[{{Pimbblet} {et~al.}(2006){Pimbblet}, {Smail}, {Edge}, {O'Hely},
  {Couch}, \& {Zabludoff}}]{2006MNRAS.366..645P}
{Pimbblet}, K.~A., {Smail}, I., {Edge}, A.~C., {et~al.} 2006, \mnras, 366, 645,
  \dodoi{10.1111/j.1365-2966.2005.09892.x}

\bibitem[{{Planck Collaboration} {et~al.}(2012){Planck Collaboration},
  {Aghanim}, {Arnaud}, {Ashdown}, {Atrio-Barandela}, {Aumont}, {Baccigalupi},
  {Balbi}, {Banday}, {Barreiro}, {Bartlett}, {Battaner}, {Benabed}, {Bernard},
  {Bersanelli}, {B{\"o}hringer}, {Bonaldi}, {Bond}, {Borrill}, {Bouchet},
  {Bourdin}, {Brown}, {Burigana}, {Butler}, {Cabella}, {Cardoso}, {Carvalho},
  {Catalano}, {Cay{\'o}n}, {Chamballu}, {Chary}, {Chiang}, {Chon},
  {Christensen}, {Clements}, {Colafrancesco}, {Colombi}, {Coulais}, {Crill},
  {Cuttaia}, {Da Silva}, {Dahle}, {Davis}, {de Bernardis}, {de Gasperis}, {de
  Zotti}, {Delabrouille}, {D{\'e}mocl{\`e}s}, {D{\'e}sert}, {Diego}, {Dolag},
  {Dole}, {Donzelli}, {Dor{\'e}}, {Douspis}, {Dupac}, {En{\ss}lin}, {Eriksen},
  {Finelli}, {Flores-Cacho}, {Forni}, {Fosalba}, {Frailis}, {Fromenteau},
  {Galeotta}, {Ganga}, {G{\'e}nova-Santos}, {Giard}, {Gonz{\'a}lez-Nuevo},
  {Gonz{\'a}lez-Riestra}, {G{\'o}rski}, {Gregorio}, {Gruppuso}, {Hansen},
  {Harrison}, {Hempel}, {Hern{\'a}ndez-Monteagudo}, {Herranz}, {Hildebrand t},
  {Hornstrup}, {Huffenberger}, {Hurier}, {Jagemann}, {Jasche}, {Juvela},
  {Keih{\"a}nen}, {Keskitalo}, {Kisner}, {Kneissl}, {Knoche}, {Knox},
  {Kurki-Suonio}, {Lagache}, {L{\"a}hteenm{\"a}ki}, {Lamarre}, {Lasenby},
  {Lawrence}, {Leach}, {Leonardi}, {Liddle}, {Lilje}, {L{\'o}pez-Caniego},
  {Luzzi}, {Mac{\'\i}as-P{\'e}rez}, {Maino}, {Mand olesi}, {Mann}, {Marleau},
  {Marshall}, {Mart{\'\i}nez-Gonz{\'a}lez}, {Masi}, {Massardi}, {Matarrese},
  {Matthai}, {Mazzotta}, {Meinhold}, {Melchiorri}, {Melin}, {Mendes},
  {Mennella}, {Miville-Desch{\^e}nes}, {Moneti}, {Montier}, {Morgante},
  {Mortlock}, {Munshi}, {Naselsky}, {Natoli}, {N{\o}rgaard-Nielsen},
  {Noviello}, {Osborne}, {Pasian}, {Patanchon}, {Perdereau}, {Perrotta},
  {Piacentini}, {Pierpaoli}, {Plaszczynski}, {Platania}, {Pointecouteau},
  {Polenta}, {Ponthieu}, {Popa}, {Poutanen}, {Pratt}, {Puget}, {Rachen},
  {Rebolo}, {Reinecke}, {Remazeilles}, {Renault}, {Ricciardi}, {Riller},
  {Ristorcelli}, {Rocha}, {Rosset}, {Rossetti}, {Rubi{\~n}o-Mart{\'\i}n},
  {Rusholme}, {Sandri}, {Savini}, {Schaefer}, {Scott}, {Smoot}, {Starck},
  {Stivoli}, {Sunyaev}, {Sutton}, {Sygnet}, {Tauber}, {Terenzi}, {Toffolatti},
  {Tomasi}, {Tristram}, {Valenziano}, {Van Tent}, {Vielva}, {Villa},
  {Vittorio}, {Wandelt}, {Weller}, {White}, {Yvon}, {Zacchei}, \&
  {Zonca}}]{2012AandA...543A.102P}
{Planck Collaboration}, {Aghanim}, N., {Arnaud}, M., {et~al.} 2012, \aap, 543,
  A102, \dodoi{10.1051/0004-6361/201118731}

\bibitem[{{Planck Collaboration} {et~al.}(2014{\natexlab{a}}){Planck
  Collaboration}, {Ade}, {Aghanim}, {Armitage-Caplan}, {Arnaud}, {Ashdown},
  {Atrio-Barandela}, {Aumont}, {Baccigalupi}, {Banday}, \&
  et~al.}]{Planck_XX_2013}
{Planck Collaboration}, {Ade}, P.~A.~R., {Aghanim}, N., {et~al.}
  2014{\natexlab{a}}, \aap, 571, A20, \dodoi{10.1051/0004-6361/201321521}

\bibitem[{{Planck Collaboration} {et~al.}(2014{\natexlab{b}}){Planck
  Collaboration}, {Ade}, {Aghanim}, {Armitage-Caplan}, {Arnaud}, {Ashdown},
  {Atrio-Barandela}, {Aumont}, {Aussel}, {Baccigalupi}, \&
  et~al.}]{Planck_XXIX_2013}
---. 2014{\natexlab{b}}, \aap, 571, A29, \dodoi{10.1051/0004-6361/201321523}

\bibitem[{{Planck Collaboration} {et~al.}(2015){Planck Collaboration}, {Ade},
  {Aghanim}, {Arnaud}, {Ashdown}, {Aumont}, {Baccigalupi}, {Banday},
  {Barreiro}, {Barrena}, {Bartolo}, {Battaner}, {Benabed}, {Benoit-L{\'e}vy},
  {Bernard}, {Bersanelli}, {Bielewicz}, {Bikmaev}, {B{\"o}hringer}, {Bonaldi},
  {Bonavera}, {Bond}, {Borrill}, {Bouchet}, {Burenin}, {Burigana}, {Butler},
  {Calabrese}, {Carvalho}, {Catalano}, {Chamballu}, {Chiang}, {Chon},
  {Christensen}, {Churazov}, {Clements}, {Colombo}, {Comis}, {Couchot},
  {Curto}, {Cuttaia}, {Dahle}, {Danese}, {Davies}, {Davis}, {de Bernardis}, {de
  Rosa}, {de Zotti}, {Delabrouille}, {Diego}, {Dole}, {Dor{\'e}}, {Douspis},
  {Ducout}, {Dupac}, {Efstathiou}, {Elsner}, {En{\ss}lin}, {Eriksen},
  {Finelli}, {Flores-Cacho}, {Forni}, {Frailis}, {Fraisse}, {Franceschi},
  {Frejsel}, {Fromenteau}, {Galeotta}, {Ganga}, {G{\'e}nova-Santos}, {Giard},
  {Gilfanov}, {Giraud-H{\'e}raud}, {Gjerl{\o}w}, {Gonz{\'a}lez-Nuevo},
  {G{\'o}rski}, {Gruppuso}, {Hansen}, {Hanson}, {Harrison}, {Hempel},
  {Henrot-Versill{\'e}}, {Hern{\'a}ndez-Monteagudo}, {Herranz}, {Hildebrand t},
  {Hivon}, {Hobson}, {Holmes}, {Hornstrup}, {Hovest}, {Huffenberger}, {Hurier},
  {Jaffe}, {Jones}, {Juvela}, {Keih{\"a}nen}, {Keskitalo}, {Khamitov},
  {Kisner}, {Kneissl}, {Knoche}, {Kunz}, {Kurki-Suonio}, {Lagache}, {Lamarre},
  {Lasenby}, {Lattanzi}, {Lawrence}, {Leonardi}, {Levrier}, {Liguori}, {Lilje},
  {Linden-V{\o}rnle}, {L{\'o}pez-Caniego}, {Lubin}, {Mac{\'\i}as-P{\'e}rez},
  {Maino}, {Mandolesi}, {Maris}, {Martin}, {Mart{\'\i}nez-Gonz{\'a}lez},
  {Masi}, {Matarrese}, {Mazzotta}, {Melin}, {Mendes}, {Mennella}, {Migliaccio},
  {Miville-Desch{\^e}nes}, {Moneti}, {Montier}, {Morgante}, {Mortlock},
  {Munshi}, {Murphy}, {Naselsky}, {Nati}, {Natoli}, {N{\o}rgaard-Nielsen},
  {Novikov}, {Novikov}, {Oxborrow}, {Pagano}, {Pajot}, {Paoletti}, {Pasian},
  {Perdereau}, {Perotto}, {Perrotta}, {Pettorino}, {Piacentini}, {Piat},
  {Pietrobon}, {Plaszczynski}, {Pointecouteau}, {Polenta}, {Popa}, {Pratt},
  {Prunet}, {Puget}, {Rachen}, {Reinecke}, {Remazeilles}, {Renault},
  {Ricciardi}, {Ristorcelli}, {Rocha}, {Roman}, {Rosset}, {Rossetti},
  {Roudier}, {Rubi{\~n}o-Mart{\'\i}n}, {Rusholme}, {Sandri}, {Scott},
  {Spencer}, {Stolyarov}, {Sudiwala}, {Sunyaev}, {Sutton}, {Suur-Uski},
  {Sygnet}, {Tauber}, {Terenzi}, {Toffolatti}, {Tomasi}, {Tristram}, {Tucci},
  {Valenziano}, {Valiviita}, {Van Tent}, {Vibert}, {Vielva}, {Villa}, {Wade},
  {Wandelt}, {Wehus}, {Yvon}, {Zacchei}, \& {Zonca}}]{2015AandA...582A..29P}
---. 2015, \aap, 582, A29, \dodoi{10.1051/0004-6361/201424674}

\bibitem[{{Planck Collaboration} {et~al.}(2016{\natexlab{a}}){Planck
  Collaboration}, {Ade}, {Aghanim}, {Arnaud}, {Ashdown}, {Aumont},
  {Baccigalupi}, {Banday}, {Barreiro}, {Bartlett}, {Bartolo}, {Battaner},
  {Battye}, {Benabed}, {Beno{\^\i}t}, {Benoit-L{\'e}vy}, {Bernard},
  {Bersanelli}, {Bielewicz}, {Bock}, {Bonaldi}, {Bonavera}, {Bond}, {Borrill},
  {Bouchet}, {Bucher}, {Burigana}, {Butler}, {Calabrese}, {Cardoso},
  {Catalano}, {Challinor}, {Chamballu}, {Chary}, {Chiang}, {Christensen},
  {Church}, {Clements}, {Colombi}, {Colombo}, {Combet}, {Comis}, {Couchot},
  {Coulais}, {Crill}, {Curto}, {Cuttaia}, {Danese}, {Davies}, {Davis}, {de
  Bernardis}, {de Rosa}, {de Zotti}, {Delabrouille}, {D{\'e}sert}, {Diego},
  {Dolag}, {Dole}, {Donzelli}, {Dor{\'e}}, {Douspis}, {Ducout}, {Dupac},
  {Efstathiou}, {Elsner}, {En{\ss}lin}, {Eriksen}, {Falgarone}, {Fergusson},
  {Finelli}, {Forni}, {Frailis}, {Fraisse}, {Franceschi}, {Frejsel},
  {Galeotta}, {Galli}, {Ganga}, {Giard}, {Giraud-H{\'e}raud}, {Gjerl{\o}w},
  {Gonz{\'a}lez-Nuevo}, {G{\'o}rski}, {Gratton}, {Gregorio}, {Gruppuso},
  {Gudmundsson}, {Hansen}, {Hanson}, {Harrison}, {Henrot-Versill{\'e}},
  {Hern{\'a}ndez-Monteagudo}, {Herranz}, {Hildebrandt}, {Hivon}, {Hobson},
  {Holmes}, {Hornstrup}, {Hovest}, {Huffenberger}, {Hurier}, {Jaffe}, {Jaffe},
  {Jones}, {Juvela}, {Keih{\"a}nen}, {Keskitalo}, {Kisner}, {Kneissl},
  {Knoche}, {Kunz}, {Kurki-Suonio}, {Lagache}, {L{\"a}hteenm{\"a}ki},
  {Lamarre}, {Lasenby}, {Lattanzi}, {Lawrence}, {Leonardi}, {Lesgourgues},
  {Levrier}, {Liguori}, {Lilje}, {Linden-V{\o}rnle}, {L{\'o}pez-Caniego},
  {Lubin}, {Mac{\'\i}as-P{\'e}rez}, {Maggio}, {Maino}, {Mand olesi},
  {Mangilli}, {Maris}, {Martin}, {Mart{\'\i}nez-Gonz{\'a}lez}, {Masi},
  {Matarrese}, {McGehee}, {Meinhold}, {Melchiorri}, {Melin}, {Mendes},
  {Mennella}, {Migliaccio}, {Mitra}, {Miville-Desch{\^e}nes}, {Moneti},
  {Montier}, {Morgante}, {Mortlock}, {Moss}, {Munshi}, {Murphy}, {Naselsky},
  {Nati}, {Natoli}, {Netterfield}, {N{\o}rgaard-Nielsen}, {Noviello},
  {Novikov}, {Novikov}, {Oxborrow}, {Paci}, {Pagano}, {Pajot}, {Paoletti},
  {Partridge}, {Pasian}, {Patanchon}, {Pearson}, {Perdereau}, {Perotto},
  {Perrotta}, {Pettorino}, {Piacentini}, {Piat}, {Pierpaoli}, {Pietrobon},
  {Plaszczynski}, {Pointecouteau}, {Polenta}, {Popa}, {Pratt}, {Pr{\'e}zeau},
  {Prunet}, {Puget}, {Rachen}, {Rebolo}, {Reinecke}, {Remazeilles}, {Renault},
  {Renzi}, {Ristorcelli}, {Rocha}, {Roman}, {Rosset}, {Rossetti}, {Roudier},
  {Rubi{\~n}o-Mart{\'\i}n}, {Rusholme}, {Sandri}, {Santos}, {Savelainen},
  {Savini}, {Scott}, {Seiffert}, {Shellard}, {Spencer}, {Stolyarov}, {Stompor},
  {Sudiwala}, {Sunyaev}, {Sutton}, {Suur-Uski}, {Sygnet}, {Tauber}, {Terenzi},
  {Toffolatti}, {Tomasi}, {Tristram}, {Tucci}, {Tuovinen}, {T{\"u}rler},
  {Umana}, {Valenziano}, {Valiviita}, {Van Tent}, {Vielva}, {Villa}, {Wade},
  {Wandelt}, {Wehus}, {Weller}, {White}, {Yvon}, {Zacchei}, \&
  {Zonca}}]{PlanckPSZ2Cosmo_2016}
---. 2016{\natexlab{a}}, \aap, 594, A24, \dodoi{10.1051/0004-6361/201525833}

\bibitem[{{Planck Collaboration} {et~al.}(2016{\natexlab{b}}){Planck
  Collaboration}, {Ade}, {Aghanim}, {Arnaud}, {Ashdown}, {Aumont},
  {Baccigalupi}, {Banday}, {Barreiro}, {Barrena}, {Bartlett}, {Bartolo},
  {Battaner}, {Battye}, {Benabed}, {Beno{\^\i}t}, {Benoit-L{\'e}vy}, {Bernard},
  {Bersanelli}, {Bielewicz}, {Bikmaev}, {B{\"o}hringer}, {Bonaldi}, {Bonavera},
  {Bond}, {Borrill}, {Bouchet}, {Bucher}, {Burenin}, {Burigana}, {Butler},
  {Calabrese}, {Cardoso}, {Carvalho}, {Catalano}, {Challinor}, {Chamballu},
  {Chary}, {Chiang}, {Chon}, {Christensen}, {Clements}, {Colombi}, {Colombo},
  {Combet}, {Comis}, {Couchot}, {Coulais}, {Crill}, {Curto}, {Cuttaia},
  {Dahle}, {Danese}, {Davies}, {Davis}, {de Bernardis}, {de Rosa}, {de Zotti},
  {Delabrouille}, {D{\'e}sert}, {Dickinson}, {Diego}, {Dolag}, {Dole},
  {Donzelli}, {Dor{\'e}}, {Douspis}, {Ducout}, {Dupac}, {Efstathiou},
  {Eisenhardt}, {Elsner}, {En{\ss}lin}, {Eriksen}, {Falgarone}, {Fergusson},
  {Feroz}, {Ferragamo}, {Finelli}, {Forni}, {Frailis}, {Fraisse}, {Franceschi},
  {Frejsel}, {Galeotta}, {Galli}, {Ganga}, {G{\'e}nova-Santos}, {Giard},
  {Giraud-H{\'e}raud}, {Gjerl{\o}w}, {Gonz{\'a}lez-Nuevo}, {G{\'o}rski},
  {Grainge}, {Gratton}, {Gregorio}, {Gruppuso}, {Gudmundsson}, {Hansen},
  {Hanson}, {Harrison}, {Hempel}, {Henrot-Versill{\'e}},
  {Hern{\'a}ndez-Monteagudo}, {Herranz}, {Hildebrandt}, {Hivon}, {Hobson},
  {Holmes}, {Hornstrup}, {Hovest}, {Huffenberger}, {Hurier}, {Jaffe}, {Jaffe},
  {Jin}, {Jones}, {Juvela}, {Keih{\"a}nen}, {Keskitalo}, {Khamitov}, {Kisner},
  {Kneissl}, {Knoche}, {Kunz}, {Kurki-Suonio}, {Lagache}, {Lamarre}, {Lasenby},
  {Lattanzi}, {Lawrence}, {Leonardi}, {Lesgourgues}, {Levrier}, {Liguori},
  {Lilje}, {Linden-V{\o}rnle}, {L{\'o}pez-Caniego}, {Lubin},
  {Mac{\'\i}as-P{\'e}rez}, {Maggio}, {Maino}, {Mak}, {Mandolesi}, {Mangilli},
  {Martin}, {Mart{\'\i}nez-Gonz{\'a}lez}, {Masi}, {Matarrese}, {Mazzotta},
  {McGehee}, {Mei}, {Melchiorri}, {Melin}, {Mendes}, {Mennella}, {Migliaccio},
  {Mitra}, {Miville-Desch{\^e}nes}, {Moneti}, {Montier}, {Morgante},
  {Mortlock}, {Moss}, {Munshi}, {Murphy}, {Naselsky}, {Nastasi}, {Nati},
  {Natoli}, {Netterfield}, {N{\o}rgaard-Nielsen}, {Noviello}, {Novikov},
  {Novikov}, {Olamaie}, {Oxborrow}, {Paci}, {Pagano}, {Pajot}, {Paoletti},
  {Pasian}, {Patanchon}, {Pearson}, {Perdereau}, {Perotto}, {Perrott},
  {Perrotta}, {Pettorino}, {Piacentini}, {Piat}, {Pierpaoli}, {Pietrobon},
  {Plaszczynski}, {Pointecouteau}, {Polenta}, {Pratt}, {Pr{\'e}zeau}, {Prunet},
  {Puget}, {Rachen}, {Reach}, {Rebolo}, {Reinecke}, {Remazeilles}, {Renault},
  {Renzi}, {Ristorcelli}, {Rocha}, {Rosset}, {Rossetti}, {Roudier}, {Rozo},
  {Rubi{\~n}o-Mart{\'\i}n}, {Rumsey}, {Rusholme}, {Rykoff}, {Sandri}, {Santos},
  {Saunders}, {Savelainen}, {Savini}, {Schammel}, {Scott}, {Seiffert},
  {Shellard}, {Shimwell}, {Spencer}, {Stanford}, {Stern}, {Stolyarov},
  {Stompor}, {Streblyanska}, {Sudiwala}, {Sunyaev}, {Sutton}, {Suur-Uski},
  {Sygnet}, {Tauber}, {Terenzi}, {Toffolatti}, {Tomasi}, {Tramonte},
  {Tristram}, {Tucci}, {Tuovinen}, {Umana}, {Valenziano}, {Valiviita}, {Van
  Tent}, {Vielva}, {Villa}, {Wade}, {Wandelt}, {Wehus}, {White}, {Wright},
  {Yvon}, {Zacchei}, \& {Zonca}}]{PlanckPSZ2_2016}
---. 2016{\natexlab{b}}, \aap, 594, A27, \dodoi{10.1051/0004-6361/201525823}

\bibitem[{{Proust} {et~al.}(2015){Proust}, {Yegorova}, {Saviane}, {Ivanov},
  {Bresolin}, {Salzer}, \& {Capelato}}]{2015MNRAS.452.3304P}
{Proust}, D., {Yegorova}, I., {Saviane}, I., {et~al.} 2015, \mnras, 452, 3304,
  \dodoi{10.1093/mnras/stv1558}

\bibitem[{{Reichardt} {et~al.}(2013){Reichardt}, {Stalder}, {Bleem}, {Montroy},
  {Aird}, {Andersson}, {Armstrong}, {Ashby}, {Bautz}, {Bayliss}, {Bazin},
  {Benson}, {Brodwin}, {Carlstrom}, {Chang}, {Cho}, {Clocchiatti}, {Crawford},
  {Crites}, {de Haan}, {Desai}, {Dobbs}, {Dudley}, {Foley}, {Forman}, {George},
  {Gladders}, {Gonzalez}, {Halverson}, {Harrington}, {High}, {Holder},
  {Holzapfel}, {Hoover}, {Hrubes}, {Jones}, {Joy}, {Keisler}, {Knox}, {Lee},
  {Leitch}, {Liu}, {Lueker}, {Luong-Van}, {Mantz}, {Marrone}, {McDonald},
  {McMahon}, {Mehl}, {Meyer}, {Mocanu}, {Mohr}, {Murray}, {Natoli}, {Padin},
  {Plagge}, {Pryke}, {Rest}, {Ruel}, {Ruhl}, {Saliwanchik}, {Saro}, {Sayre},
  {Schaffer}, {Shaw}, {Shirokoff}, {Song}, {Spieler}, {Staniszewski}, {Stark},
  {Story}, {Stubbs}, {{\v S}uhada}, {van Engelen}, {Vanderlinde}, {Vieira},
  {Vikhlinin}, {Williamson}, {Zahn}, \& {Zenteno}}]{Reichardt_2013}
{Reichardt}, C.~L., {Stalder}, B., {Bleem}, L.~E., {et~al.} 2013, \apj, 763,
  127, \dodoi{10.1088/0004-637X/763/2/127}

\bibitem[{{Romer} {et~al.}(2000){Romer}, {Nichol}, {Holden}, {Ulmer}, {Pildis},
  {Merrelli}, {Adami}, {Burke}, {Collins}, {Metevier}, {Kron}, \&
  {Commons}}]{2000ApJS..126..209R}
{Romer}, A.~K., {Nichol}, R.~C., {Holden}, B.~P., {et~al.} 2000, \apjs, 126,
  209, \dodoi{10.1086/313302}

\bibitem[{{Romero} {et~al.}(2020){Romero}, {Sievers}, {Ghirardini}, {Dicker},
  {Giacintucci}, {Mroczkowski}, {Mason}, {Sarazin}, {Devlin}, {Gaspari},
  {Battaglia}, {Hilton}, {Bulbul}, {Lowe}, \& {Stanchfield}}]{Romero_2020}
{Romero}, C.~E., {Sievers}, J., {Ghirardini}, V., {et~al.} 2020, \apj, 891, 90,
  \dodoi{10.3847/1538-4357/ab6d70}

\bibitem[{{Rykoff} {et~al.}(2014){Rykoff}, {Rozo}, {Busha}, {Cunha},
  {Finoguenov}, {Evrard}, {Hao}, {Koester}, {Leauthaud}, {Nord}, {Pierre},
  {Reddick}, {Sadibekova}, {Sheldon}, \& {Wechsler}}]{Rykoff_2014}
{Rykoff}, E.~S., {Rozo}, E., {Busha}, M.~T., {et~al.} 2014, \apj, 785, 104,
  \dodoi{10.1088/0004-637X/785/2/104}

\bibitem[{{Rykoff} {et~al.}(2016){Rykoff}, {Rozo}, {Hollowood}, {Bermeo-Hernand
  ez}, {Jeltema}, {Mayers}, {Romer}, {Rooney}, {Saro}, {Vergara Cervantes},
  {Wechsler}, {Wilcox}, {Abbott}, {Abdalla}, {Allam}, {Annis},
  {Benoit-L{\'e}vy}, {Bernstein}, {Bertin}, {Brooks}, {Burke}, {Capozzi},
  {Carnero Rosell}, {Carrasco Kind}, {Castander}, {Childress}, {Collins},
  {Cunha}, {D'Andrea}, {da Costa}, {Davis}, {Desai}, {Diehl}, {Dietrich},
  {Doel}, {Evrard}, {Finley}, {Flaugher}, {Fosalba}, {Frieman}, {Glazebrook},
  {Goldstein}, {Gruen}, {Gruendl}, {Gutierrez}, {Hilton}, {Honscheid}, {Hoyle},
  {James}, {Kay}, {Kuehn}, {Kuropatkin}, {Lahav}, {Lewis}, {Lidman}, {Lima},
  {Maia}, {Mann}, {Marshall}, {Martini}, {Melchior}, {Miller}, {Miquel},
  {Mohr}, {Nichol}, {Nord}, {Ogando}, {Plazas}, {Reil}, {Sahl{\'e}n},
  {Sanchez}, {Santiago}, {Scarpine}, {Schubnell}, {Sevilla-Noarbe}, {Smith},
  {Soares-Santos}, {Sobreira}, {Stott}, {Suchyta}, {Swanson}, {Tarle},
  {Thomas}, {Tucker}, {Uddin}, {Viana}, {Vikram}, {Walker}, {Zhang}, \& {DES
  Collaboration}}]{Rykoff_2016}
{Rykoff}, E.~S., {Rozo}, E., {Hollowood}, D., {et~al.} 2016, \apjs, 224, 1,
  \dodoi{10.3847/0067-0049/224/1/1}

\bibitem[{{Salvato} {et~al.}(2011){Salvato}, {Ilbert}, {Hasinger}, {Rau},
  {Civano}, {Zamorani}, {Brusa}, {Elvis}, {Vignali}, {Aussel}, {Comastri},
  {Fiore}, {Le Floc'h}, {Mainieri}, {Bardelli}, {Bolzonella}, {Bongiorno},
  {Capak}, {Caputi}, {Cappelluti}, {Carollo}, {Contini}, {Garilli}, {Iovino},
  {Fotopoulou}, {Fruscione}, {Gilli}, {Halliday}, {Kneib}, {Kakazu},
  {Kartaltepe}, {Koekemoer}, {Kovac}, {Ideue}, {Ikeda}, {Impey}, {Le Fevre},
  {Lamareille}, {Lanzuisi}, {Le Borgne}, {Le Brun}, {Lilly}, {Maier},
  {Manohar}, {Masters}, {McCracken}, {Messias}, {Mignoli}, {Mobasher}, {Nagao},
  {Pello}, {Puccetti}, {Perez-Montero}, {Renzini}, {Sargent}, {Sanders},
  {Scodeggio}, {Scoville}, {Shopbell}, {Silvermann}, {Taniguchi}, {Tasca},
  {Tresse}, {Trump}, \& {Zucca}}]{Salvato_2011}
{Salvato}, M., {Ilbert}, O., {Hasinger}, G., {et~al.} 2011, \apj, 742, 61,
  \dodoi{10.1088/0004-637X/742/2/61}

\bibitem[{{Schmidt} \& {Allen}(2007)}]{2007MNRAS.379..209S}
{Schmidt}, R.~W., \& {Allen}, S.~W. 2007, \mnras, 379, 209,
  \dodoi{10.1111/j.1365-2966.2007.11928.x}

\bibitem[{{Schwope} {et~al.}(2000){Schwope}, {Hasinger}, {Lehmann}, {Schwarz},
  {Brunner}, {Neizvestny}, {Ugryumov}, {Balega}, {Tr{\"u}mper}, \&
  {Voges}}]{2000AN....321....1S}
{Schwope}, A., {Hasinger}, G., {Lehmann}, I., {et~al.} 2000, Astronomische
  Nachrichten, 321, 1,
  \dodoi{10.1002/(SICI)1521-3994(200003)321:1<1::AID-ASNA1>3.0.CO;2-C}

\bibitem[{{Scodeggio} {et~al.}(2018){Scodeggio}, {Guzzo}, {Garilli}, {Granett},
  {Bolzonella}, {de la Torre}, {Abbas}, {Adami}, {Arnouts}, {Bottini}, {Cappi},
  {Coupon}, {Cucciati}, {Davidzon}, {Franzetti}, {Fritz}, {Iovino}, {Krywult},
  {Le Brun}, {Le F{\`e}vre}, {Maccagni}, {Ma{\l}ek}, {Marchetti}, {Marulli},
  {Polletta}, {Pollo}, {Tasca}, {Tojeiro}, {Vergani}, {Zanichelli}, {Bel},
  {Branchini}, {De Lucia}, {Ilbert}, {McCracken}, {Moutard}, {Peacock},
  {Zamorani}, {Burden}, {Fumana}, {Jullo}, {Marinoni}, {Mellier}, {Moscardini},
  \& {Percival}}]{2018AandA...609A..84S}
{Scodeggio}, M., {Guzzo}, L., {Garilli}, B., {et~al.} 2018, \aap, 609, A84,
  \dodoi{10.1051/0004-6361/201630114}

\bibitem[{{Sehgal} {et~al.}(2011)}]{Sehgal_2011}
{Sehgal}, N., {et~al.} 2011, ApJ, 732, 44, \dodoi{10.1088/0004-637X/732/1/44}

\bibitem[{{Sereno}(2015)}]{Sereno_2015}
{Sereno}, M. 2015, \mnras, 450, 3665, \dodoi{10.1093/mnras/stu2505}

\bibitem[{{Sharon} {et~al.}(2010){Sharon}, {Gal-Yam}, {Maoz}, {Filippenko},
  {Foley}, {Silverman}, {Ebeling}, {Ma}, {Ofek}, {Kneib}, {Donahue}, {Ellis},
  {Freedman}, {Kirshner}, {Mulchaey}, {Sarajedini}, \&
  {Voit}}]{2010ApJ...718..876S}
{Sharon}, K., {Gal-Yam}, A., {Maoz}, D., {et~al.} 2010, \apj, 718, 876,
  \dodoi{10.1088/0004-637X/718/2/876}

\bibitem[{{Shectman} {et~al.}(1996){Shectman}, {Landy}, {Oemler}, {Tucker},
  {Lin}, {Kirshner}, \& {Schechter}}]{1996ApJ...470..172S}
{Shectman}, S.~A., {Landy}, S.~D., {Oemler}, A., {et~al.} 1996, \apj, 470, 172,
  \dodoi{10.1086/177858}

\bibitem[{{Sif{\'o}n} {et~al.}(2016){Sif{\'o}n}, {Battaglia}, {Hasselfield},
  {Menanteau}, {Barrientos}, {Bond}, {Crichton}, {Devlin}, {D{\"u}nner},
  {Hilton}, {Hincks}, {Hlozek}, {Huffenberger}, {Hughes}, {Infante},
  {Kosowsky}, {Marsden}, {Marriage}, {Moodley}, {Niemack}, {Page}, {Spergel},
  {Staggs}, {Trac}, \& {Wollack}}]{2016MNRAS.461..248S}
{Sif{\'o}n}, C., {Battaglia}, N., {Hasselfield}, M., {et~al.} 2016, \mnras,
  461, 248, \dodoi{10.1093/mnras/stw1284}

\bibitem[{{Simet} {et~al.}(2017){Simet}, {McClintock}, {Mandelbaum}, {Rozo},
  {Rykoff}, {Sheldon}, \& {Wechsler}}]{Simet_2017}
{Simet}, M., {McClintock}, T., {Mandelbaum}, R., {et~al.} 2017, \mnras, 466,
  3103, \dodoi{10.1093/mnras/stw3250}

\bibitem[{{Smith} {et~al.}(2016){Smith}, {Mazzotta}, {Okabe}, {Ziparo},
  {Mulroy}, {Babul}, {Finoguenov}, {McCarthy}, {Lieu}, {Bah{\'e}}, {Bourdin},
  {Evrard}, {Futamase}, {Haines}, {Jauzac}, {Marrone}, {Martino}, {May},
  {Taylor}, \& {Umetsu}}]{Smith_2016}
{Smith}, G.~P., {Mazzotta}, P., {Okabe}, N., {et~al.} 2016, \mnras, 456, L74,
  \dodoi{10.1093/mnrasl/slv175}

\bibitem[{{Smith} {et~al.}(2004){Smith}, {Hudson}, {Nelan}, {Moore}, {Quinney},
  {Wegner}, {Lucey}, {Davies}, {Malecki}, {Schade}, \&
  {Suntzeff}}]{2004AJ....128.1558S}
{Smith}, R.~J., {Hudson}, M.~J., {Nelan}, J.~E., {et~al.} 2004, \aj, 128, 1558,
  \dodoi{10.1086/423915}

\bibitem[{{Somboonpanyakul} {et~al.}(2018){Somboonpanyakul}, {McDonald}, {Lin},
  {Stalder}, \& {Stark}}]{Somboonpanyakul_2018}
{Somboonpanyakul}, T., {McDonald}, M., {Lin}, H.~W., {Stalder}, B., \& {Stark},
  A. 2018, \apj, 863, 122, \dodoi{10.3847/1538-4357/aace55}

\bibitem[{{Song} {et~al.}(2012){Song}, {Zenteno}, {Stalder}, {Desai}, {Bleem},
  {Aird}, {Armstrong}, {Ashby}, {Bayliss}, {Bazin}, {Benson}, {Bertin},
  {Brodwin}, {Carlstrom}, {Chang}, {Cho}, {Clocchiatti}, {Crawford}, {Crites},
  {de Haan}, {Dobbs}, {Dudley}, {Foley}, {George}, {Gettings}, {Gladders},
  {Gonzalez}, {Halverson}, {Harrington}, {High}, {Holder}, {Holzapfel},
  {Hoover}, {Hrubes}, {Joy}, {Keisler}, {Knox}, {Lee}, {Leitch}, {Liu},
  {Lueker}, {Luong-Van}, {Marrone}, {McDonald}, {McMahon}, {Mehl}, {Meyer},
  {Mocanu}, {Mohr}, {Montroy}, {Natoli}, {Nurgaliev}, {Padin}, {Plagge},
  {Pryke}, {Reichardt}, {Rest}, {Ruel}, {Ruhl}, {Saliwanchik}, {Saro}, {Sayre},
  {Schaffer}, {Shaw}, {Shirokoff}, {{\v{S}}uhada}, {Spieler}, {Stanford},
  {Staniszewski}, {Stark}, {Story}, {Stubbs}, {van Engelen}, {Vanderlinde},
  {Vieira}, {Williamson}, \& {Zahn}}]{2012ApJ...761...22S}
{Song}, J., {Zenteno}, A., {Stalder}, B., {et~al.} 2012, \apj, 761, 22,
  \dodoi{10.1088/0004-637X/761/1/22}

\bibitem[{{Sonnenfeld} {et~al.}(2018){Sonnenfeld}, {Chan}, {Shu}, {More},
  {Oguri}, {Suyu}, {Wong}, {Lee}, {Coupon}, {Yonehara}, {Bolton}, {Jaelani},
  {Tanaka}, {Miyazaki}, \& {Komiyama}}]{Sonnenfeld_2018}
{Sonnenfeld}, A., {Chan}, J. H.~H., {Shu}, Y., {et~al.} 2018, \pasj, 70, S29,
  \dodoi{10.1093/pasj/psx062}

\bibitem[{{Stanford} {et~al.}(2014){Stanford}, {Gonzalez}, {Brodwin},
  {Gettings}, {Eisenhardt}, {Stern}, \& {Wylezalek}}]{2014ApJS..213...25S}
{Stanford}, S.~A., {Gonzalez}, A.~H., {Brodwin}, M., {et~al.} 2014, \apjs, 213,
  25, \dodoi{10.1088/0067-0049/213/2/25}

\bibitem[{{Staniszewski} {et~al.}(2009)}]{Staniszewski_2009}
{Staniszewski}, Z., {et~al.} 2009, ApJ, 701, 32,
  \dodoi{10.1088/0004-637X/701/1/32}

\bibitem[{{Stein} {et~al.}(2020){Stein}, {Alvarez}, {Bond}, {van Engelen}, \&
  {Battaglia}}]{Stein_2020}
{Stein}, G., {Alvarez}, M.~A., {Bond}, J.~R., {van Engelen}, A., \&
  {Battaglia}, N. 2020, arXiv e-prints, arXiv:2001.08787.
\newblock \doarXiv{2001.08787}

\bibitem[{{Stocke} {et~al.}(1991){Stocke}, {Morris}, {Gioia}, {Maccacaro},
  {Schild}, {Wolter}, {Fleming}, \& {Henry}}]{1991ApJS...76..813S}
{Stocke}, J.~T., {Morris}, S.~L., {Gioia}, I.~M., {et~al.} 1991, \apjs, 76,
  813, \dodoi{10.1086/191582}

\bibitem[{{Struble} \& {Rood}(1991)}]{1991ApJS...77..363S}
{Struble}, M.~F., \& {Rood}, H.~J. 1991, \apjs, 77, 363, \dodoi{10.1086/191608}

\bibitem[{{Struble} \& {Rood}(1999)}]{1999ApJS..125...35S}
---. 1999, \apjs, 125, 35, \dodoi{10.1086/313274}

\bibitem[{{Sunyaev} \& {Zel'dovich}(1970)}]{SZ_1970}
{Sunyaev}, R.~A., \& {Zel'dovich}, Y.~B. 1970, Comments on Astrophysics and
  Space Physics, 2, 66

\bibitem[{{Sunyaev} \& {Zeldovich}(1972)}]{SZ_1972}
{Sunyaev}, R.~A., \& {Zeldovich}, Y.~B. 1972, Comments on Astrophysics and
  Space Physics, 4, 173

\bibitem[{{Swetz} {et~al.}(2011)}]{Swetz_2011}
{Swetz}, D.~S., {et~al.} 2011, ApJS, 194, 41,
  \dodoi{10.1088/0067-0049/194/2/41}

\bibitem[{{Thornton} {et~al.}(2016){Thornton}, {Ade}, {Aiola}, {Angil{\`e}},
  {Amiri}, {Beall}, {Becker}, {Cho}, {Choi}, {Corlies}, {Coughlin}, {Datta},
  {Devlin}, {Dicker}, {D{\"u}nner}, {Fowler}, {Fox}, {Gallardo}, {Gao},
  {Grace}, {Halpern}, {Hasselfield}, {Henderson}, {Hilton}, {Hincks}, {Ho},
  {Hubmayr}, {Irwin}, {Klein}, {Koopman}, {Li}, {Louis}, {Lungu}, {Maurin},
  {McMahon}, {Munson}, {Naess}, {Nati}, {Newburgh}, {Nibarger}, {Niemack},
  {Niraula}, {Nolta}, {Page}, {Pappas}, {Schillaci}, {Schmitt}, {Sehgal},
  {Sievers}, {Simon}, {Staggs}, {Tucker}, {Uehara}, {van Lanen}, {Ward}, \&
  {Wollack}}]{Thornton_2016}
{Thornton}, R.~J., {Ade}, P.~A.~R., {Aiola}, S., {et~al.} 2016, \apjs, 227, 21,
  \dodoi{10.3847/1538-4365/227/2/21}

\bibitem[{{Tinker} {et~al.}(2008){Tinker}, {Kravtsov}, {Klypin}, {Abazajian},
  {Warren}, {Yepes}, {Gottl{\"o}ber}, \& {Holz}}]{Tinker_2008}
{Tinker}, J., {Kravtsov}, A.~V., {Klypin}, A., {et~al.} 2008, \apj, 688, 709,
  \dodoi{10.1086/591439}

\bibitem[{{Tucker} {et~al.}(1998){Tucker}, {Blanco}, {Rappoport}, {David},
  {Fabricant}, {Falco}, {Forman}, {Dressler}, \&
  {Ramella}}]{1998ApJ...496L...5T}
{Tucker}, W., {Blanco}, P., {Rappoport}, S., {et~al.} 1998, \apjl, 496, L5,
  \dodoi{10.1086/311234}

\bibitem[{{Valtchanov} {et~al.}(2004){Valtchanov}, {Pierre}, {Willis}, {Dos
  Santos}, {Jones}, {Andreon}, {Adami}, {Altieri}, {Bolzonella}, {Bremer},
  {Duc}, {Gosset}, {Jean}, \& {Surdej}}]{2004AandA...423...75V}
{Valtchanov}, I., {Pierre}, M., {Willis}, J., {et~al.} 2004, \aap, 423, 75,
  \dodoi{10.1051/0004-6361:20040162}

\bibitem[{{Vanderlinde} {et~al.}(2010)}]{Vanderlinde_2010}
{Vanderlinde}, K., {et~al.} 2010, ApJ, 722, 1180,
  \dodoi{10.1088/0004-637X/722/2/1180}

\bibitem[{{Vikhlinin} {et~al.}(1998){Vikhlinin}, {McNamara}, {Forman}, {Jones},
  {Quintana}, \& {Hornstrup}}]{Vikhlinin_1998}
{Vikhlinin}, A., {McNamara}, B.~R., {Forman}, W., {et~al.} 1998, \apj, 502,
  558, \dodoi{10.1086/305951}

\bibitem[{{von der Linden} {et~al.}(2014){von der Linden}, {Mantz}, {Allen},
  {Applegate}, {Kelly}, {Morris}, {Wright}, {Allen}, {Burchat}, {Burke},
  {Donovan}, \& {Ebeling}}]{vonDerLinden_2014}
{von der Linden}, A., {Mantz}, A., {Allen}, S.~W., {et~al.} 2014, \mnras, 443,
  1973, \dodoi{10.1093/mnras/stu1423}

\bibitem[{{Wen} \& {Han}(2015)}]{WH_2015}
{Wen}, Z.~L., \& {Han}, J.~L. 2015, \apj, 807, 178,
  \dodoi{10.1088/0004-637X/807/2/178}

\bibitem[{{Wen} {et~al.}(2012){Wen}, {Han}, \& {Liu}}]{WHL_2012}
{Wen}, Z.~L., {Han}, J.~L., \& {Liu}, F.~S. 2012, \apjs, 199, 34,
  \dodoi{10.1088/0067-0049/199/2/34}

\bibitem[{{White}(2000)}]{2000MNRAS.312..663W}
{White}, D.~A. 2000, \mnras, 312, 663, \dodoi{10.1046/j.1365-8711.2000.03163.x}

\bibitem[{{Williamson} {et~al.}(2011){Williamson}, {Benson}, {High}, {Vand
  erlinde}, {Ade}, {Aird}, {Andersson}, {Armstrong}, {Ashby}, {Bautz}, {Bazin},
  {Bertin}, {Bleem}, {Bonamente}, {Brodwin}, {Carlstrom}, {Chang}, {Chapman},
  {Clocchiatti}, {Crawford}, {Crites}, {de Haan}, {Desai}, {Dobbs}, {Dudley},
  {Fazio}, {Foley}, {Forman}, {Garmire}, {George}, {Gladders}, {Gonzalez},
  {Halverson}, {Holder}, {Holzapfel}, {Hoover}, {Hrubes}, {Jones}, {Joy},
  {Keisler}, {Knox}, {Lee}, {Leitch}, {Lueker}, {Luong-Van}, {Marrone},
  {McMahon}, {Mehl}, {Meyer}, {Mohr}, {Montroy}, {Murray}, {Padin}, {Plagge},
  {Pryke}, {Reichardt}, {Rest}, {Ruel}, {Ruhl}, {Saliwanchik}, {Saro},
  {Schaffer}, {Shaw}, {Shirokoff}, {Song}, {Spieler}, {Stalder}, {Stanford},
  {Staniszewski}, {Stark}, {Story}, {Stubbs}, {Vieira}, {Vikhlinin}, \&
  {Zenteno}}]{Williamson_2011}
{Williamson}, R., {Benson}, B.~A., {High}, F.~W., {et~al.} 2011, \apj, 738,
  139, \dodoi{10.1088/0004-637X/738/2/139}

\bibitem[{{Willis} {et~al.}(2013{\natexlab{a}}){Willis}, {Clerc}, {Bremer},
  {Pierre}, {Adami}, {Ilbert}, {Maughan}, {Maurogordato}, {Pacaud},
  {Valtchanov}, {Chiappetti}, {Thanjavur}, {Gwyn}, {Stanway}, \&
  {Winkworth}}]{2013MNRAS.430..134W}
{Willis}, J.~P., {Clerc}, N., {Bremer}, M.~N., {et~al.} 2013{\natexlab{a}},
  \mnras, 430, 134, \dodoi{10.1093/mnras/sts540}

\bibitem[{{Willis} {et~al.}(2013{\natexlab{b}}){Willis}, {Clerc}, {Bremer},
  {Pierre}, {Adami}, {Ilbert}, {Maughan}, {Maurogordato}, {Pacaud},
  {Valtchanov}, {Chiappetti}, {Thanjavur}, {Gwyn}, {Stanway}, \&
  {Winkworth}}]{Willis_2013}
---. 2013{\natexlab{b}}, \mnras, 430, 134, \dodoi{10.1093/mnras/sts540}

\bibitem[{{Wong} {et~al.}(2018){Wong}, {Sonnenfeld}, {Chan}, {Rusu}, {Tanaka},
  {Jaelani}, {Lee}, {More}, {Oguri}, {Suyu}, \& {Komiyama}}]{Wong_2018}
{Wong}, K.~C., {Sonnenfeld}, A., {Chan}, J. H.~H., {et~al.} 2018, \apj, 867,
  107, \dodoi{10.3847/1538-4357/aae381}

\bibitem[{{Wright} {et~al.}(2019){Wright}, {Hildebrandt}, {Kuijken}, {Erben},
  {Blake}, {Buddelmeijer}, {Choi}, {Cross}, {de Jong}, {Edge},
  {Gonzalez-Fernandez}, {Gonz{\'a}lez Solares}, {Grado}, {Heymans}, {Irwin},
  {Kupcu Yoldas}, {Lewis}, {Mann}, {Napolitano}, {Radovich}, {Schneider},
  {Sif{\'o}n}, {Sutherland}, {Sutorius}, \& {Verdoes Kleijn}}]{Wright_2019}
{Wright}, A.~H., {Hildebrandt}, H., {Kuijken}, K., {et~al.} 2019, \aap, 632,
  A34, \dodoi{10.1051/0004-6361/201834879}

\bibitem[{{Wright} {et~al.}(2010){Wright}, {Eisenhardt}, {Mainzer}, {Ressler},
  {Cutri}, {Jarrett}, {Kirkpatrick}, {Padgett}, {McMillan}, {Skrutskie},
  {Stanford}, {Cohen}, {Walker}, {Mather}, {Leisawitz}, {Gautier}, {McLean},
  {Benford}, {Lonsdale}, {Blain}, {Mendez}, {Irace}, {Duval}, {Liu}, {Royer},
  {Heinrichsen}, {Howard}, {Shannon}, {Kendall}, {Walsh}, {Larsen}, {Cardon},
  {Schick}, {Schwalm}, {Abid}, {Fabinsky}, {Naes}, \& {Tsai}}]{Wright_2010}
{Wright}, E.~L., {Eisenhardt}, P. R.~M., {Mainzer}, A.~K., {et~al.} 2010, \aj,
  140, 1868, \dodoi{10.1088/0004-6256/140/6/1868}

\bibitem[{{Wuyts} {et~al.}(2010){Wuyts}, {Barrientos}, {Gladders}, {Sharon},
  {Bayliss}, {Carrasco}, {Gilbank}, {Yee}, {Koester}, \&
  {Mu{\~n}oz}}]{2010ApJ...724.1182W}
{Wuyts}, E., {Barrientos}, L.~F., {Gladders}, M.~D., {et~al.} 2010, \apj, 724,
  1182, \dodoi{10.1088/0004-637X/724/2/1182}

\bibitem[{{Zaritsky} {et~al.}(2006){Zaritsky}, {Gonzalez}, \&
  {Zabludoff}}]{2006ApJ...638..725Z}
{Zaritsky}, D., {Gonzalez}, A.~H., \& {Zabludoff}, A.~I. 2006, \apj, 638, 725,
  \dodoi{10.1086/498672}

\end{thebibliography}

\end{document}